\documentclass[preprint,aps ,nofootinbib,UTF8]{revtex4}
\usepackage{graphicx}
\pdfoutput=1
\usepackage{epsfig}
\usepackage{amsmath}
\usepackage{amsfonts}
\usepackage{amssymb}
\usepackage{color}
\usepackage{dcolumn}
\usepackage{hyperref}
\usepackage{multirow}
\usepackage{booktabs}
\usepackage{subfigure}
\usepackage{CJKutf8}
\usepackage{mathrsfs}
\usepackage{graphics}
\usepackage{float}
\usepackage{indentfirst}
\providecommand{\U}[1]{\protect\rule{.1in}{.1in}}
\newcolumntype{L}{>{$}l<{$}}
\newcolumntype{R}{>{$}r<{$}}

\newcommand{\f}{\begin{equation}}
\newcommand{\ff}{\end{equation}}
\newcommand{\fa}{\begin{eqnarray}}
\newcommand{\ffa}{\end{eqnarray}}

\begin{document}
\title{Accretion Disk For Regular Black Holes with sub-Planckian Curvature}
\author{Wei Zeng$^{1}$}
\email{cengwei0702@stu.cwnu.edu.cn} 
\author{Yi Ling $^{2,3,1}$}
\email{lingy@ihep.ac.cn}
\author{Qing-Quan Jiang$^{1}$}
\email{qqjiangphys@yeah.net}
\author{Guo-Ping Li$^{1}$}
\email{gpliphys@yeah.net} \affiliation{$^1$ School of Physics and Astronomy, China West Normal University, Nanchong 637002, China \\$^2$ Institute of High Energy
Physics, Chinese Academy of Sciences, Beijing 100049, China\\$^3$
School of Physics, University of Chinese Academy of Sciences,
Beijing 100049, China}
\begin{abstract}  
    We investigate the accretion disk for a sort of regular black holes which are characterized by sub-Planckian curvature and Minkowskian core. We derive null geodesics outside the horizon of such regular black holes and analyze the feature of the light rays from the accretion disk which can be classified into direct emission, lensed rings, and photon rings. We find that the observed brightness under different emission models is mainly determined by direct emission, while the contribution from the flux of the lensed and photon rings is limited. By comparing with Bardeen black hole with a dS core, it is found that 
    the black hole with a Minkowskian core 
    exhibits distinct astronomical optical features when surrounded by accretion disk, which potentially provides a way to distinguish these two sorts of black holes by astronomical observation.
\end{abstract}
\maketitle
\section{Introduction}
Black hole is a celestial body predicted by general relativity, but suffers from the notorious singularity problem, which states that  the scalar curvature becomes divergent at the center of the black hole\cite{Hawking:1966sx,Penrose:1964wq,Joshi:2011rlc,Goswami:2005fu,Janis:1968zz}. In particular, once the quantum effect of matter is taken into account, the black hole can evaporate by Hawking radiation  and ultimately leads to the information loss paradox\cite{Hawking:1976ra,Giddings:1992hh,Hawking:1975vcx,Preskill:1992tc,Xiang:2013sza,Chen:2014jwq,Casadio:2000py,Unruh:1976db,Page:1993wv}. 
People widely believe that completely solving these problems and understanding the involved puzzles rely on a complete theory of quantum gravity\cite{Garay:1994en,tHooft:1984kcu,Han:2004wt,Donoghue:1993eb,Callan:1992rs,DeWitt:1967yk,DeWitt:1967ub,DeWitt:1967uc,Calmet:2017qqa,Ali:2015tva}. Before such a theory can be established, one way to walk around the singularity theorem is that one could attempt to construct the regular black holes without singularity at the phenomenological level by introducing some exotic matter which violates the ordinary energy conditions or the quantum correction to spacetime geometry. Such exotic matter or quantum geometry could result from some nonlinear electrodynamics in classical theory, or from the consideration of quantum gravity effects. For recent review, we refer to \cite{Ashtekar:2023cod,Lan:2023cvz}. Traditionally, Bardeen black hole, Hayward black hole, and Frolov black hole are some prominent examples of regular black holes which are singularity free\cite{Bardeen:1968,Hayward:2005gi,Frolov:2014jva}. They are characterized by a de-Sitter core at the center of black hole. Recently, a new class of regular black holes  was constructed in \cite{Ling:2021olm} based on the work in \cite{Xiang:2013sza}. These black holes contain an exponentially suppressing Newton potential such that the spacetime near the center of the black hole is Minkowskian rather than  de-Sitter(dS). In addition, their Kretschmann scalar curvature is sub-Planckian everywhere. In literature, other regular black holes with Minkowskian core can be found, for instance in \cite{Culetu:2013fsa,Culetu:2014lca,Rodrigues:2015ayd,Simpson:2019mud,Ghosh:2014pba,Ghosh:2018bxg,Berry:2020ntz,Ling:2022vrv,Zeng:2022yrm,LingLingYi:2021rfn}. 
The black hole with dS core and the black hole with Minkowskian core  exhibit similar asymptotic behavior at infinity and a one-to-one correspondence between these two sorts of black holes is established in \cite{Ling:2021olm}. 

With the development of astronomical observation, one may wonder if we could distinguish these two different black holes by investigating the light rays emitted from the accretion disk surrounding black holes. In particular, after the Event Horizon Telescope (EHT) collaboration has successfully captured the images of M87* black hole and the supermassive black hole at the center of galaxy Sgr A*, one is facing the great opportunity of testing the theory of black holes with observational data\cite{EventHorizonTelescope:2019dse}\cite{EventHorizonTelescope:2022xnr}. These two images are the direct evidence for the existence of black holes, and exhibit a common feature: a black disk at the center surrounded by bright rings. Usually the region of black disk is called the black hole shadow, representing the region where light rays radiated from the accretion matter are completely absorbed by the black hole and thus can not be received by the observer at infinity, so it is an absolute dark region. If light rays can escape to infinity after orbiting the black hole several times under its strong gravity, then it has a chance to be detected by observers, and these rays form the bright rings in the image of a black hole\cite{Luminet:1979nyg,Younsi:2021dxe,Zeng:2020dco,Zeng:2020vsj,Peng:2020wun,Saurabh:2020zqg,Li:2021ypw,Li:2021riw,Hu:2022lek}. The marker that distinguishes these two types of light rays is the critical curve called the photon sphere, which corresponds to an orbital radius called the critical impact parameter. For photons with orbital radius larger than the critical impact parameter, they may escape to infinity, corresponding to the light captured by the observer, while for photons with orbital radius smaller than the critical impact parameter, they are completely absorbed by the black hole\cite{Claudel:2000yi,Boonserm:2018orb,Berry:2020ntz,Decanini:2009mu,Stefanov:2010xz,Wei:2011zw,Wei:2013mda,Cunha:2018acu}. In particular, for Schwarzschild black hole with mass $M$, the radius of photon sphere is $3M$ and the corresponding critical impact parameter is $3\sqrt{3}M$ \cite{Gralla:2019xty}.
\par
The observational characteristics of black holes is an important clue to the nature of black holes. In this paper  we intend to investigate the shadow of regular black holes with Minkowskian core and compare it with the shadow of black holes with dS core. In \cite{Zeng:2022yrm}, we have investigated their photon spheres and the marginally stable circular orbits, attempting to distinguish these two types of black holes based on these observables. It is found that the difference between these two sorts of regular black holes is limited within the black hole phase, but is more pronounced within the phase of compact massive object (CMO). In \cite{Ling:2022vrv}, the regular black hole solution with Minkowskian core has been extended to the rotating spacetime and the shape of the shadow is studied.
It is found that the black hole with sub-Planckian curvature has a larger shadow deformation and a smaller shadow region in comparison with regular black holes with dS core. In \cite{He:2021htq}, the shadow of Bardeen black hole with different accretion matter is investigated, and in \cite{Guo:2021bhr}, the shadow of Hayward black hole illuminated by different accretion matter is investigated. The above work demonstrate the observed phenomena of a regular black hole with dS core. In this paper, we intend to explore the shadow and optical features of regular black holes with Minkowskian core after the accretion disk is considered as a radiation source and try to compare them with those obtained for regular black holes with dS core. \par
The paper is organized as follows. In next section, we will discuss the null geodesics around a specific black holes with Minkowskian core and sub-Planckian curvature, and analyze the behavior of light rays around such black holes by using ray tracing code, after which we consider the optical characteristics under three emission models. In section three, we will compare the images of two different regular black holes by investigating the behavior of light rays over Bardeen black hole with dS core and over a black hole with Minkowskian core with the same deviation parameter. Our conclusions and discussions are presented in section four.

\section{The regular black holes with $x=2/3$ and $n=2$}
We start with the spherically symmetric spacetime and suppose the metric of black hole has the following general form
\begin{equation}\label{Eq.metric}
    ds^{2}=-f(r)dt^2+f(r)dr^2+r^2 \left(d\theta^2+\sin^2\theta d\phi^2\right).
\end{equation}
In Ref.\cite{Ling:2021olm}, a new sort of regular black holes with sub-Planckain curvature  is proposed with an exponentially suppressing Newton potential
\begin{equation}\label{Eq.NP}
    f(r)=1+2\psi(r)=1-\frac{2M}{r}e^{\frac{-\alpha_0 M^x}{r^n}},
\end{equation}
where $M$ is the mass of the black hole and $\alpha_0$ is a dimensionless deviation parameter arising from the quantum effects of gravity and characterizes the modification of the standard Heisenberg uncertainty principle due to quantum gravity.  Obviously, as $r \rightarrow 0$, the spacetime is characterized by a Minkowskian core. It has also been shown in Ref.\cite{Ling:2021olm} that when two factors satisfy $n \geq x \geq n/3$ and $n\geq 2$, then the spacetime is a regular black hole with sub-Planckian curvature, independent of the black hole mass $M$. 

Without loss of generality, in this paper we will focus on a specific metric of black hole with $x=2/3$ and $n=2$
\begin{equation}\label{Eq.Min}
    \psi(r)=-\frac{M}{r}e^{\frac{-\alpha_0 M^{2/3}}{r^2}}.
\end{equation}
In \cite{Zeng:2022yrm}, it is found that to guarantee the existence of the horizon, the deviation parameter $\alpha_0$ should fall into the interval  $0 \leq \alpha_0 \leq 0.73$. \par

Next we consider  the dynamics of null massless particles around the black hole. 
The motion of photons satisfies the Euler-Lagrangian equation,
\begin{equation}
    \frac{d}{d\lambda}\left(\frac{\partial\mathscr{L} }{\partial \dot{x}^{\mu}} \right)= \frac{\partial \mathscr{L}}{\partial x^{\mu}},
\end{equation}
where $\lambda$ is an affine parameter and $\dot{x}^{\mu}$ is the four-velocity of the photon. The Lagrangian density $\mathscr{L}$ is given by
\begin{equation}\label{Eq.L}
    \mathscr{L}=-\frac{1}{2}g_{\mu \nu} \frac{dx^\mu}{d\lambda} \frac{dx^\nu}{d\lambda}=-\frac{1}{2}\left(-f(r) \dot{t}^2 +f^{-1}(r) \dot{r}^2 +r^2\left(\dot{\theta}^2+\sin^2\theta\dot{\phi}^2\right) \right).
\end{equation}
For photons, $\mathscr{L}=0$. Without loss of generality  we only consider the motion of photons on the equatorial plane by setting $\theta=0$ and $\ddot{\theta}=0$. Since there are two Killing vector fields in this spacetime, namely $\frac{\partial}{\partial t}$ and $\frac{\partial}{\partial \phi}$, one obtains two conserved quantities for the trajectory of light, namely the energy $E$ and the angular momentum $L$, which are separately given by
\begin{equation}
    E=f(r)\left(\frac{dt}{d\lambda} \right);\quad\quad\quad L=r^2 \left(\frac{d\phi}{d\lambda}\right).
\end{equation}
\par In this spherically symmetric spacetime, the four-velocity is given by 
\begin{subequations}
    \begin{align}
     &\frac{dt}{d\lambda}=\frac{1}{b} \left(1-\frac{2M}{r} e^{\frac{-\alpha_{0}M^{2/3}}{r^2}}\right) 
     \label{eqn-6a} \\
     &\frac{d\phi}{d\lambda}=\pm \frac{1}{r^2}\label{eqn-6b} \\
     &\frac{dr}{d\lambda}=\sqrt{\frac{1}{b^2}-\frac{1}{r^2}\left(1-\frac{2M}{r} e^{\frac{-\alpha_{0}M^{2/3}}{r^2}}\right)}\label{eqn-6c},
    \end{align}
\end{subequations}
where  ``$\pm$'' corresponds the directions of photon on the equatorial plane (clockwise``$+$'', counterclockwise``$-$''). $b$ is the impact parameter, defined by
\begin{equation}
    b=\frac{\mid L \mid}{E}=\frac{r^2 \dot{\phi}}{f(r)\dot{t}}.
\end{equation}
Furthermore, we define effect potential $V_{eff}$ by
\begin{equation}
    V_{eff}=\frac{1}{r^2}\left(1-\frac{2M}{r}e^{\frac{-\alpha_{0}M^{2/3}}{r^2}} \right),
\end{equation}
then Eq.(\ref{eqn-6c}) can be rewritten as
\begin{equation}
    \dot{r}^2=\frac{1}{b^2}-V_{eff}.
\end{equation}
Obviously, if photons take a circular motion and form a photon sphere, then they must satisfy the following conditions
\begin{equation}\label{Eq.EP}
   V'_{eff}(r_c)=0;\quad\quad\quad V_{eff}(r_{c})=\frac{1}{b^2_c},
\end{equation}
where $r_{c}$ locates the radius of the photon sphere, and $b_{c}$ is the corresponding impact parameter. For an observer at infinity, $b_c$ indicates the radius of black hole shadow that can be observed.
\par 
One can obtain numerical results for $r_c$ and $b_c$ corresponding to different values of $\alpha_0$, as we illustrate in table.\ref{PR}.
\begin{table}[htbp]
\centering
\caption{The locations of the outer horizon ($r_h$), photon sphere ($r_c$), and impact parameter ($b_c$) with the variation of the deviation parameter $\alpha_0$ ($M=1$).}  
\label{PR}
    \begin{tabular}{cccccc}\hline 
$\alpha_0$&\quad\quad\quad $r_{h}$&\quad\quad\quad $r_c$ & \quad\quad\quad$b_c$   \\ \hline
0 & \quad\quad\quad2.00000 & \quad\quad\quad3.00000 & \quad\quad\quad5.19615  \\ \hline
$0.1$& \quad\quad\quad1.94798 &\quad\quad\quad2.94273 &\quad\quad\quad5.13696 \\ \hline
$0.3$& \quad\quad\quad1.82833 &\quad\quad\quad2.81574&\quad\quad\quad5.00833\\ \hline
$0.5$&\quad\quad\quad1.67271&\quad\quad\quad2.66478&\quad\quad\quad4.86117\\ \hline
\end{tabular}
\end{table}
We notice that when $\alpha_0=0$, it is the standard  results of Schwarzschild black hole. While with the increase of  $\alpha_0$, both $r_c$ and $b_c$ become smaller.  
\par 
\subsection{Light ray around the black hole with $x=2/3$ and $n=2$}
Now we are concerned with the motion of photons around the black hole, which can be described by the change of the radial coordinate with the azimuthal angle $\phi$. By setting $u=1/r$, one can combine Eqs.(\ref{eqn-6b}) and (\ref{eqn-6c}) conveniently and obtain
\begin{equation}
    \frac{du}{d\phi}=\sqrt{\frac{1}{b^2}-u^2 \left(1-2Mue^{-u^2 \alpha_0 M^{2/3}}\right)},
\end{equation}
The azimuthal angle $\phi$ can be integrated out as
\begin{equation}\label{angle}
    \phi=\int \frac{du}{\sqrt{\frac{1}{b^2}-u^2 \left(1-2Mue^{-u^2 \alpha_0 M^{2/3}}\right)}}.
\end{equation} 
Suppose that a thin disk is located in the equatorial plane and the observer faces the north pole of black hole. From \cite{Gralla:2019xty}, the trajectory of light rays near the black hole can be classified into three types according to the number of times that it passes through the thin disk, which in general can be evaluated by the number of occurrences $n$
\begin{equation}\label{Eq.TM}
    n(b)=\frac{\phi}{2\pi}.
\end{equation}
Explicitly, the trajectory of light rays can be divided into following three types:
\begin{itemize}
    \item [1)]
    Direct Emission $(n<3/4)$: The light trajectories intersect the thin accretion disk just once. 
    \item [2)]
    Lensed Ring $(3/4<n<5/4)$: The light trajectories intersect the thin accretion disk twice.  
    \item [3)]
    Photon Ring $(n>5/4)$: The light trajectories intersect the thin accretion disk at least three times.
\end{itemize}
\par
\begin{figure*}
	\centering
	\subfigure[$\alpha_0=0.1$]{
		\begin{minipage}[t]{0.33\linewidth}
			\centering
			\includegraphics[width=2.1in]{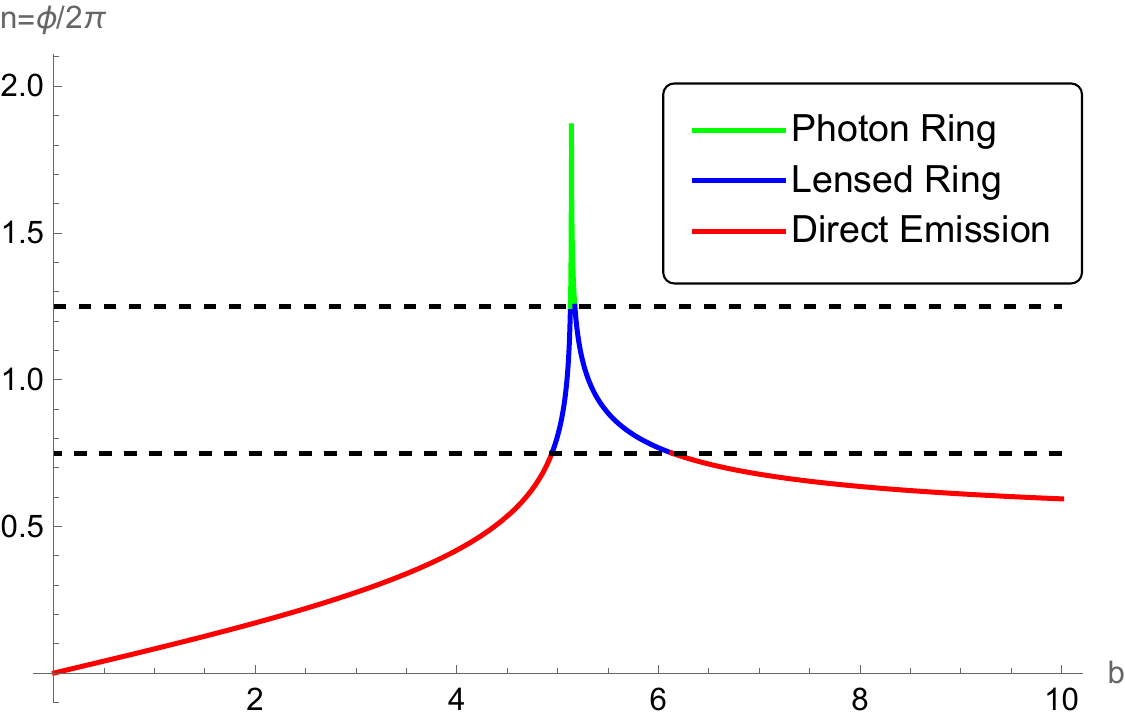}\\
			\vspace{0.1cm}
			\includegraphics[width=2.1in]{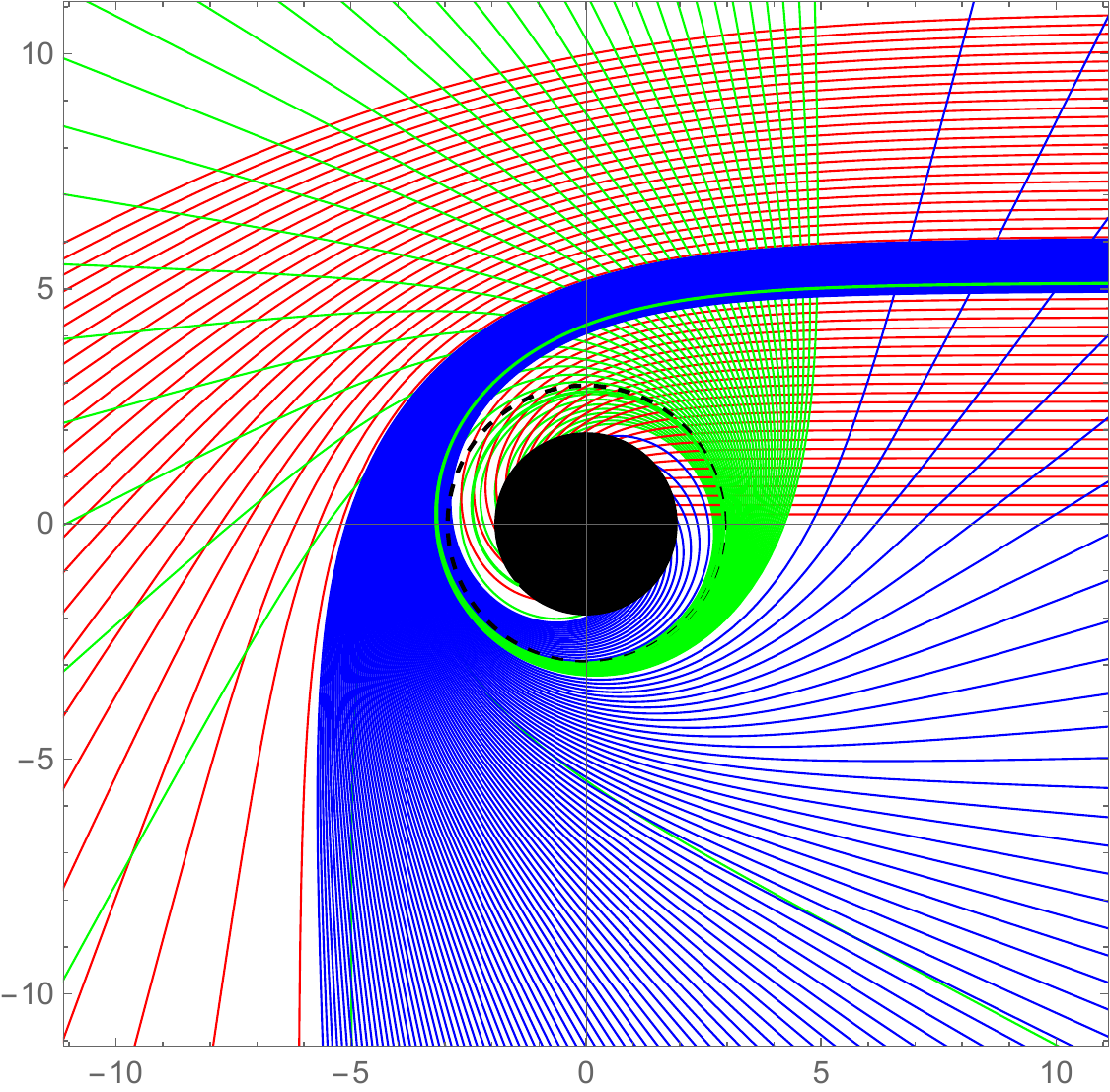}\\
			\vspace{0.1cm}
		\end{minipage}%
	}%
	\subfigure[$\alpha_0=0.3$]{
		\begin{minipage}[t]{0.33\linewidth}
			\centering
			\includegraphics[width=2.1in]{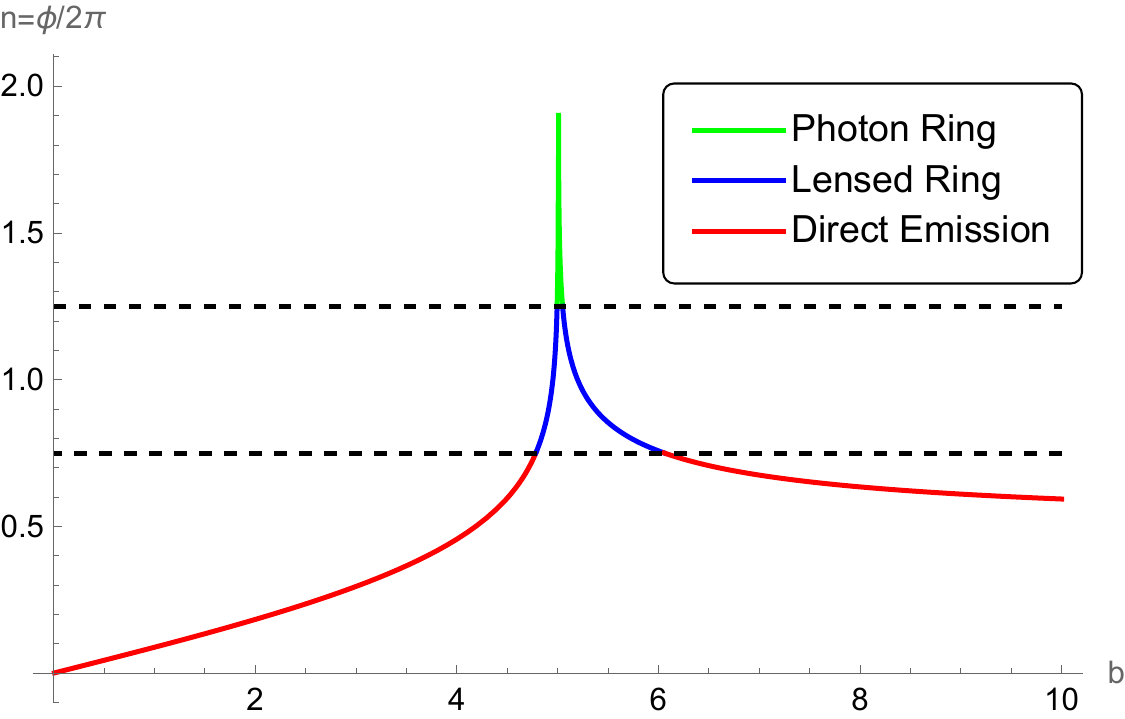}\\
			\vspace{0.1cm}
			\includegraphics[width=2.1in]{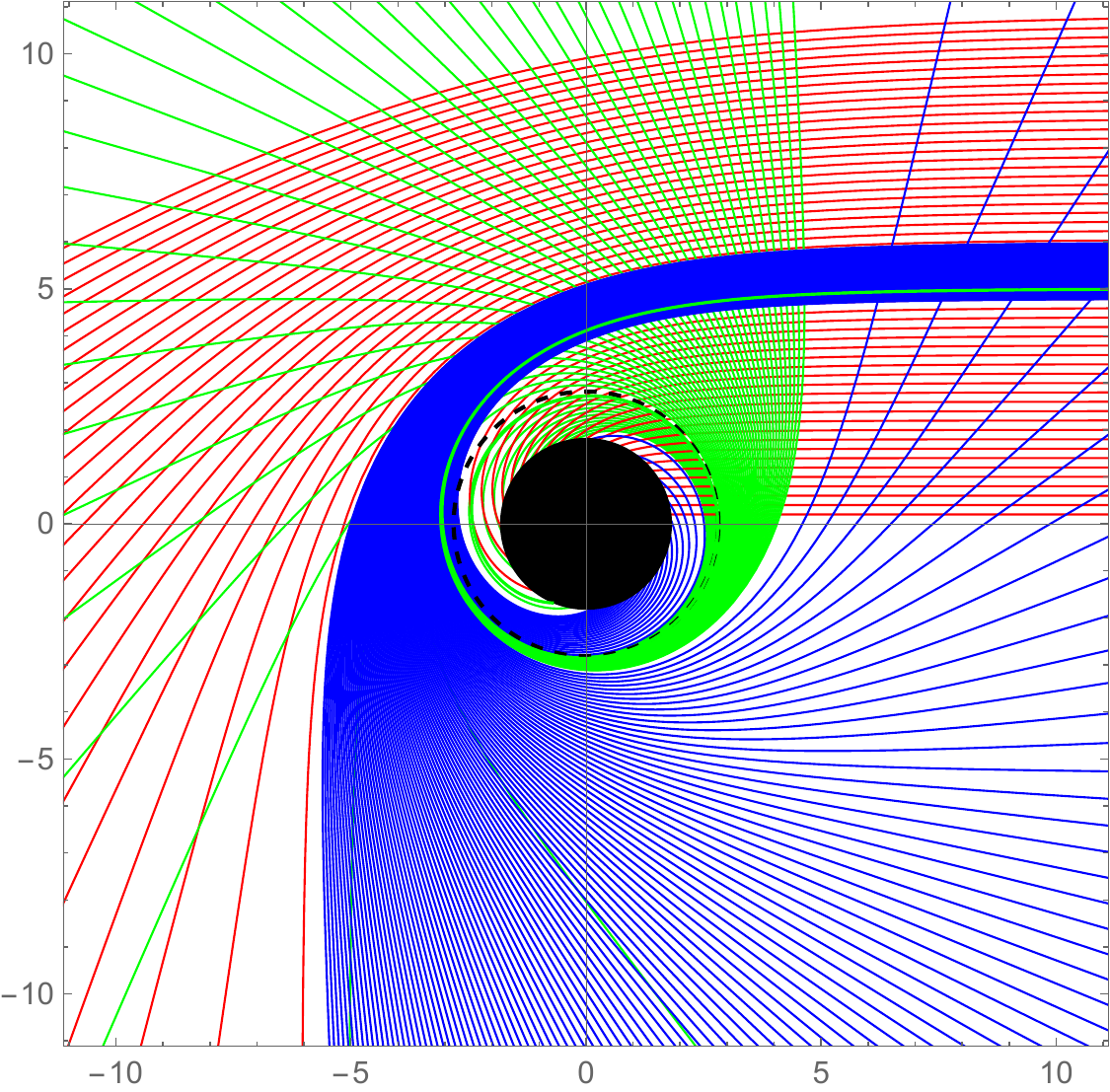}\\
			\vspace{0.1cm}
		\end{minipage}%
	}%
	\subfigure[$\alpha_0=0.5$]{
		\begin{minipage}[t]{0.33\linewidth}
			\centering
			\includegraphics[width=2.1in]{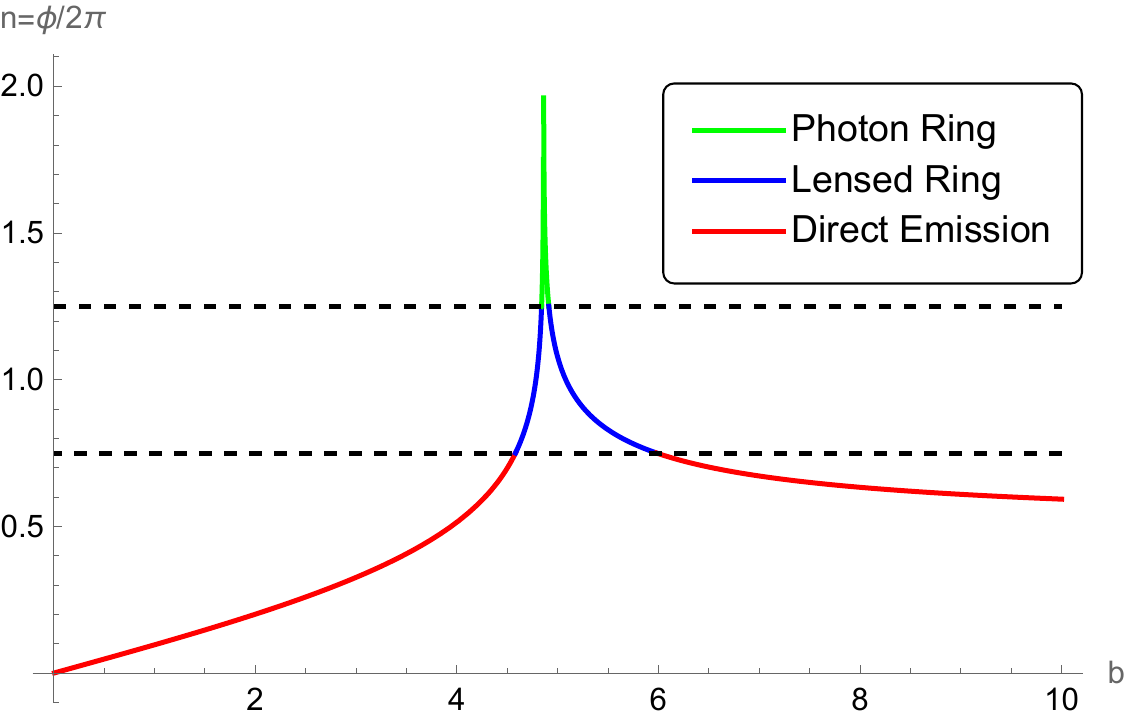}\\
			\vspace{0.1cm}
			\includegraphics[width=2.1in]{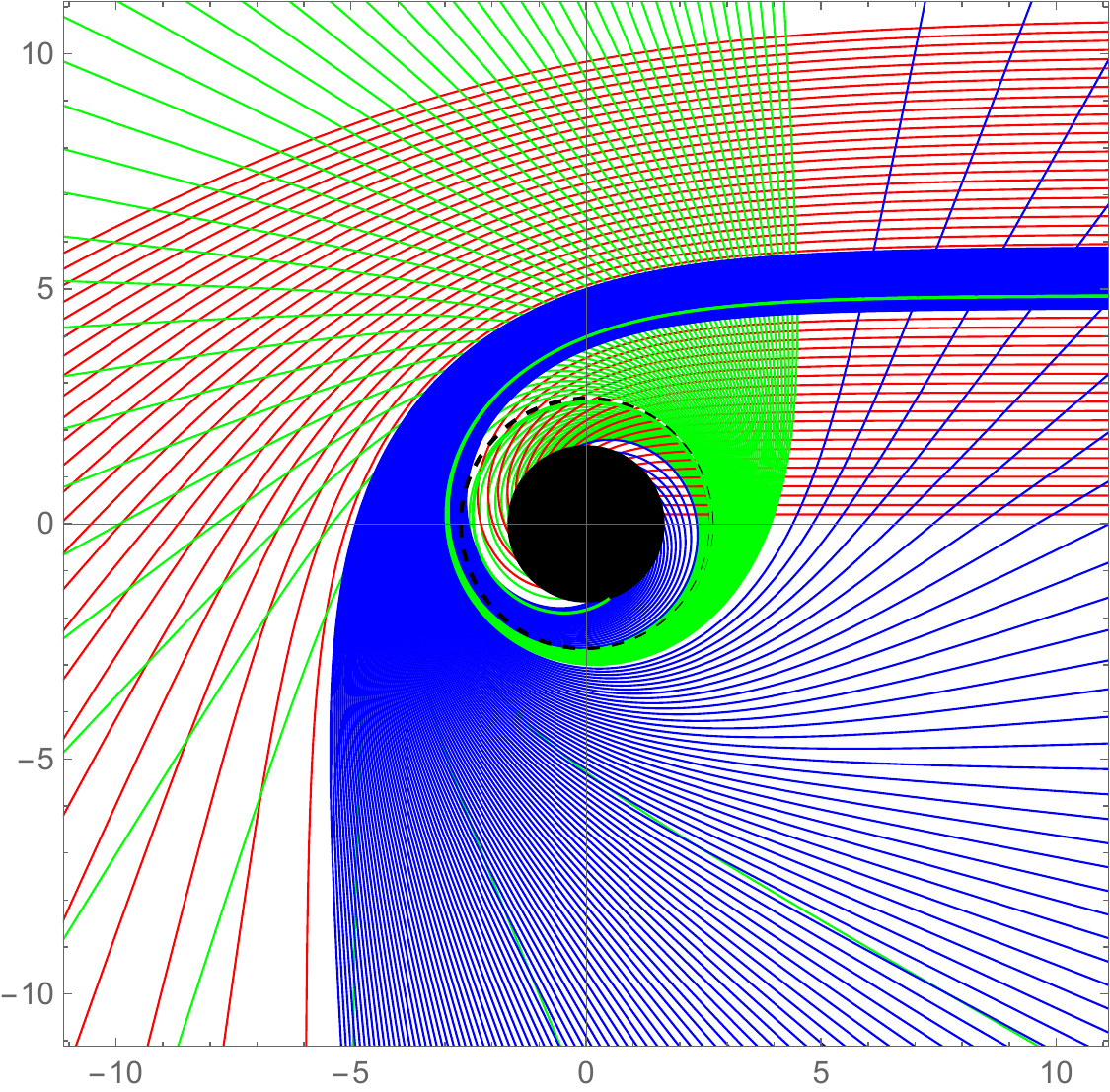}\\
			\vspace{0.1cm}
		\end{minipage}
	}
	\centering
	\caption{The plots on the top shows the relation   between the orbit number $n$ and impact parameter $b$ for different $\alpha_0$ with $M=1$, while the plots at the bottom demonstrate the trajectory of light rays. The  curves in red, blue and green correspond to the direct emission, the lensed ring and the photon ring, respectively. The black hole region is marked as the black solid disk, and the black dashed curve denotes the photon sphere. }
	\vspace{-0.2cm}
	\label{fig.PH}
\end{figure*}

We illustrate the range of the impact parameter $b$ which gives rise to the direct emission, lensed ring emission and photon ring emission respectively for various $\alpha_0$  in Table (\ref{LR}). In order to distinguish these three types of light trajectories more clearly, we adopt the polar coordinate $(b,\phi)$ and employ ray tracing code to plot the light trajectories around the black hole, which is illustrated in Fig. (\ref{fig.PH}). During this process, one needs to determine the turning point $u_0$ of light trajectory which is given by
 \begin{equation}\label{Eq.Turn}
     \frac{1}{b^2}-u_{0}^2(1-2Mu_{0}e^{-{u_0}^2 \alpha_{0} M^{2/3}})=0.
 \end{equation}
 In general one has no analytical solution to  
 Eq.(\ref{Eq.Turn}). Therefore, we perform the numerical calculation to determine the turning point $u_0$. Secondly, for the light trajectories with $b>b_{c}$, one needs to perform the integration in Eq. (\ref{angle}) separately for the segments before and after the turning point, and then combine two integral results together. 
 In Fig.(\ref{fig.PH}), the black dashed curve denotes the trajectory of light rays with the critical parameter $b=b_c$ which finally form  the photon sphere. For $b>b_c$, light trajectory will escape to infinity after passing through the turning point. While for light rays with $b<b_c$, they will fall into the black hole and can not be received by the distant observer. Additionally, we notice that with the increase of $\alpha_0$, the region where light falls into black holes will shrink, as shown in Table (\ref{PR}).
 \par 
 
\begin{table}[htbp]
\centering
\caption{The ranges of impact parameter $b$ corresponding to the direct emission, lensed ring emission and photon ring emission of light rays for various $\alpha_0$.} 
\label{LR}
\resizebox{\textwidth}{15mm}{
    \begin{tabular}{|c|c|c|c|c|c}\hline 
$\alpha_0$& Direct$(n<3/4)$& lensed ring$(3/4<n<5/4)$ & Photon ring$(n>5/4)$   \\ \hline
$0.1$& $4.94441<b$ and $b>6.13190$&$4.94441<b<5.12750$ and $5.17136<b<6.13190$ &$5.12750<b<5.17136$ \\ \hline
$0.3$& $4.78429<b$ and $b>6.05831$&$4.78429<b<4.99549$ and $5.04965<b<6.05831$ &$4.99549<b<5.04965$\\ \hline
$0.5$& $4.58252<b$ and $b>5.98140$&$4.58252<b<4.84148$ and $4.91329<b<5.98140$ &$4.84148<b<4.91329$\\ \hline
\end{tabular}}
\end{table}

\subsection{Transfer Function of Black Hole With $x=2/3$ And $n=2$}
The light ray extracts energy from the thin accretion disk each time when passing through it. Thus, the total amount of energy extracted depends on the number of times passing through the thin disk. Consequently, the intensity of light ray obtained by an observer at infinity is also proportional to the number of times it passes through.\par
\begin{figure*}
	\centering
	\subfigure[$\alpha_0=0.1$]{
		\begin{minipage}[t]{0.33\linewidth}
			\centering
			\includegraphics[width=2.1in]{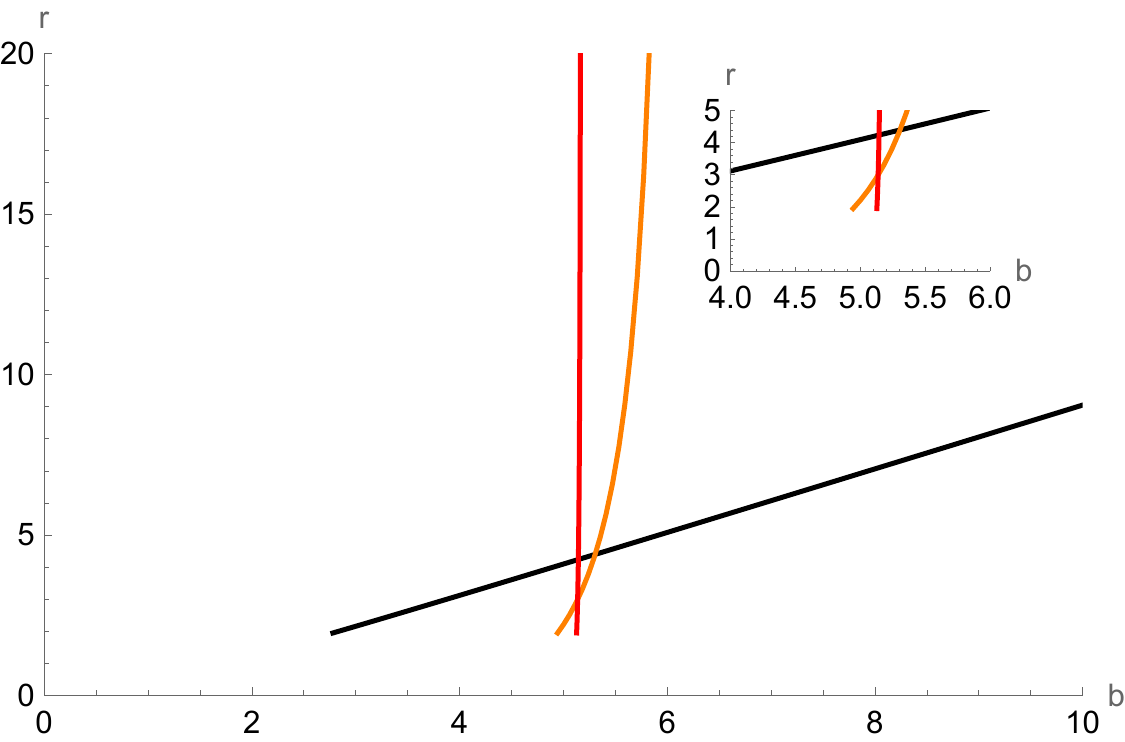}\\
			\vspace{0.1cm}
		\end{minipage}%
	}%
	\subfigure[$\alpha_0=0.3$]{
		\begin{minipage}[t]{0.33\linewidth}
			\centering
			\includegraphics[width=2.1in]{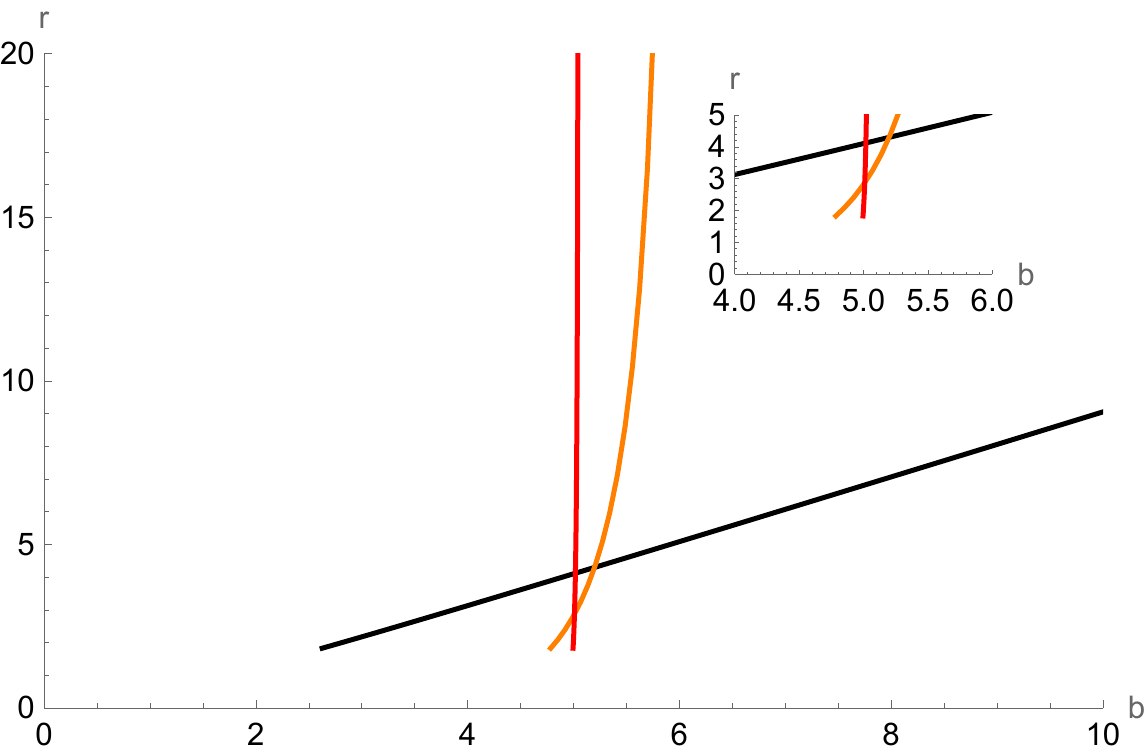}\\
			\vspace{0.1cm}
		\end{minipage}%
	}%
	\subfigure[$\alpha_0=0.5$]{
		\begin{minipage}[t]{0.33\linewidth}
			\centering
			\includegraphics[width=2.1in]{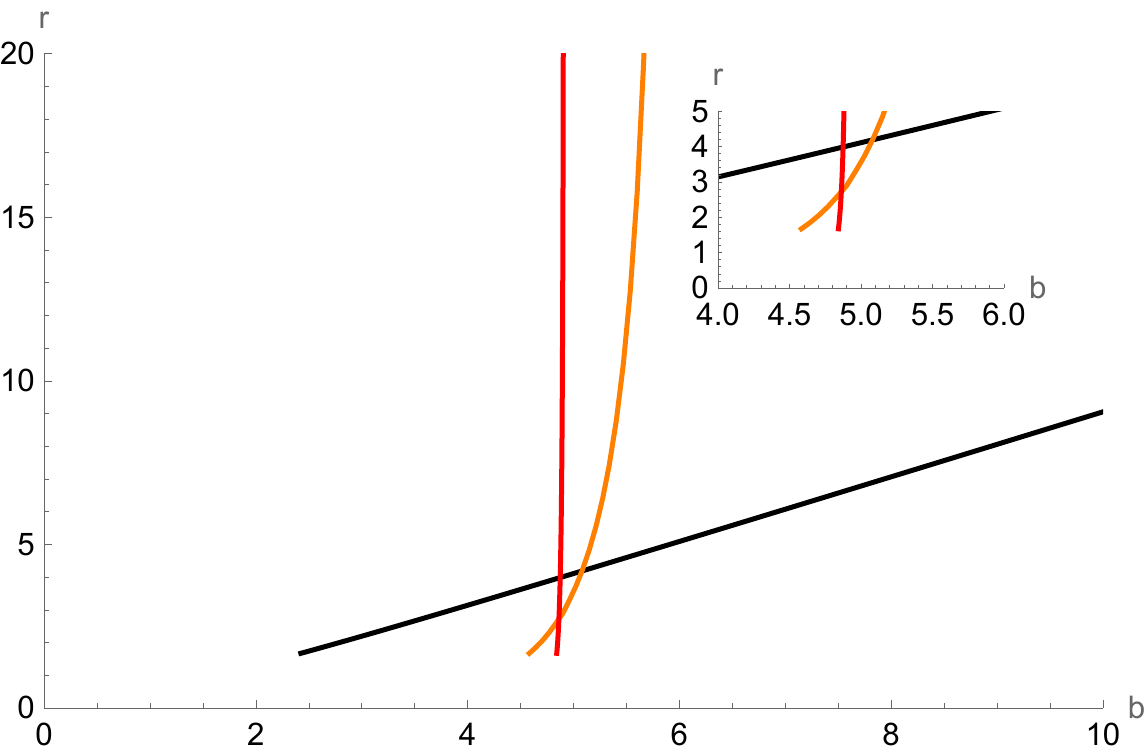}\\
			\vspace{0.1cm}
		\end{minipage}
	}
	\centering
	\caption{The transfer function for different $\alpha_0$ with $M=1$. The black solid line represents the first transfer function which corresponds to the direct emission, while the orange solid line represents the second transfer function corresponding to the lensed ring, and the red line is the third transfer function corresponding to the photon ring.}
	\vspace{-0.2cm}
	\label{fig.TF}
\end{figure*}

Next we consider the relation between the observed intensity of light and the emitted intensity. We assume that the lights emitted from thin accretion disk is isotropic for the static observer, and denote the specific emitted intensity by $I^{em}_{\nu}(r)=I(r)$, where $\nu$ is the emission frequency in a static frame. Ignoring the absorption, then based on the Liouvilles theorem we know $I^{em}_{\nu}(r)/\nu^3$ is a conserved quantity along a ray. Therefore, we obtain the single observed intensity $I_{\nu^{'}}^{obs}(r)$ as the function of $I^{em}_{\nu}(r)$ as
\begin{equation}
    I_{\nu^{'}}^{obs}(r)=g^3I^{em}_{\nu}(r)=f(r)^{3/2}I^{em}_{\nu}(r),
\end{equation}
where $g=f(r)^{1/2}$ is the redshift factor. As a result, the total observed intensity of one light trajectory $I^{obs}(r)$ is
\begin{equation}
    I^{obs}(r)=\int I_{\nu^{'}}^{obs}(r)d\nu=f(r)^2 I^{em}_{\nu}(r).
\end{equation}
As mentioned earlier, the light extracts energy whenever passing through the thin disk. Therefore, the intensity received by the observer should be the sum of the brightness at each intersection point, yielding
\begin{equation}
    I^{obs}(r)=\sum_{m} f(r)^2I^{em}_{\nu}(r)|_{r=r_{m}(b)},
\end{equation}
where $r_{m}(b)$ is the transfer function, denoting the radial coordinate at which a trajectory of light with impact parameter $b$ intersects the disk. We plot the first three transfer functions $(m=1,2,3)$ in Fig.(\ref{fig.TF}). 
The curves in Fig.(\ref{fig.TF}) have different slopes $dr/db$, which represent the demagnification factor and indicate the degree of demagnified of the observed image. The black curve represents the transfer function for direct emission and approximately has an average slope of one, implying that it is a direct image of the redshift source. The orange curve corresponds to the transfer function for lensed ring emission, and its average slope is much greater than one, implying that it appears to the observer as a highly demagnified image. The red curve corresponds to the transfer function for photon ring emission, with a slope that tends to infinity, indicating that the observed photon ring is a so demagnified image. 
The contribution of three types of light rays to the total flux can be seen from the average slope of their transfer functions. The largest contribution comes from the direct emission rays with the smallest average slope, and then from  the lensed ring, which has a large average slope, and finally from the photon ring, whose average slope tends to infinity and therefore its contribution to the total flux can be neglected.
Furthermore, as the parameter $\alpha_0$ increases, we notice that the transfer function shifts to the left. In addition, the average slope of the transfer functions for the lensed and photon rings decreases slightly, indicating that the contribution of these two types of light rays to the total flux increases with the increase of $\alpha_0$.
\subsection{The astronomical appearance of the black hole with $x=2/3$ and $n=2$ surrounded by the thin accretion disk}
One adopts different toy models to describe the radiation with different starting positions. The first toy model is to consider the radiation starting from the inner most stable circular orbit (ISCO). In this case one usually assumes that no radiation is emitted inside the ISCO, thus the intensity is given by
\begin{equation}\label{EQ.Mod1}
I^{{em}}_{IS}(r)=\left\{
	\begin{aligned}
	\left(\frac{1}{r-(r_{ISCO}-1)}\right)^2\quad r>r_{ISCO} ,\\
	0 \quad\quad\quad\quad\quad r\leq r_{ISCO} ,\\
	\end{aligned}
	\right	.
\end{equation}
For the spherically symmetric spacetime, $r_{ISCO}$ is given by
\begin{equation}
    r_{ISCO}=\frac{3 f(r_{ISCO})f'(r_{ISCO})}{2f'(r_{ISCO})^2-f(r_{ISCO})f''(r_{ISCO})}.
\end{equation}
\par
\begin{figure*}
	\centering
	\subfigure[$\alpha_0=0.1$]{
		\begin{minipage}[t]{0.33\linewidth}
			\centering
			\includegraphics[width=2.1in]{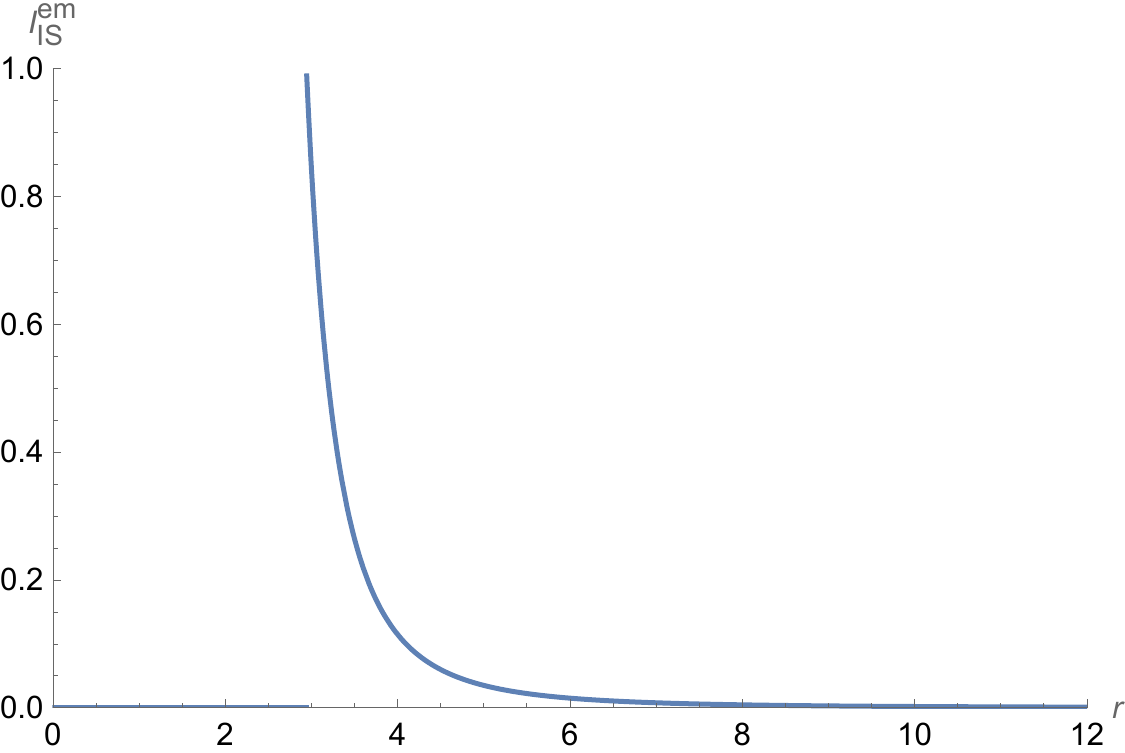}\\
			\vspace{0.1cm}
			\includegraphics[width=2.1in]{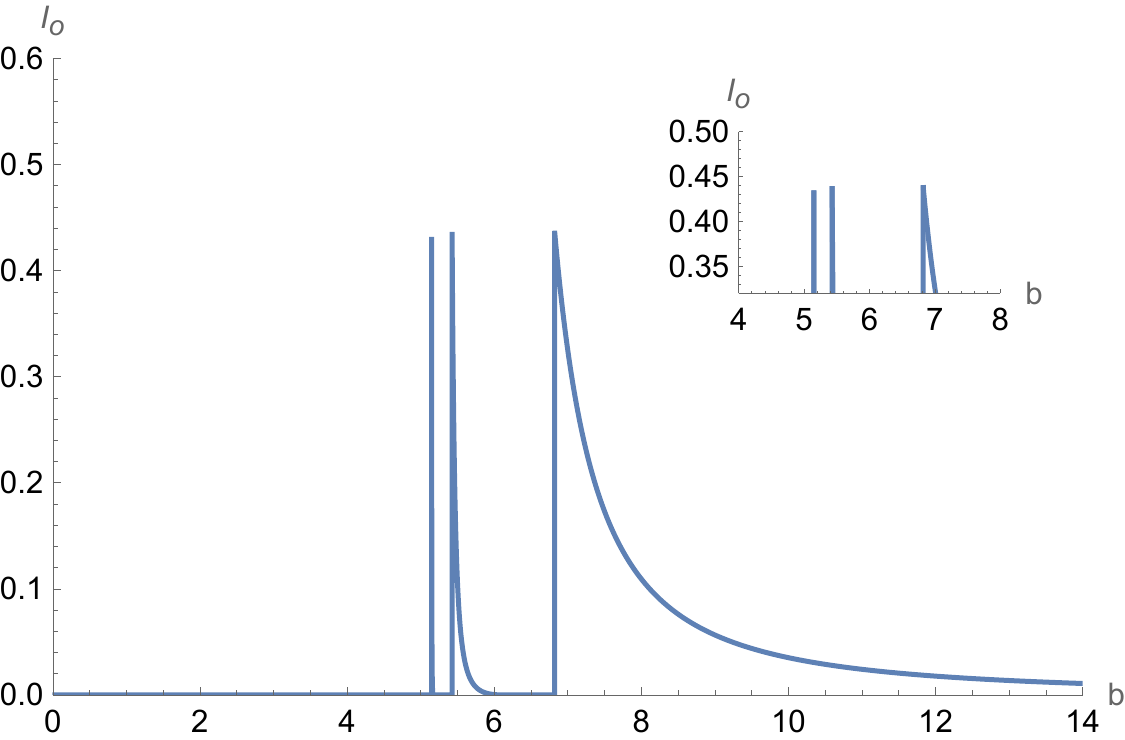}\\
			\vspace{0.1cm}
                \includegraphics[width=2.1in]{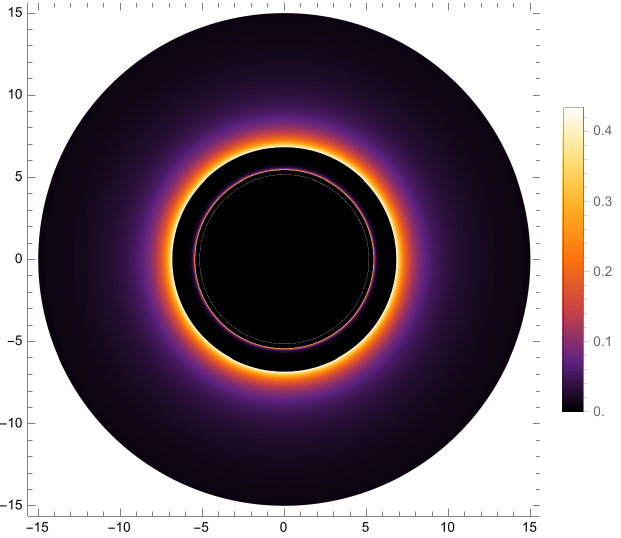}\\
			\vspace{0.1cm}
		\end{minipage}%
	}%
	\subfigure[$\alpha_0=0.3$]{
		\begin{minipage}[t]{0.33\linewidth}
			\centering
			\includegraphics[width=2.1in]{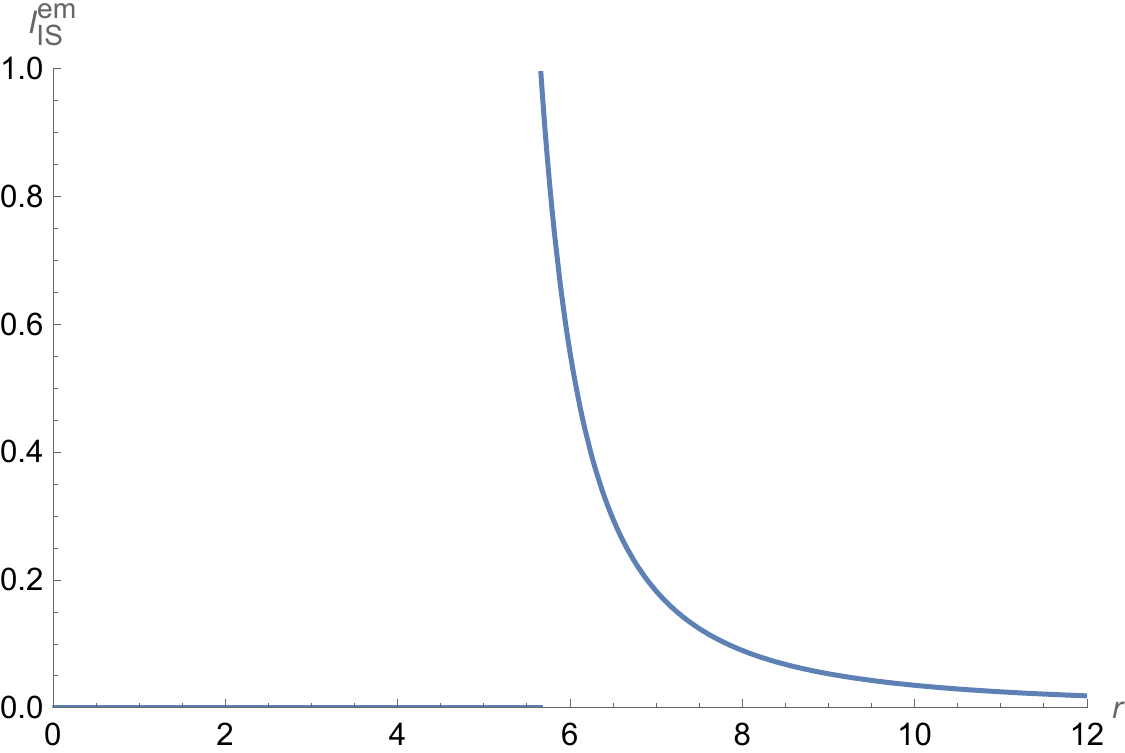}\\
			\vspace{0.1cm}
			\includegraphics[width=2.1in]{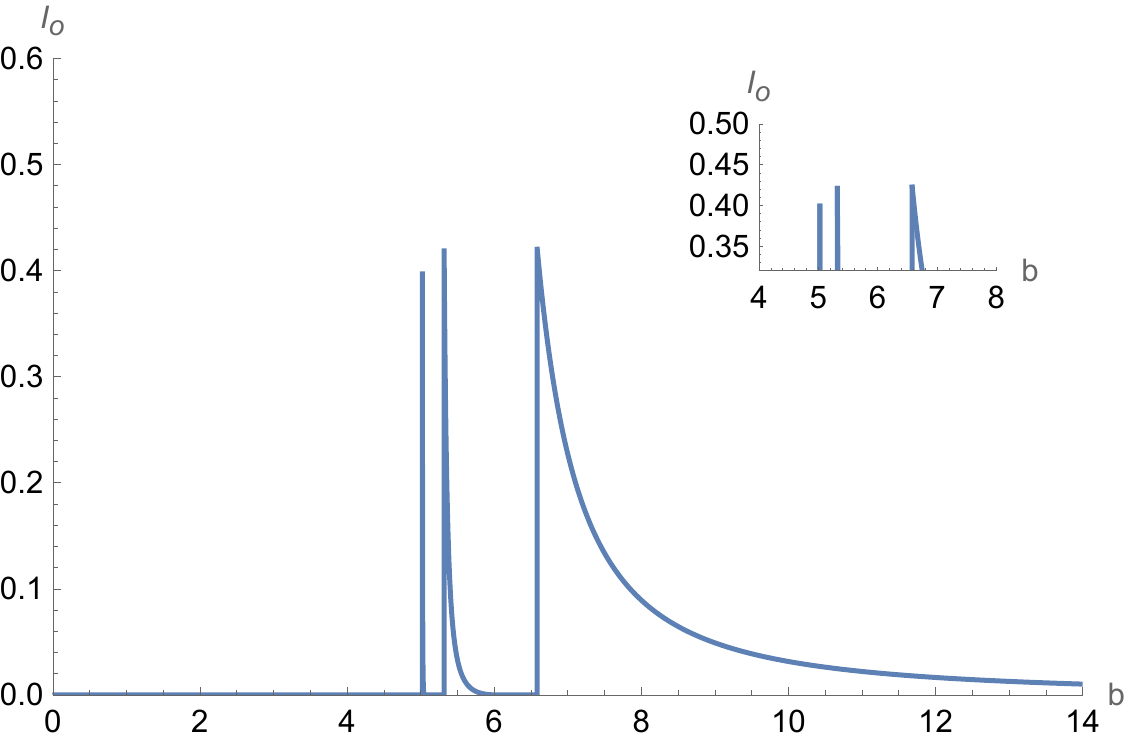}\\
			\vspace{0.1cm}
                \includegraphics[width=2.1in]{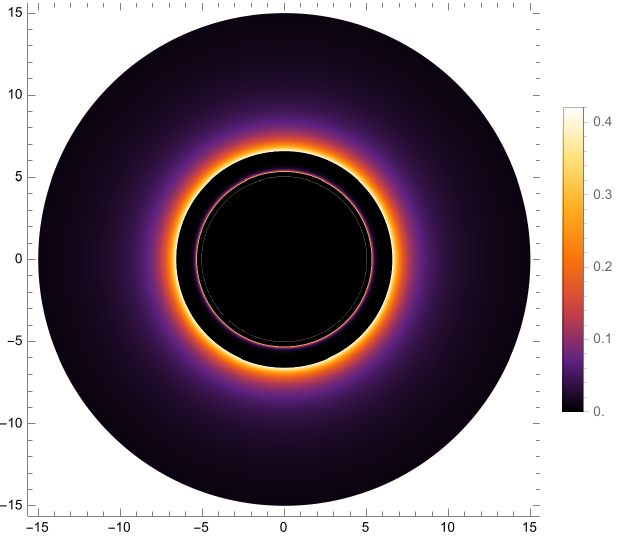}\\
			\vspace{0.1cm}
		\end{minipage}%
	}%
	\subfigure[$\alpha_0=0.5$]{
		\begin{minipage}[t]{0.33\linewidth}
			\centering
			\includegraphics[width=2.1in]{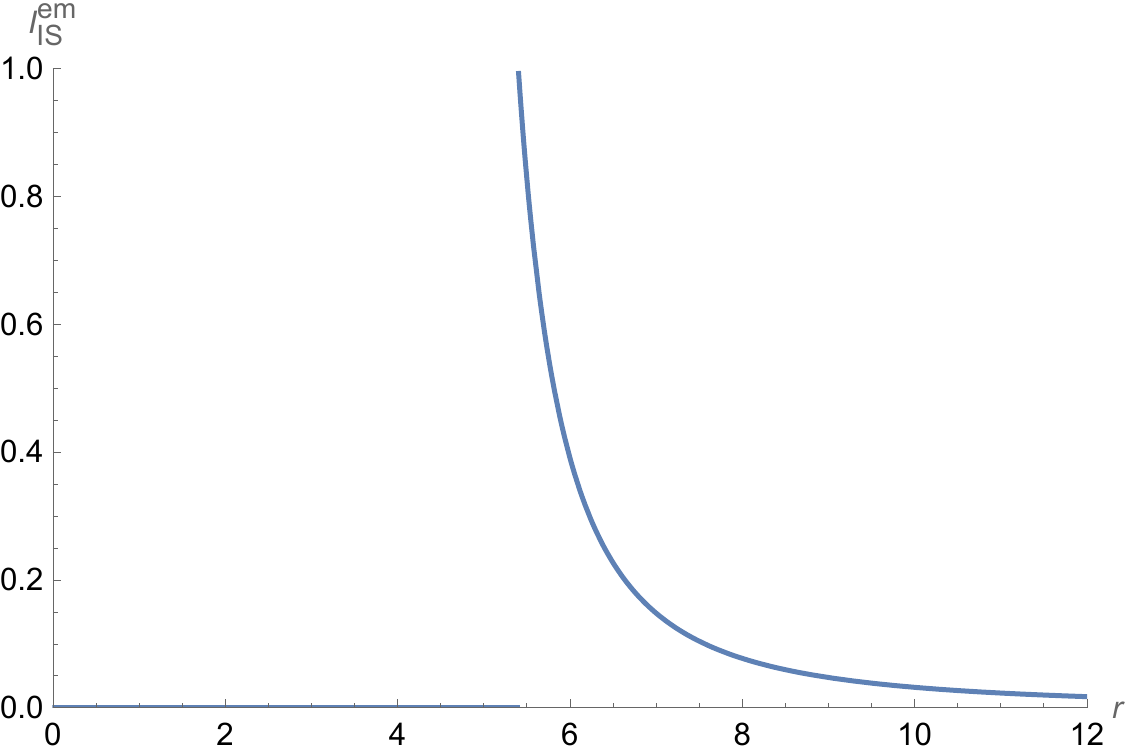}\\
			\vspace{0.1cm}
			\includegraphics[width=2.1in]{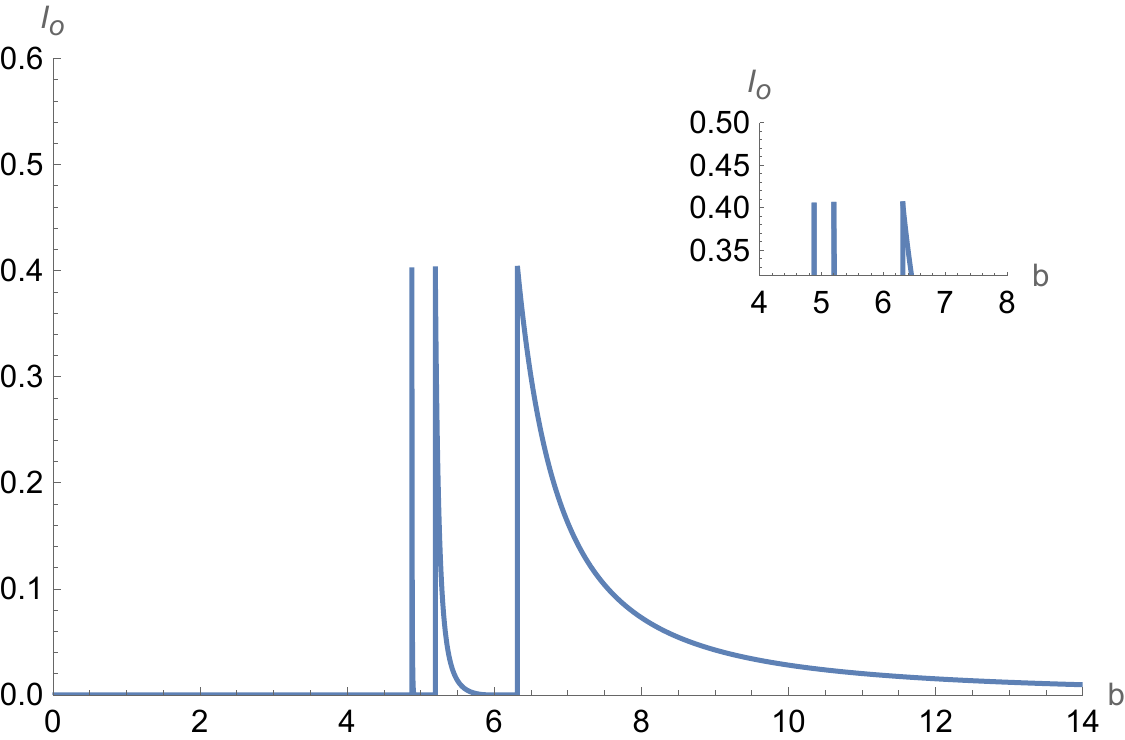}\\
			\vspace{0.1cm}
                \includegraphics[width=2.1in]{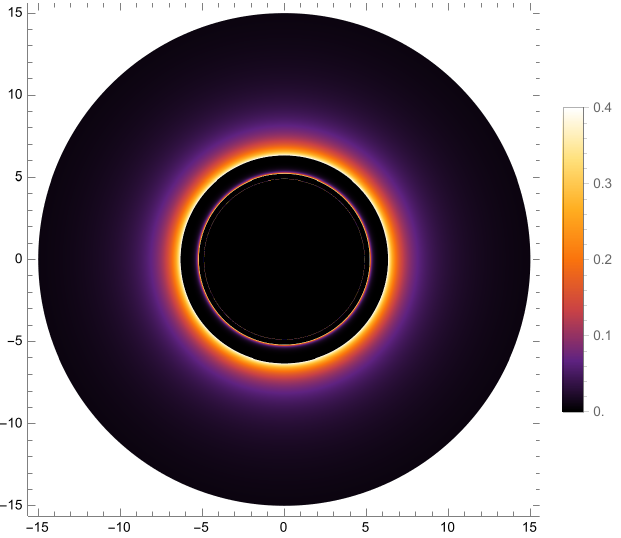}\\
			\vspace{0.1cm}
		\end{minipage}
	}
	\centering
	\caption{The observational appearances of the ISCO emission model of the thin accretion disk for different $\alpha_0$ with $M=1$. The plot on the top shows the variation of the emission intensity with respect to $r$ for the ISCO, while the middle plot shows the intensity received by the observer during the ISCO emission, and the plot at the bottom shows the two-dimensional observational images under the ISCO emission model.  The first column represents the observational characteristics of the black hole when $\alpha_0=0.1$. The second column represents the observational characteristics when $\alpha_0=0.3$, and the third column represents the characteristics when $\alpha_0=0.5$. }
	\vspace{-0.2cm}
	\label{fig.IS}
\end{figure*}

 We apply the expression $(I^{{em}}_{IS}(r))$ to plot the theoretical image of a black hole detected by an observer in Fig.(\ref{fig.IS}). 
The top row of plots shows the emission intensity as a function of the radius $r$. It is noticed that there is no emission within $r<r_{ISCO}$, and the emission intensity reaches its peak at $r=r_{ISCO}$ before rapidly decreasing thereafter. The plots at the second row show the  intensity received by an observer as a function of the impact parameter $b$. It exhibits three peaks in each plot. The source of the first peak is the photon ring, which corresponds to a very small region, resulting in a very small contribution to the total flux. The second peak corresponds to the lensed ring, whose region is slightly larger than that of the photon ring, indicating that the contribution of the lensed ring is slightly larger than that of the photon ring. The rightmost peak is the result of direct emission, which covers a large area, meaning that it has the largest contribution to the total flux. In addition, because the radiation source is far from the photon sphere, the redshift effect is not as severe, and their overall peak is higher than that of the other two models. The bottom row demonstrates the theoretical two-dimensional images, each of which contains three rings. The innermost ring is the photon ring, which contributes very little to the total flux, and it is so demagnified, resulting in a very small and faint ring.  Outside the photon ring is the lensed ring. It is highly demagnified and slightly brighter than the photon ring. The outermost ring, which is the result of direct emission and direct image of the radiation source, is the main contributor to the total flux, making it the brightest and widest of the three rings.  As $\alpha_0$ increases, three peaks shift to the left as a whole, and the peak heights decrease slightly. This means that the observed area of central dark region and its overall brightness will decrease slightly.

\par
\begin{figure*}
	\centering
	\subfigure[$\alpha_0=0.1$]{
		\begin{minipage}[t]{0.33\linewidth}
			\centering
			\includegraphics[width=2.1in]{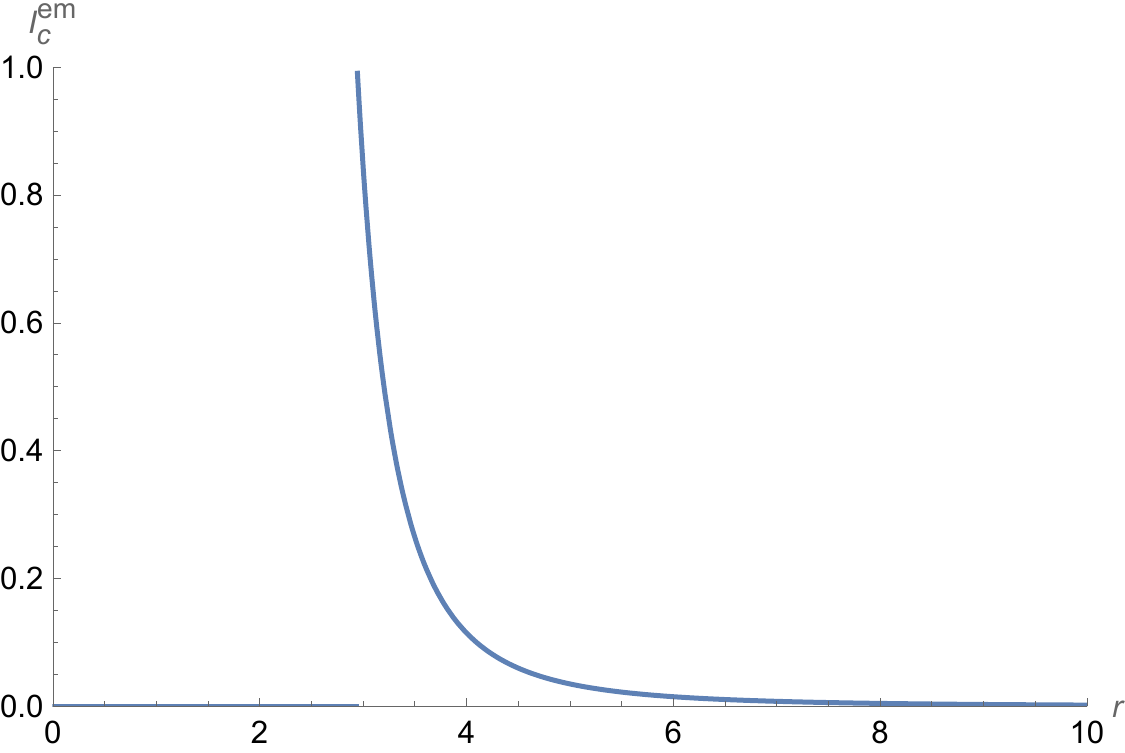}\\
			\vspace{0.1cm}
			\includegraphics[width=2.1in]{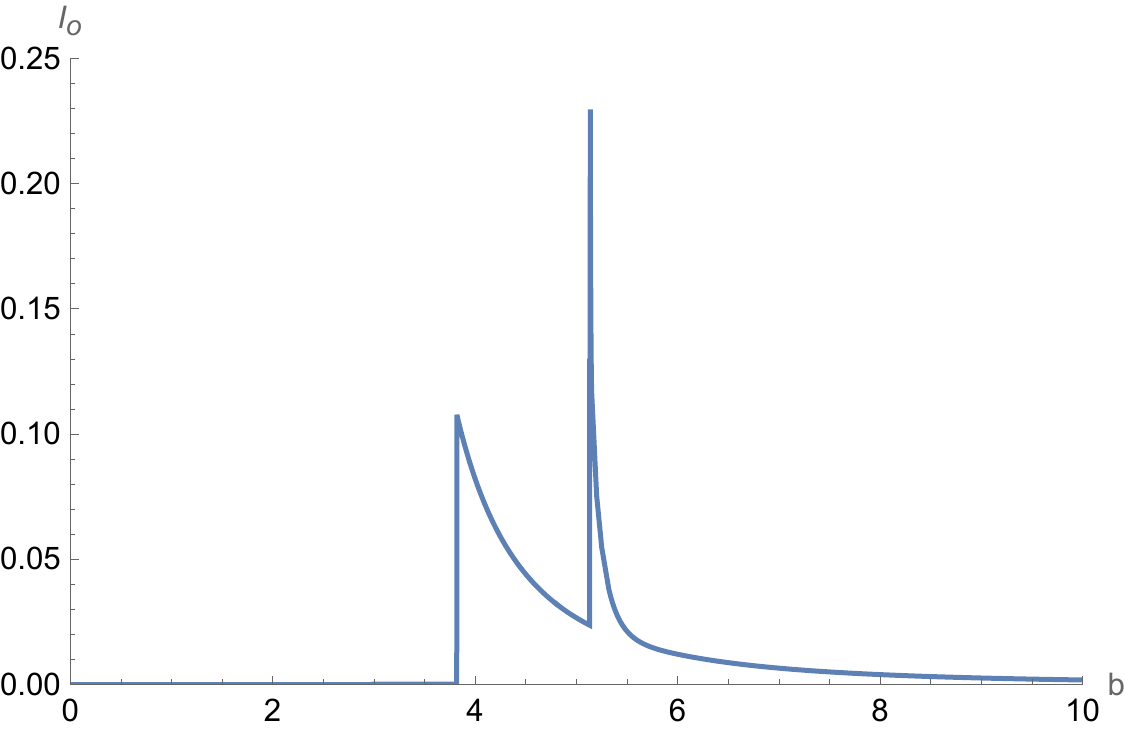}\\
			\vspace{0.1cm}
                \includegraphics[width=2.1in]{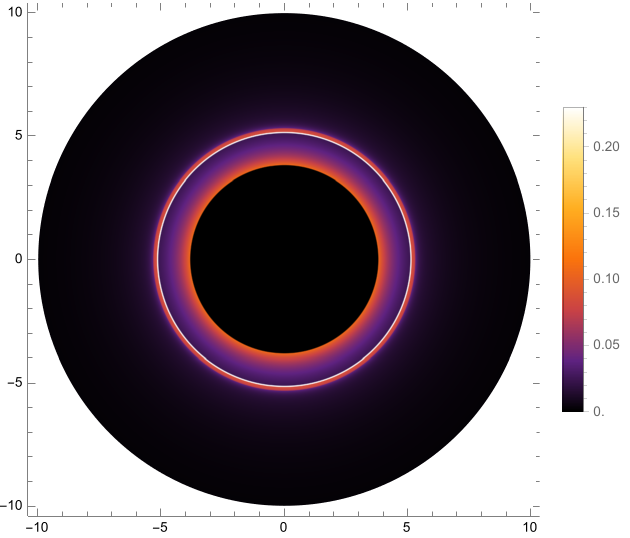}\\
			\vspace{0.1cm}
		\end{minipage}%
	}%
	\subfigure[$\alpha_0=0.3$]{
		\begin{minipage}[t]{0.33\linewidth}
			\centering
			\includegraphics[width=2.1in]{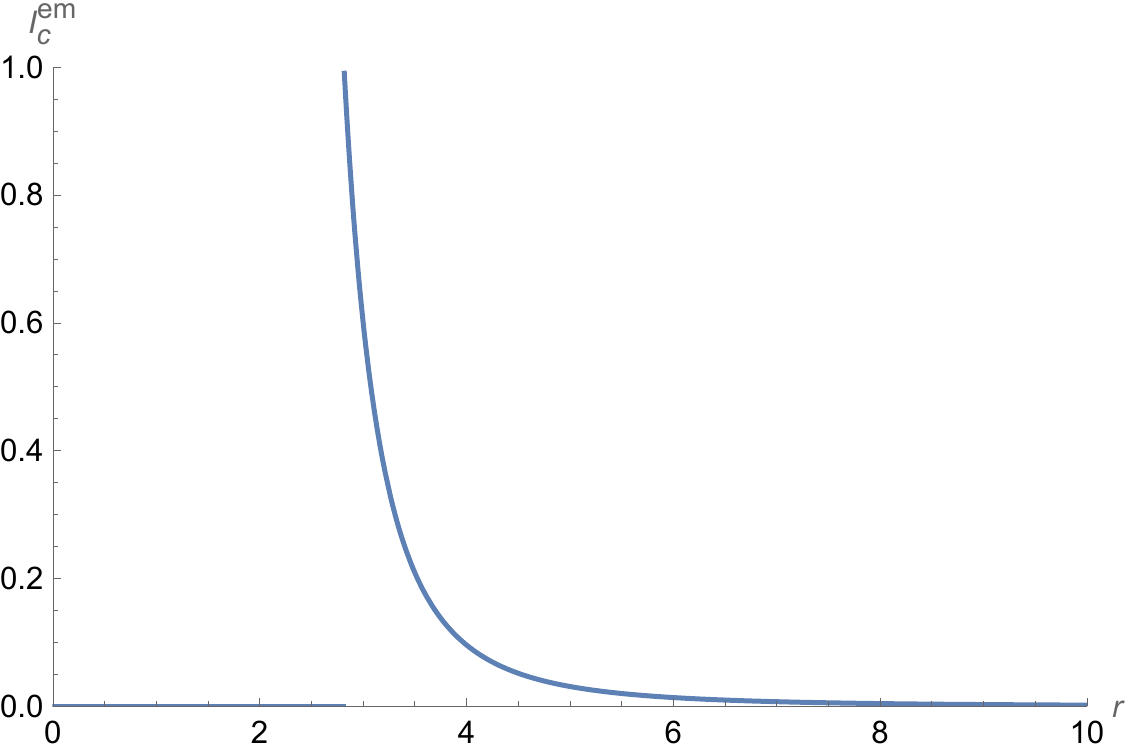}\\
			\vspace{0.1cm}
			\includegraphics[width=2.1in]{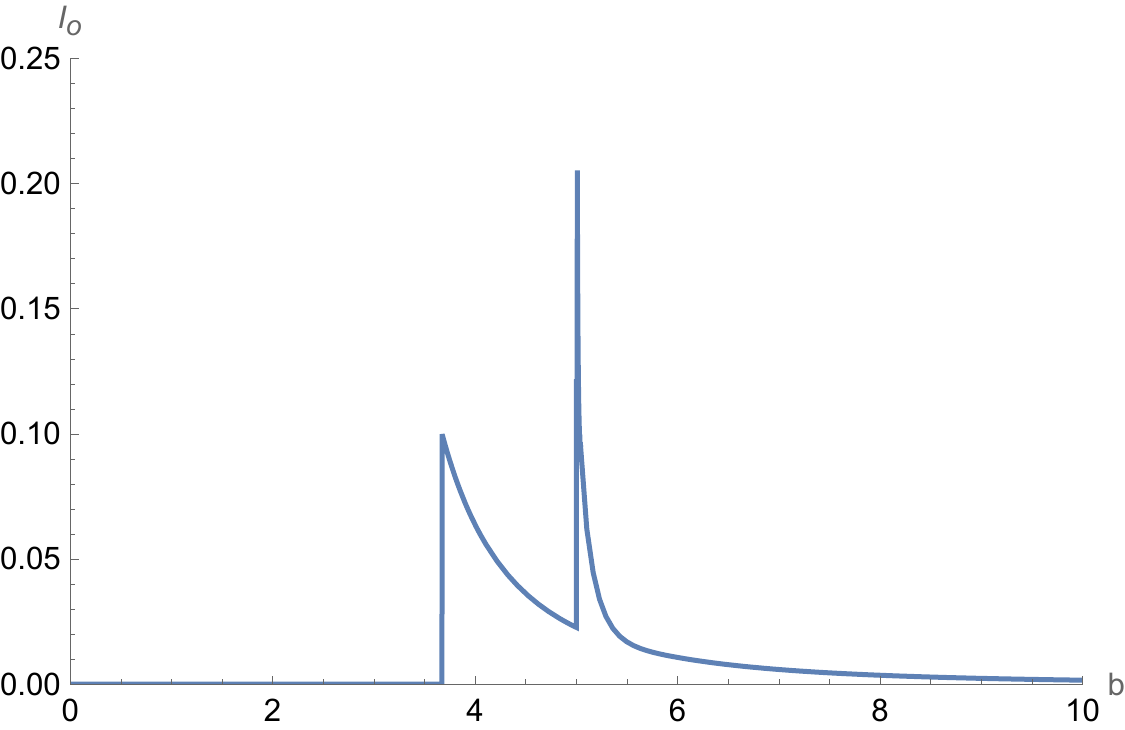}\\
			\vspace{0.1cm}
                \includegraphics[width=2.1in]{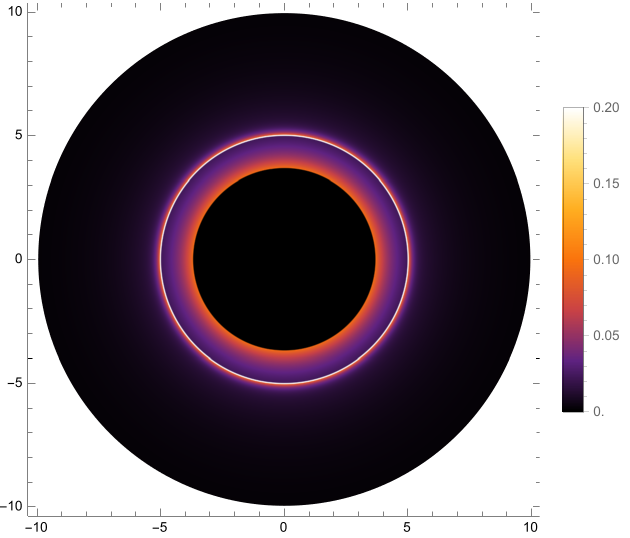}\\
			\vspace{0.1cm}
		\end{minipage}%
	}%
	\subfigure[$\alpha_0=0.5$]{
		\begin{minipage}[t]{0.33\linewidth}
			\centering
			\includegraphics[width=2.1in]{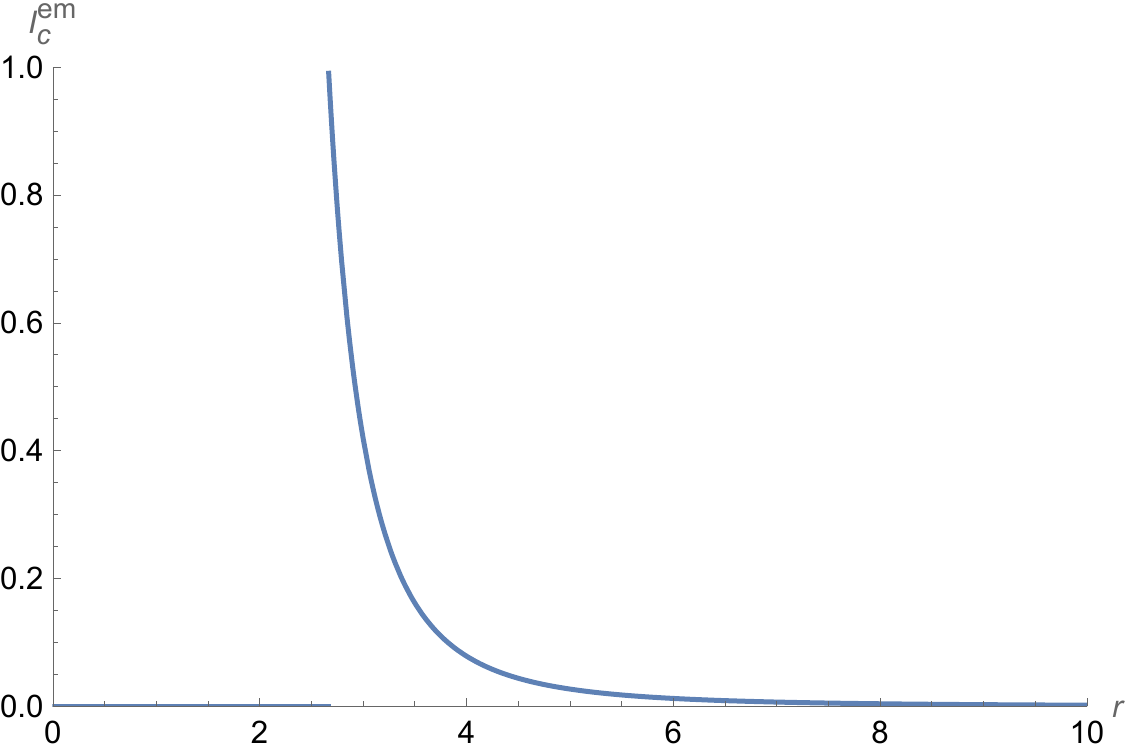}\\
			\vspace{0.1cm}
			\includegraphics[width=2.1in]{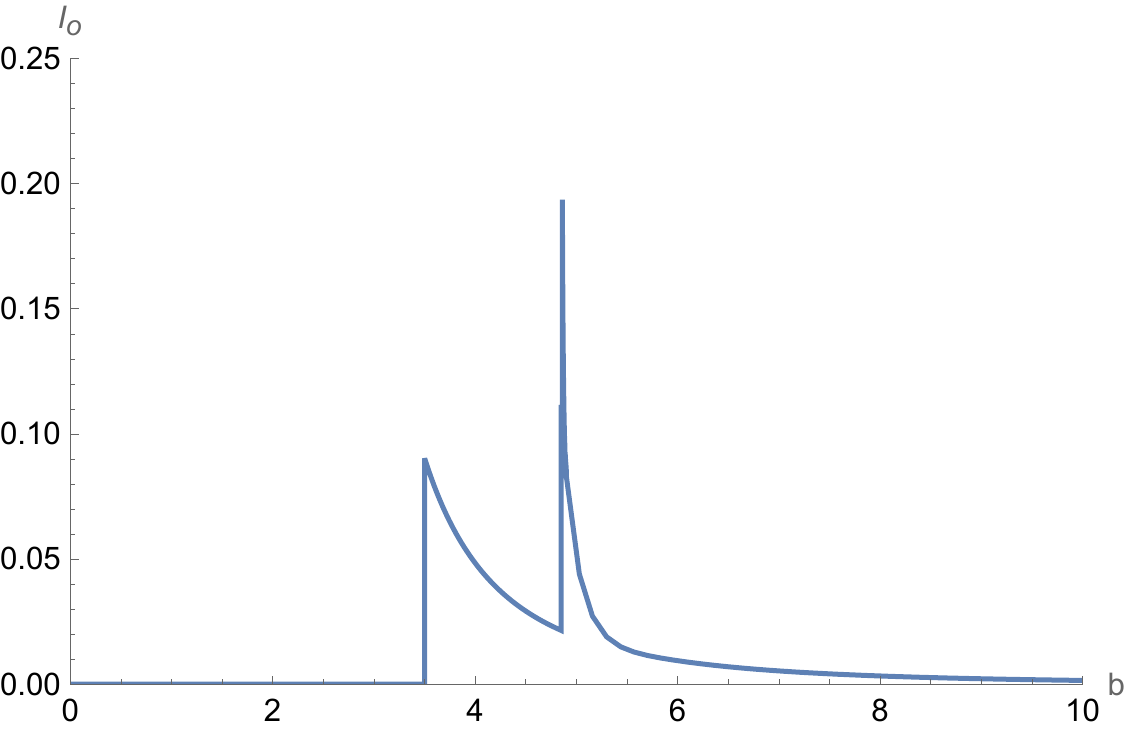}\\
			\vspace{0.1cm}
                \includegraphics[width=2.1in]{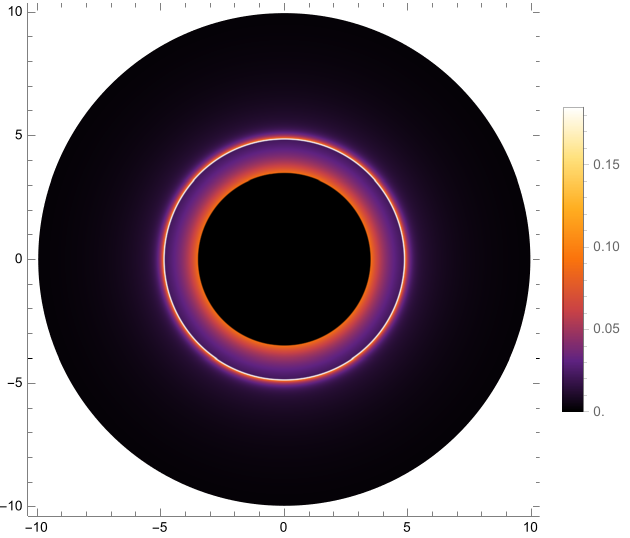}\\
			\vspace{0.1cm}
		\end{minipage}
	}
	\centering
	\caption{The observational appearances of the photon sphere emission model of the thin accretion disk for different $\alpha_0$ with $M=1$. The plot on the top shows the variation of the emission intensity with respect to $r$ for the photon sphere, while the middle plot shows the intensity received by the observer during the photon sphere emission, and the plot at the bottom shows the two-dimensional observational map under the photon sphere emission model. Each  column from left to right represents the observational characteristics of the black hole for $\alpha_0=0.1,0.3,0.5$, respectively}. 
 
	\vspace{-0.2cm}
	\label{fig.PE}
\end{figure*}

 The second toy model is to consider the radiation starting from the photon sphere, in which the intensity is given by 
\begin{equation}\label{Eq.Mod2}
    I^{{em}}_{c}(r)=\left\{
	\begin{aligned}
	\left(\frac{1}{r-(r_{c}-1)}\right)^3\quad r>r_{c},\\
	0 \quad\quad\quad\quad\quad r\leq r_{c},\\
	\end{aligned}
	\right	.
\end{equation}
where $r_{c}$ locates the cut-off position of the radiation. Specifically, there is no radiation in the region with $r<r_{c}$, and the emission intensity gradually decreases after $r>r_{c}$.  We illustrate the emission intensity $I^{em}_{c}$,the observation intensity $I_{o}$ and the corresponding two-dimensional image of the black hole in Fig.(\ref{fig.PE}).
In this toy model, it is found that the peak of the direct emission is located inside the peaks of the photon ring and the lensed ring, which are overlapped and hard to distinct. 
Furthermore, since the emission starts from the photon sphere, the overall observed peaks are significantly lower due to the redshift effect, and the observed lensed ring and photon ring are superimposed on the direct emission. In addition, with the increase of  $\alpha_0$, we notice that the peaks of the observed intensity shift to the left, accompanying with the decreasing of the peak height. The rightmost peaks corresponding to the photon and lensed rings are higher, but the corresponding range of $b$ becomes smaller, indicating that the observed flux still mainly comes from the direct emission. In two-dimensional image of black holes, one sees a white thin ring outside of the bright ring, which corresponds to the photon ring. Inside are the lensed ring and the direct emission. As $\alpha_0$ increases, the overall brightness of the bright ring decreases and the area of central dark region decreases as well.
 \par
 \begin{figure}[H]
	\centering
	\subfigure[$\alpha_0=0.1$]{
		\begin{minipage}[t]{0.33\linewidth}
			\centering
			\includegraphics[width=2.1in]{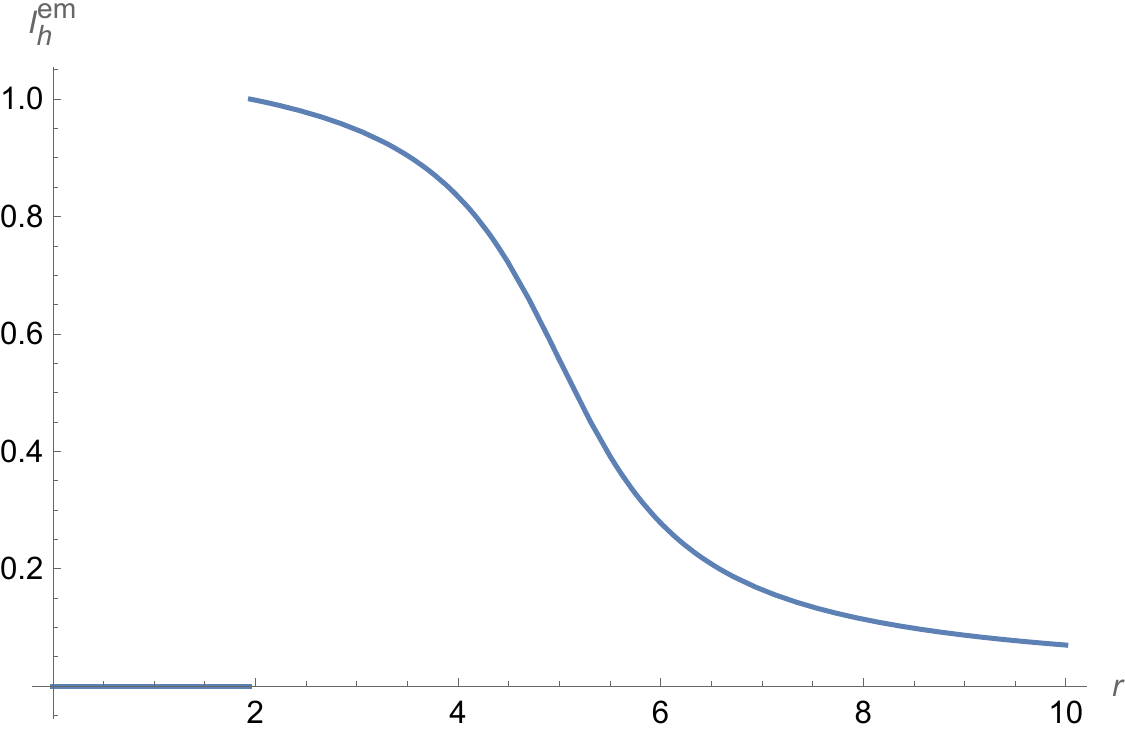}\\
			\vspace{0.1cm}
			\includegraphics[width=2.1in]{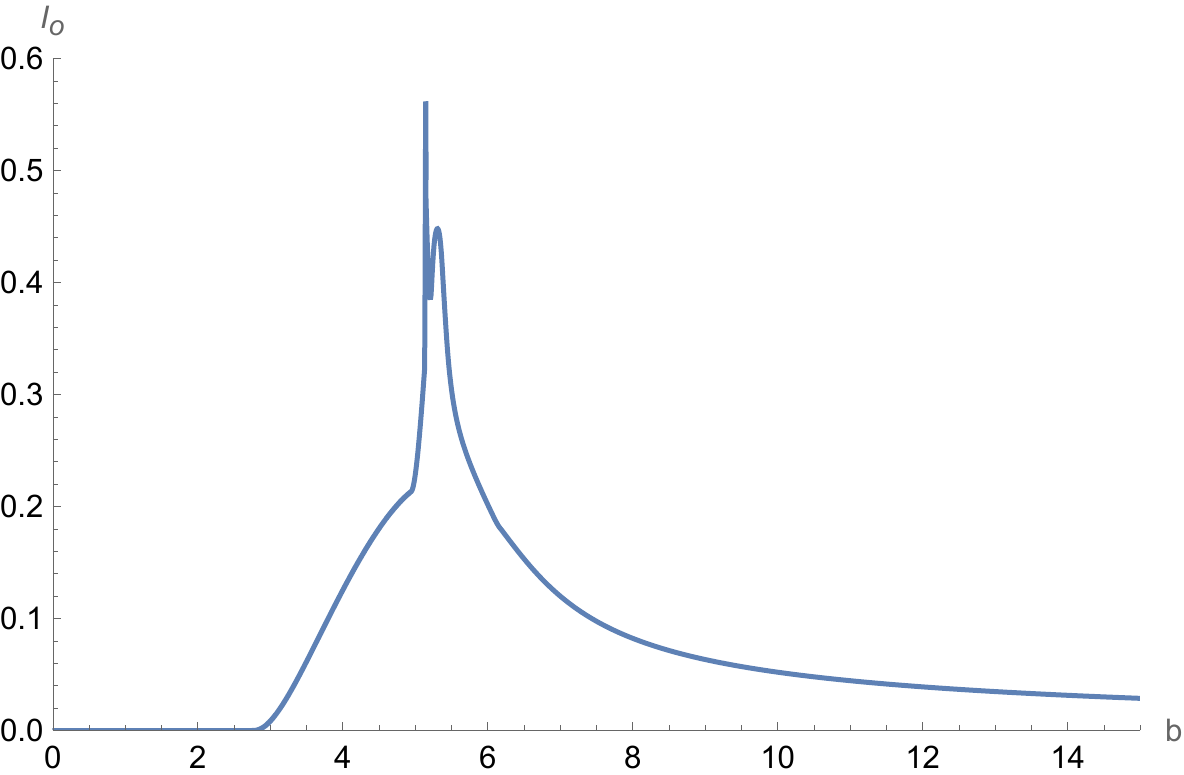}\\
			\vspace{0.1cm}
                \includegraphics[width=2.1in]{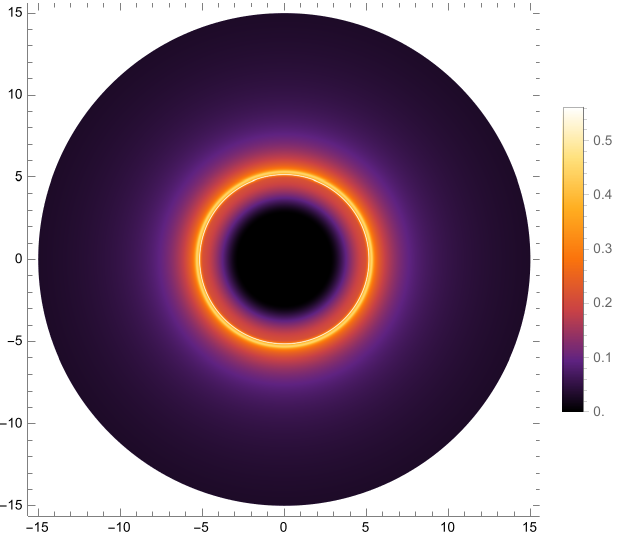}\\
			\vspace{0.1cm}
		\end{minipage}%
	}%
	\subfigure[$\alpha_0=0.3$]{
		\begin{minipage}[t]{0.33\linewidth}
			\centering
			\includegraphics[width=2.1in]{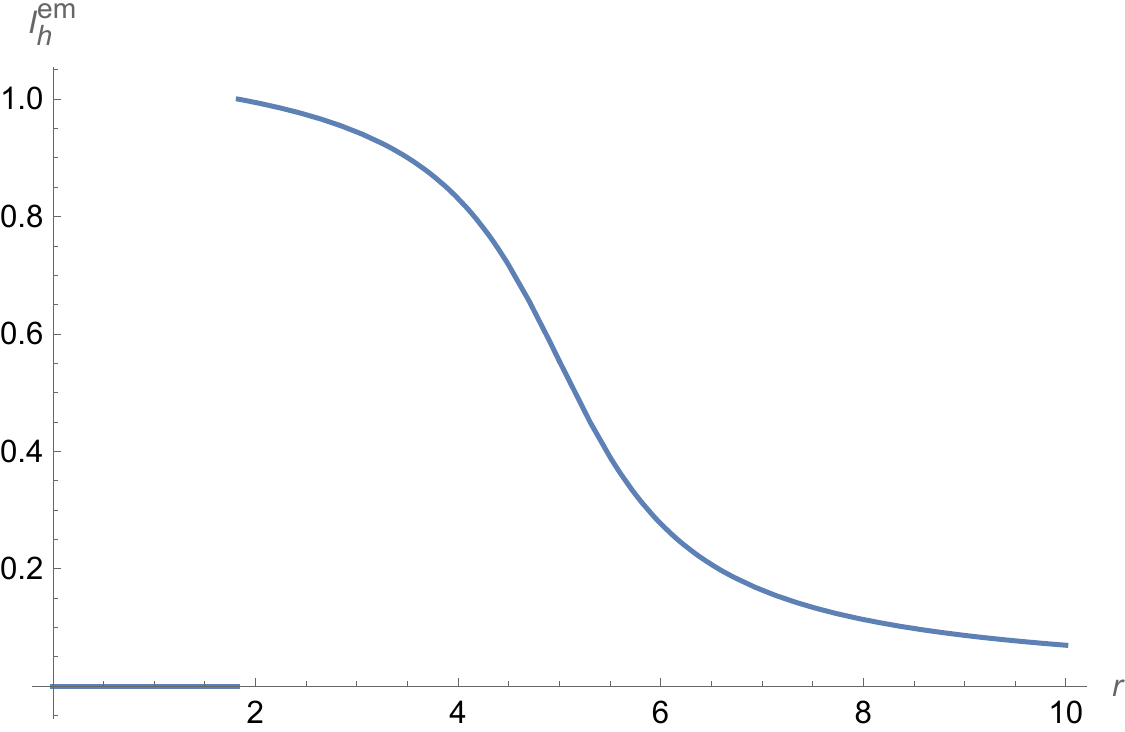}\\
			\vspace{0.1cm}
			\includegraphics[width=2.1in]{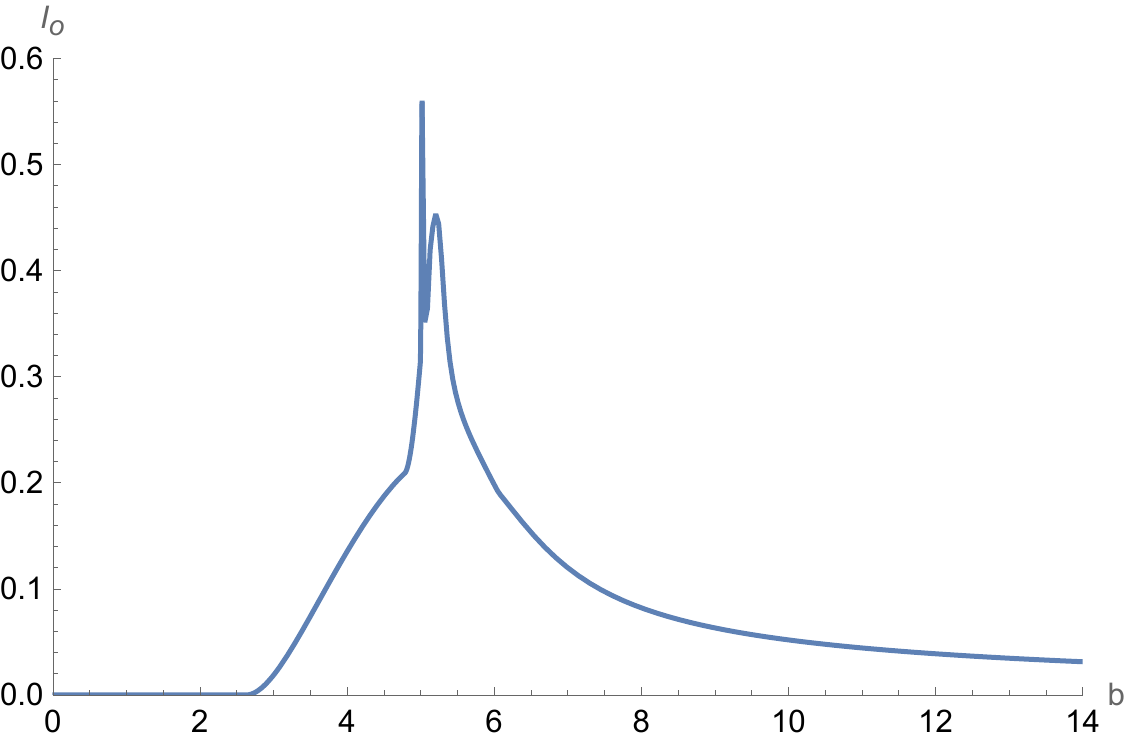}\\
			\vspace{0.1cm}
                \includegraphics[width=2.1in]{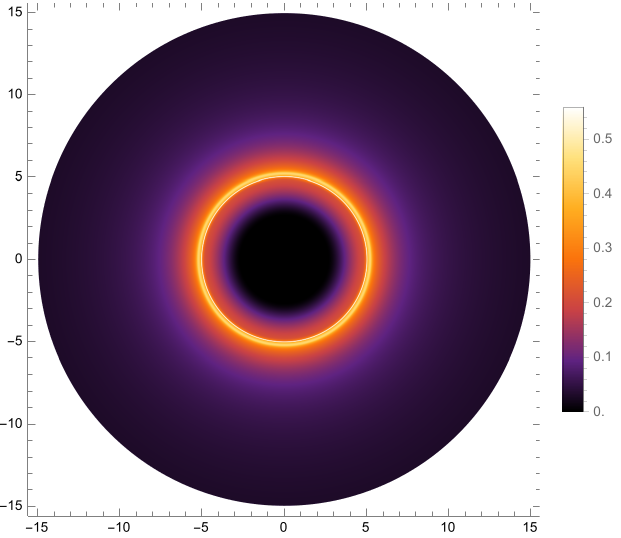}\\
			\vspace{0.1cm}
		\end{minipage}%
	}%
	\subfigure[$\alpha_0=0.5$]{
		\begin{minipage}[t]{0.33\linewidth}
			\centering
			\includegraphics[width=2.1in]{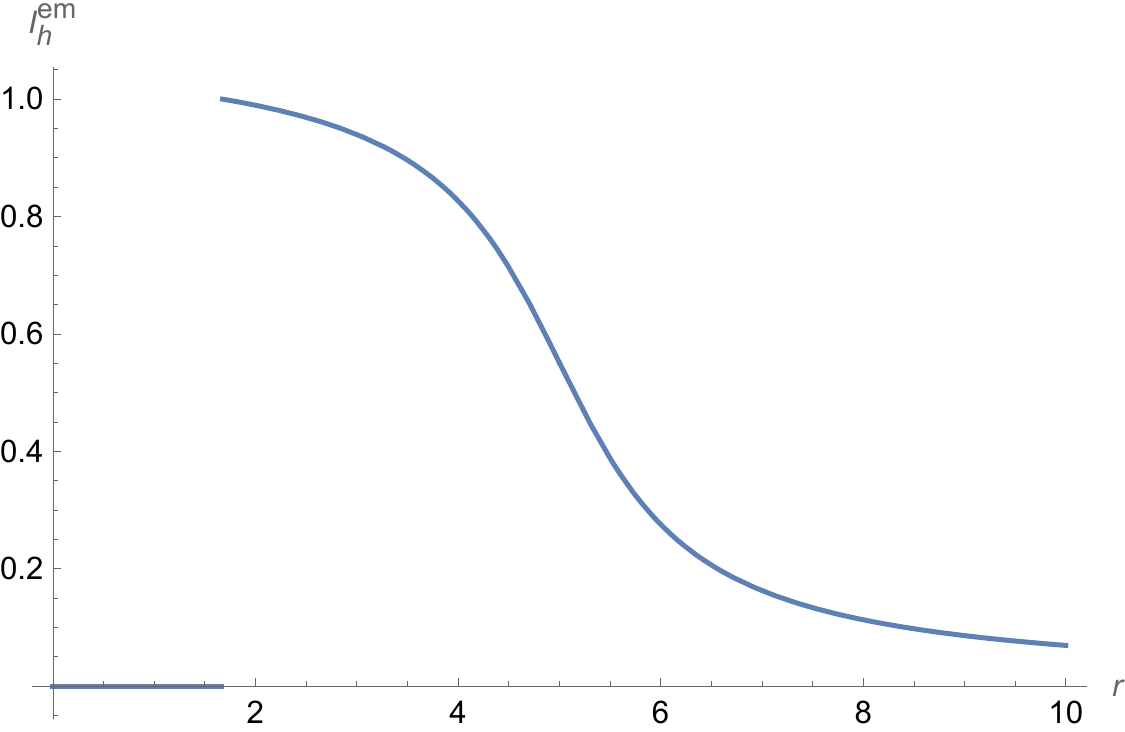}\\
			\vspace{0.1cm}
			\includegraphics[width=2.1in]{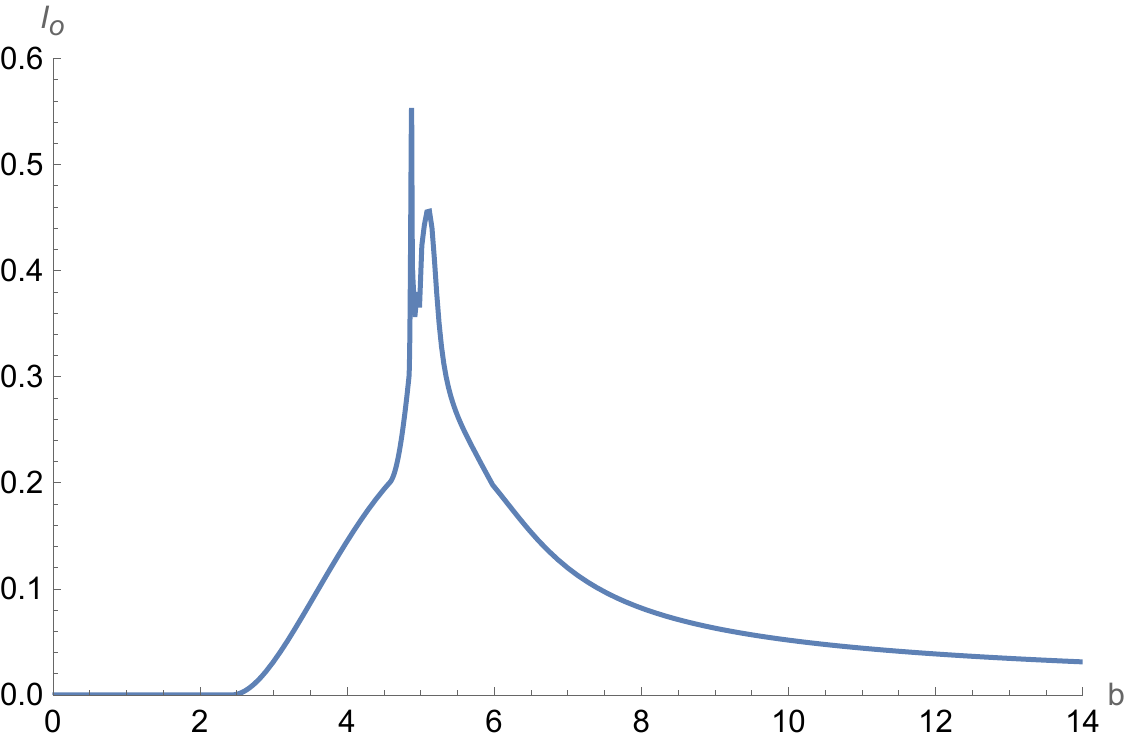}\\
			\vspace{0.1cm}
                \includegraphics[width=2.1in]{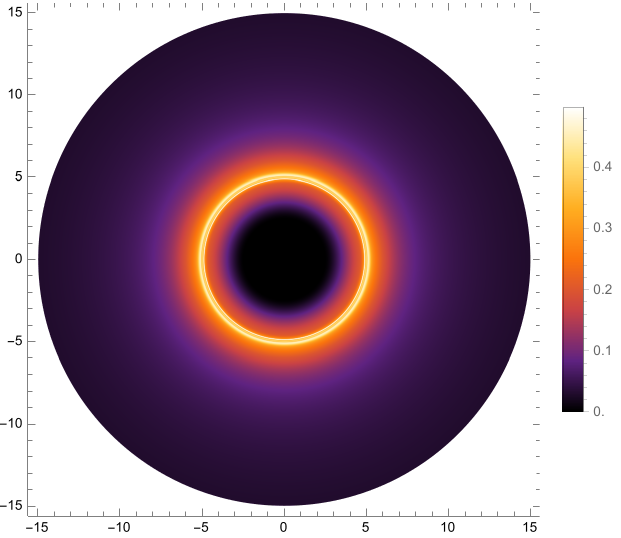}\\
			\vspace{0.1cm}
		\end{minipage}
	}
	\centering
	\caption{The observational appearances of the horizon emission model of the thin accretion disk for different $\alpha_0$ with $M=1$. The plot on the top shows the variation of the emission intensity with respect to $r$ for the photon sphere, while the middle plot shows the intensity received by the observer during the photon sphere emission, and the plot at the bottom shows the two-dimensional observational map under the photon sphere emission model. Each  column from left to right represents the observational characteristics of the black hole for $\alpha_0=0.1,0.3,0.5$, respectively}.
 
	\vspace{-0.2cm}
	\label{fig.EV}
\end{figure}
 Now we turn to the third toy model which assumes the starting position of the emission is the horizon. The radiation intensity exhibits its maximum at the horizon, and then gradually attenuates as it moves away from the horizon. In particular, the degree of attenuation is more gradual than the other two emission models. Specifically, the radiation intensity is given by
 \begin{equation}\label{Eq.Mod3}
    I^{{em}}_{h}(r)=\left\{
	\begin{aligned}
	\frac{1-\tan^{-1}(r-5)}{1-\tan^{-1}(r_{h}-5)}\quad r>r_{h},\\
	0 \quad\quad\quad\quad\quad r\leq r_{h}.\\
	\end{aligned}
	\right	.
\end{equation}
Fig.(\ref{fig.EV}) provides more details on our findings. In this model, the peak of the lensed ring is broader than that of the photon ring, indicating that the lensed ring contributes more to the observed flux. Nevertheless, the direct emission still makes a main contribution. The bright region in the two-dimensional image corresponds to the combined region of the photon ring and the lensed ring, and it is difficult to distinguish them in observation because of the overlapping of their peaks. Beyond the bright region, the other bright regions are contributed by the direct emission. As the parameter $\alpha_0$ increases, the peak of the photon ring gradually decreases. This implies that the brightness of the photon ring slightly decreases in observation, and the area of central dark region also decreases.
\section{In compraison with Bardeen black hole} 
In this section, we compare the black hole shadow and optical characteristic of Bardeen spacetime and the regular black hole with $x=2/3$ and $n=2$ surrounded by a thin accretion disk. In \cite{Ling:2021olm}, one-to-one correspondence between regular black holes with a dS core and those with a Minkowskian core has been established based on their asymptotic behavior outside the horizon. Corresponding to the regular black hole with a Minkowskian core described by Eq.(\ref{Eq.NP}), the Newtonian potential of regular black holes with a dS core is given by
\begin{equation}\label{Eq.dS}
   \psi^{d}(r)=-\frac{M r^{\frac{n}{x}-1}}{(r^n + x \alpha_0 M^x)^{1/x}},
\end{equation}
Specifically, when $x=2/3$ and $n=2$, it corresponds to the well-known Bardeen black hole, and $f^{B}(r)$ is given by
\begin{equation}
     f^{B}(r)=1-\frac{2 Mr^2}{(r^2 +\frac{2}{3} \alpha_{0} M^{2/3})^2}.
\end{equation}
 Previously the shadow of Bardeen black hole has been investigated in \cite{He:2021htq}. Here we will compare the shadow of these two black holes with $\alpha_0=0.73$, which is the maximal value of allowed $\alpha_0$. Because with larger deviation parameter $\alpha_0$, more light rays will be received and the difference of images between these two black holes is expected to become more distinct. 
 
\begin{figure*}
	\centering
	\subfigure[Schwarzschild]{
		\begin{minipage}[t]{0.33\linewidth}
			\centering
			\includegraphics[width=2.1in]{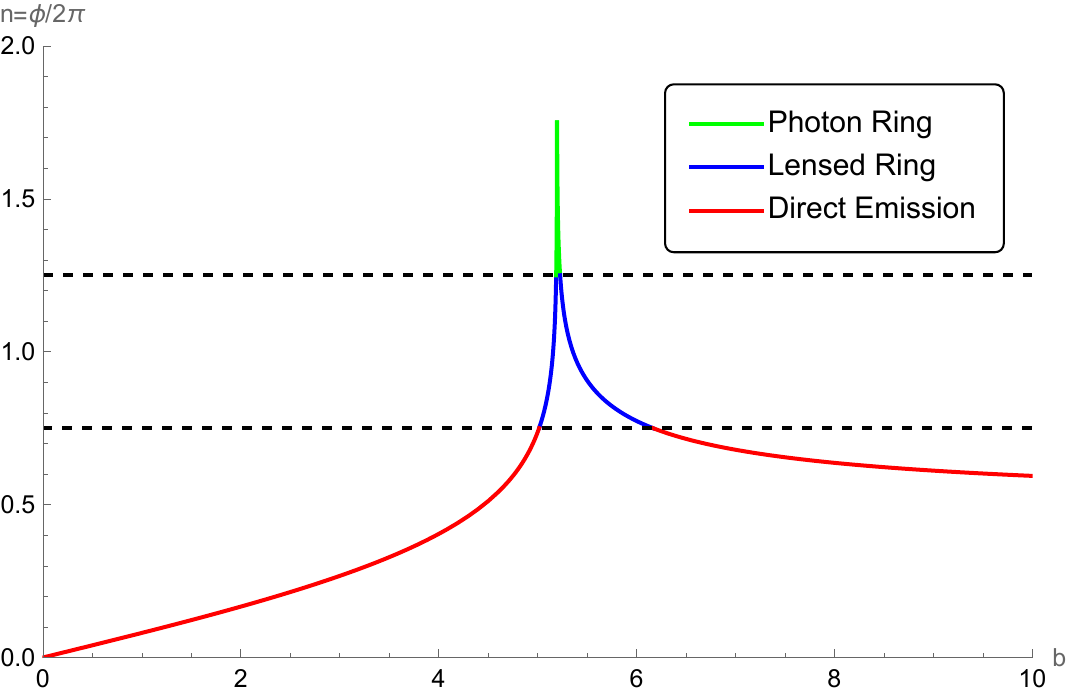}\\
			\vspace{0.1cm}
                \includegraphics[width=2.1in]{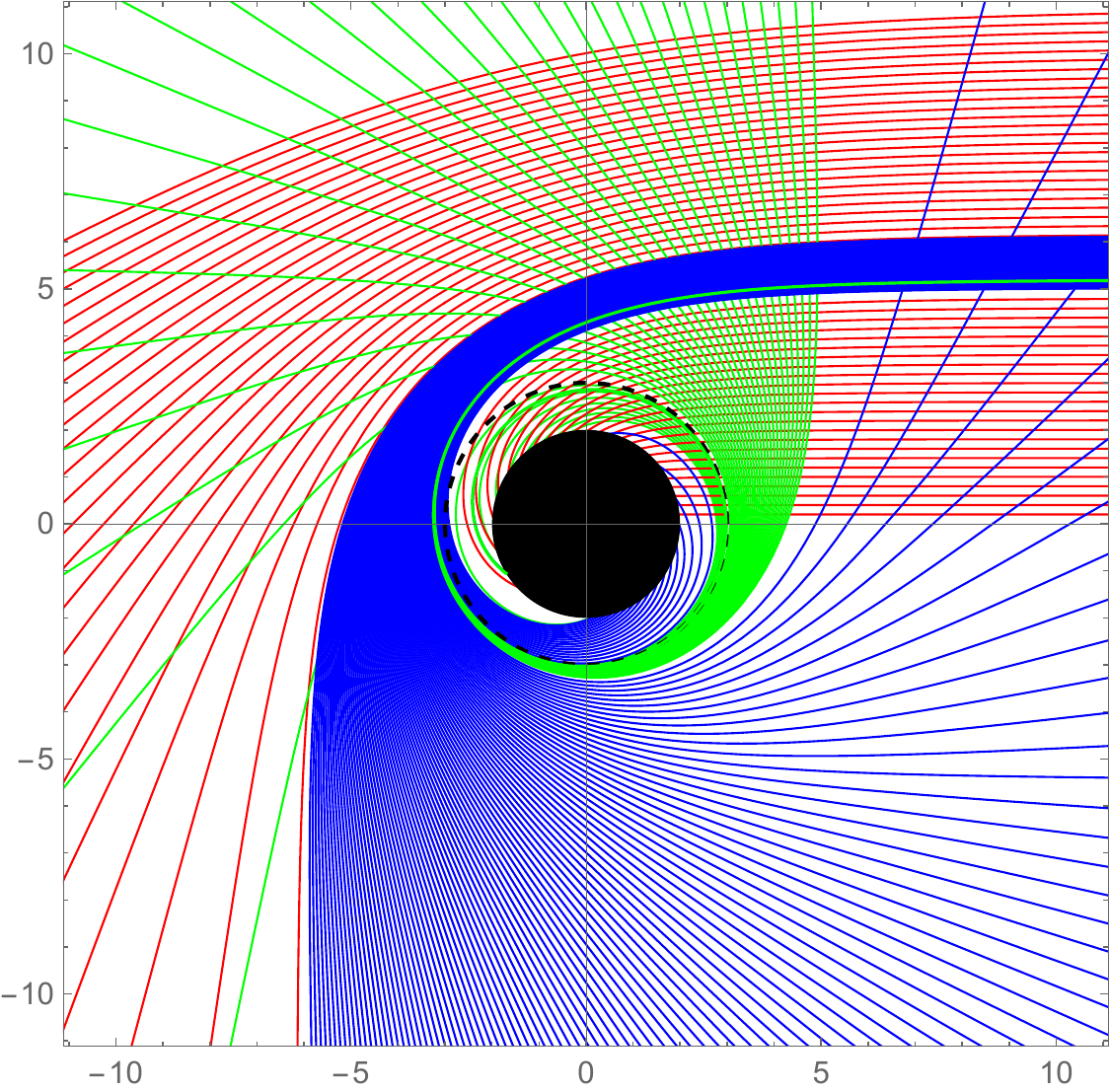}\\
			\vspace{0.1cm}
                \includegraphics[width=2.1in]{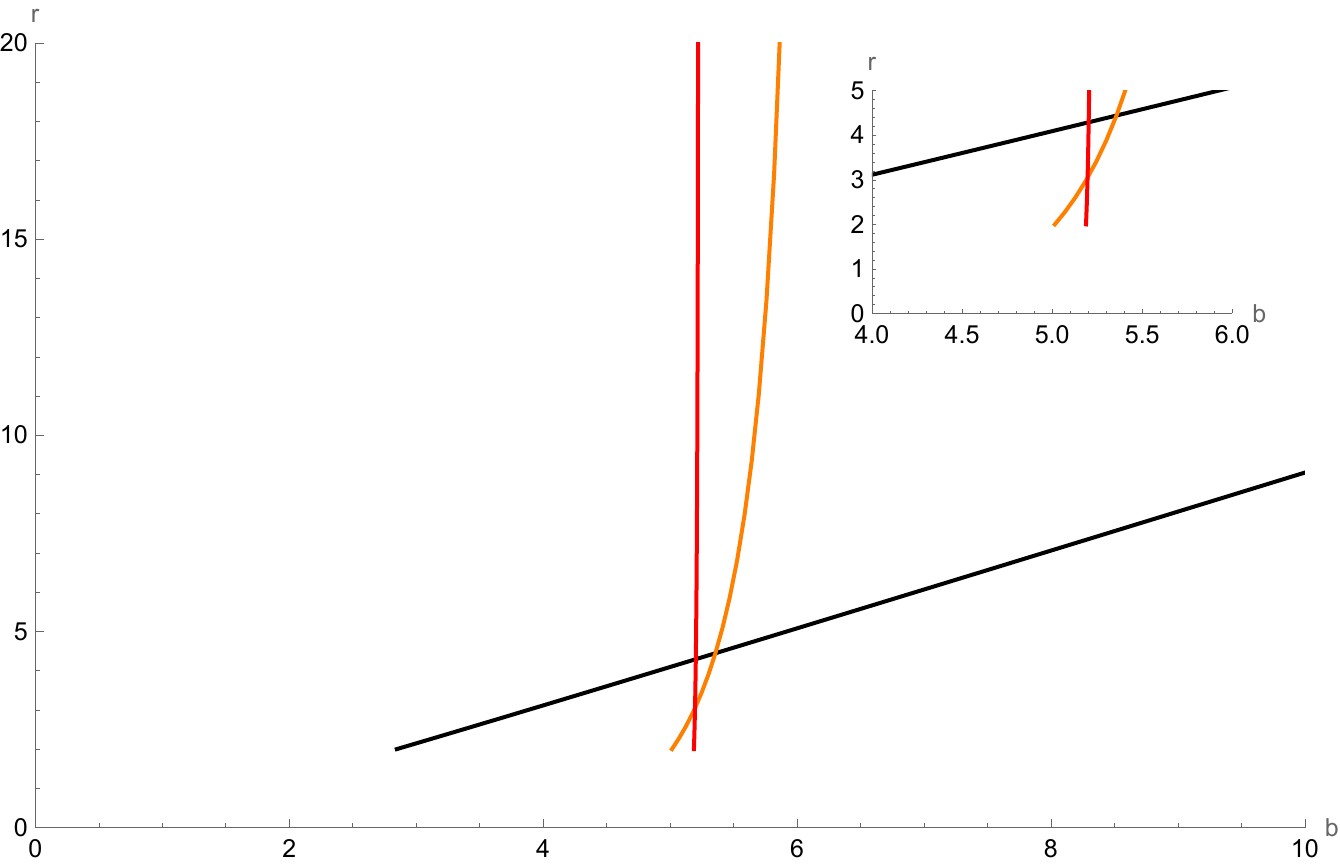}\\
			\vspace{0.1cm}
		\end{minipage}%
	}%
	\subfigure[dS core]{
		\begin{minipage}[t]{0.33\linewidth}
			\centering
			\includegraphics[width=2.1in]{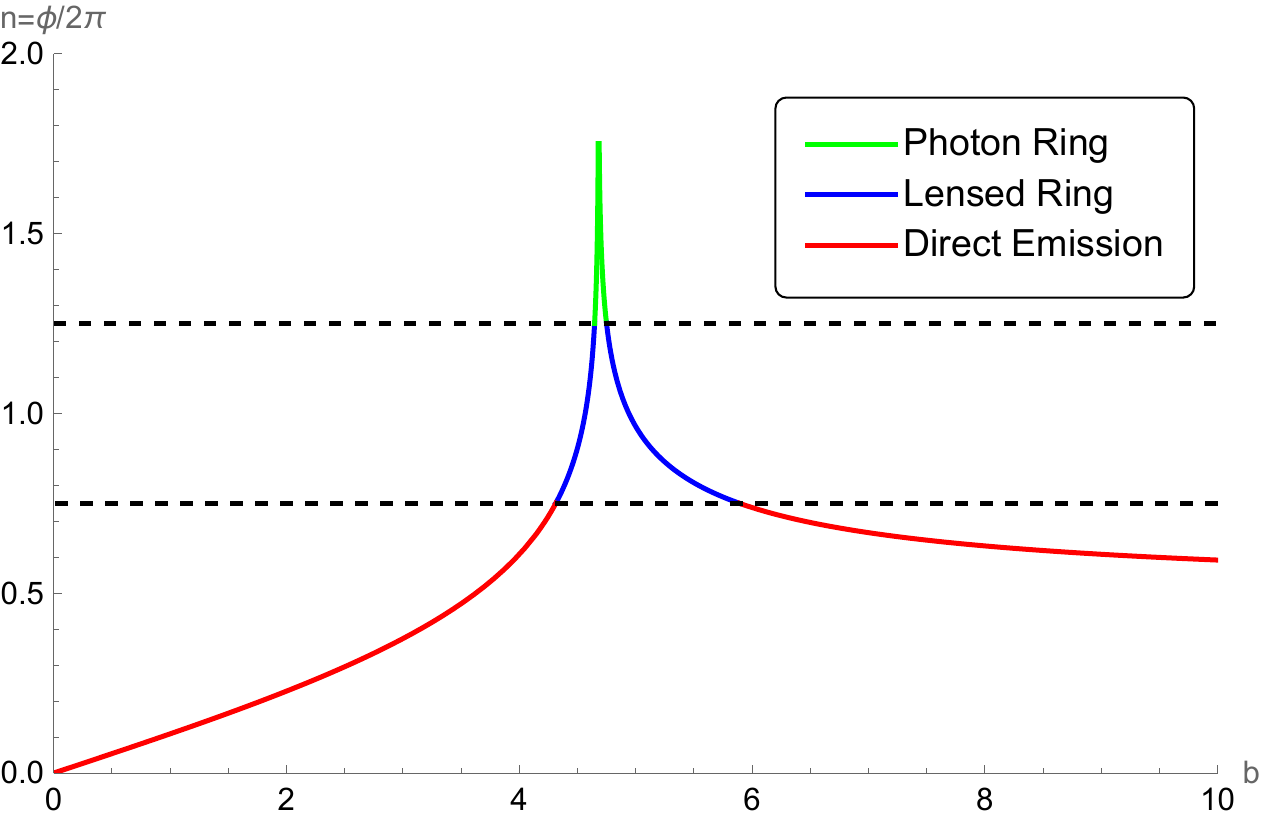}\\
			\vspace{0.1cm}
                \includegraphics[width=2.1in]{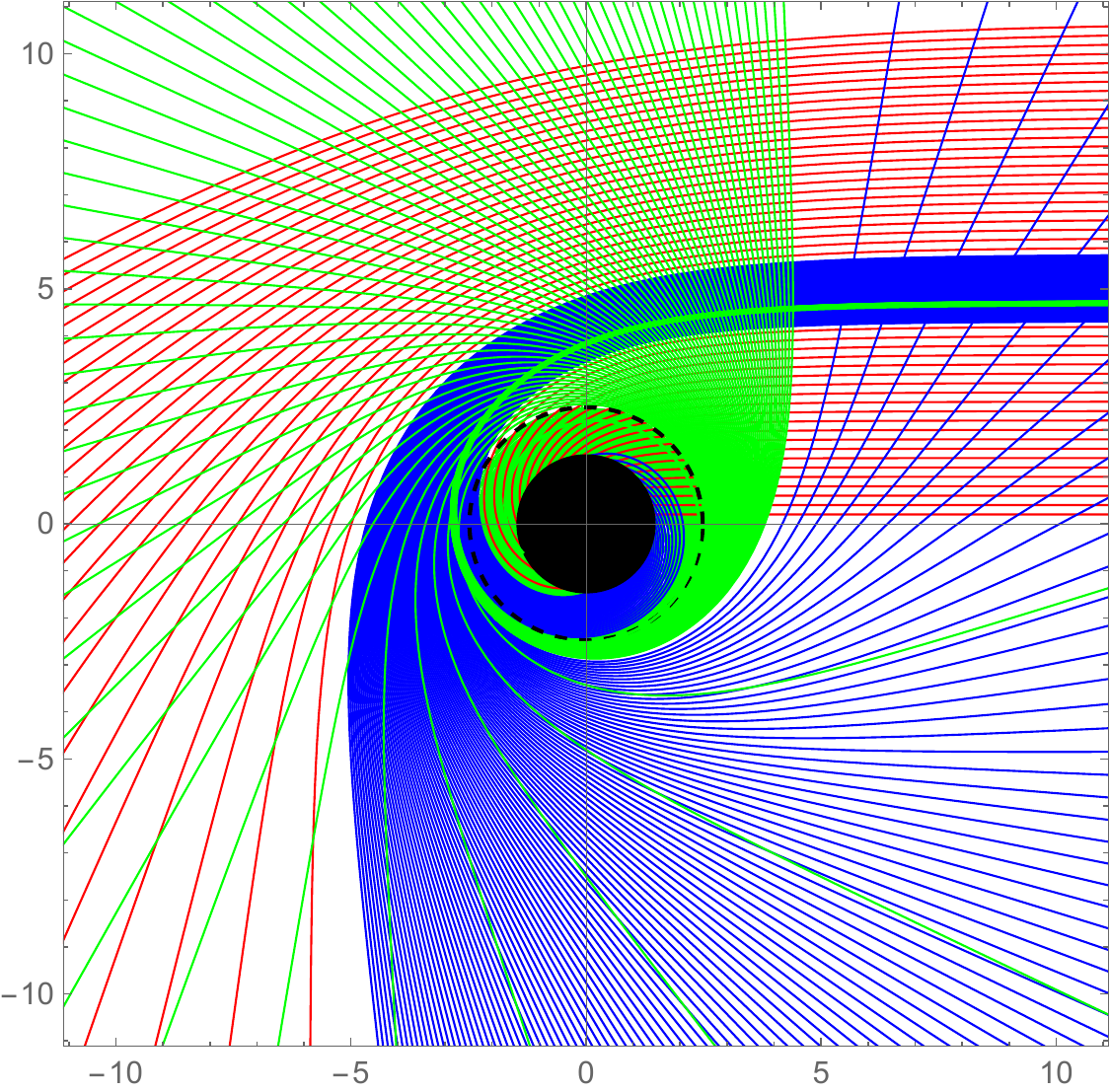}\\
			\vspace{0.1cm}
                \includegraphics[width=2.1in]{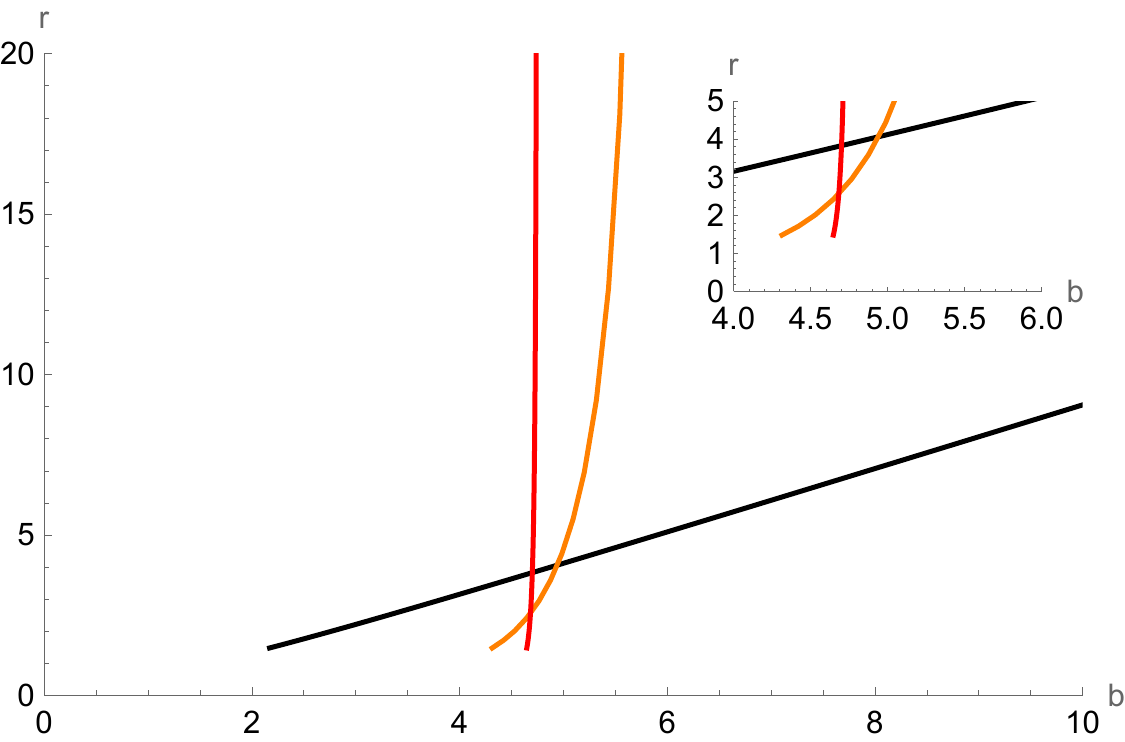}\\
			\vspace{0.1cm}
		\end{minipage}%
	}%
	\subfigure[Minkowskian core]{
		\begin{minipage}[t]{0.33\linewidth}
			\centering
			\includegraphics[width=2.1in]{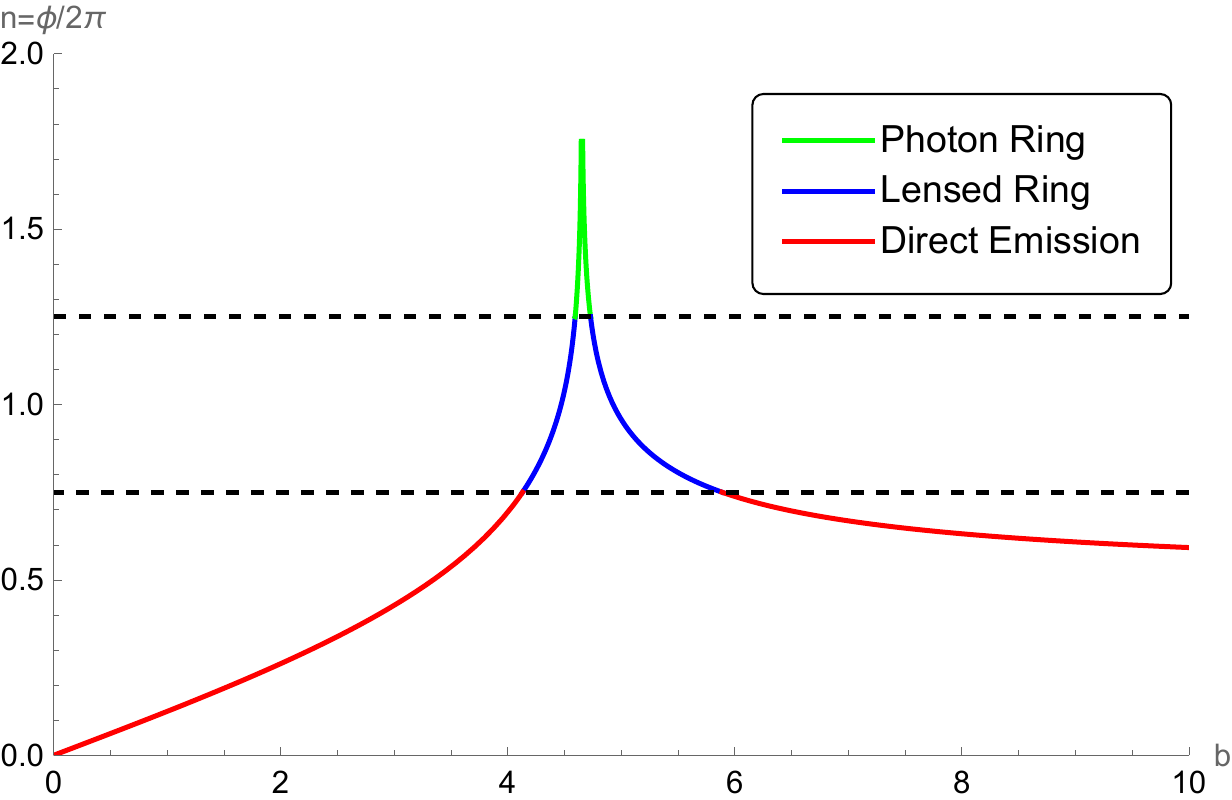}\\
			\vspace{0.1cm}
                \includegraphics[width=2.1in]{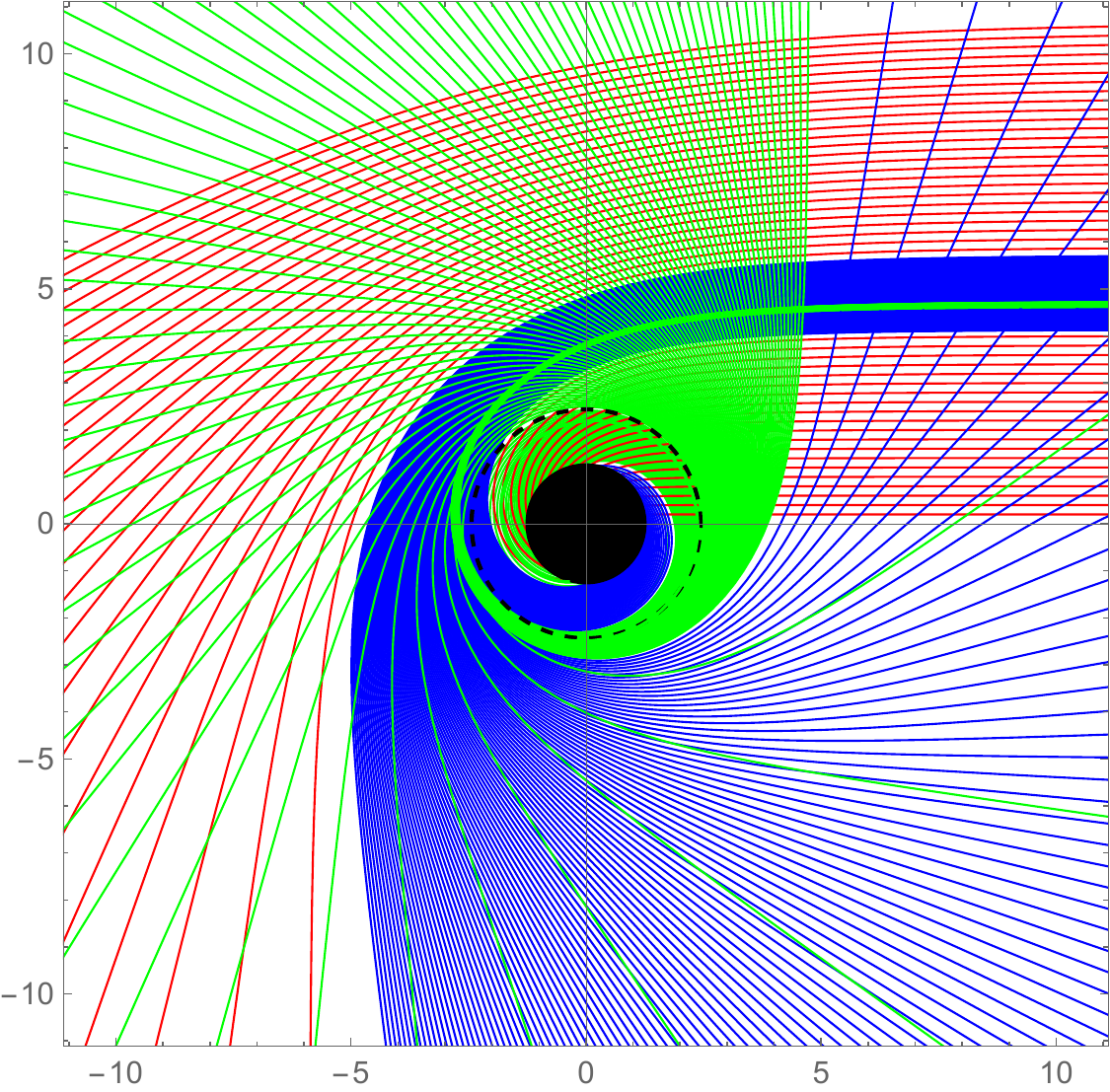}\\
			\vspace{0.1cm}
                \includegraphics[width=2.1in]{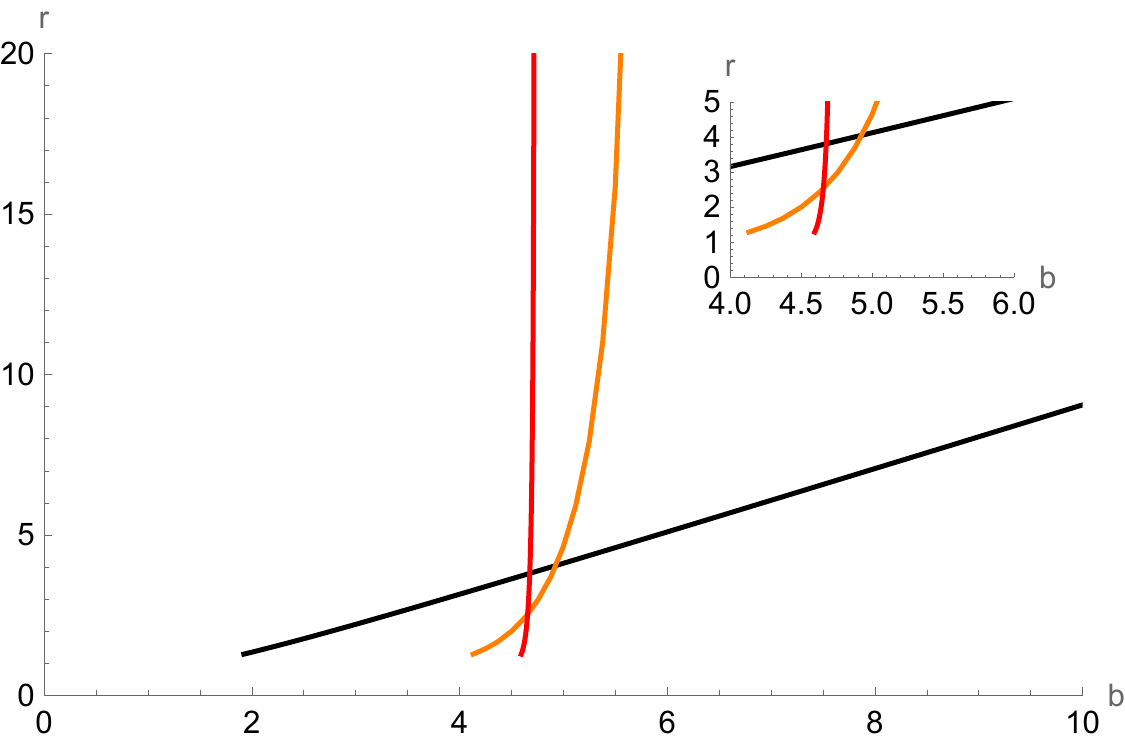}\\
			\vspace{0.1cm}
		\end{minipage}
	}
	\centering
	\caption{Light rays around the black hole in different spacetime and transfer functions. The first row shows the relationship between the number of times the light passes through a thin disk and the impact parameter. The second row shows the motion of light around the black hole, and the last row shows the images of the transfer function. And the first column shows the Schwarzschild black hole, the second column shows Bardeen black hole with dS core with $\alpha_0=0.73$, and the third column shows the black holes with Minkowskian core with $x=2/3$, $n=2$, and $\alpha_0=0.73$.}
	\vspace{-0.2cm}
	\label{fig.TFBD}
\end{figure*}

First of all, using Eq.(\ref{Eq.TM}) one can obtain the range of photon ring, lensed ring and direct emission ray in these three different spacetimes, as summarized in table.(\ref{LR.BD}). In parallel,  we  plot the light trajectories and transfer functions around three types of black holes in Fig.(\ref{fig.TFBD}),  which includes the Schwarzschild black hole. From table.(\ref{LR.BD}) and Fig.(\ref{fig.TFBD}), we notice that the width of photon ring and lensed ring for the black hole with a Minkowskian core is wider than that of the black hole with a dS core. As described in \cite{Ling:2022vrv}, for black holes with a Minkowskian core, the maximal value of the Kretschmann scalar moves toward the horizon as the deviation parameter $\alpha_0$ increases, while for black holes with a dS core, the maximal value of the Kretschmann scalar always locates at the center of the black hole. This implies that when $\alpha_0$ takes a larger value, the attraction of black holes with a Minkowskian core for photons becomes stronger, so the width of the photon ring is wider than that of black holes with a dS core. Therefore, for observation, the photon ring of a black hole with a Minkowskian core is comparably easier to detect.
\begin{table}[htbp]
\centering
\caption{The ranges of impact parameter $b$ corresponding to the direct emission, lensed ring emission and photon ring emission of black holes with $x=2/3$ and $n=2$}
\label{LR.BD}
\resizebox{\textwidth}{15mm}{
    \begin{tabular}{|c|c|c|c|c|c}\hline 
\quad\quad& Direct$(n<3/4)$& lensed ring$(3/4<n<5/4)$ & Photon ring$(n>5/4)$   \\ \hline
 Schwarzschild& $b<5.01514$ and $b>6.16757$&$5.01514<b<5.18781$ and $5.22794<b<6.16757$ &$5.18781<b<5.22794$\\ \hline
Bardeen& $b<4.31464$ and $b>5.89258$&$4.31464<b<4.64795$ and $4.75061<b<5.89258$ &$4.64795<b<4.75061$ \\ \hline
$x=2/3$,$n=3$& $b<4.13119$ and $b>5.88825$&$4.13119<b<4.59487$ and $4.72978<b<5.88825$ &$4.59487<b<4.72978$\\ \hline
\end{tabular}}
\end{table}
\par  We can see more clues from the transfer function of these three types of black holes. In the bottom row of Fig.(\ref{fig.TFBD}), we notice that comparing with that of the Schwarzschild black hole, the demagnification factor of both regular black holes is suppressed by the parameter $\alpha_0$, and the suppression of black holes with Minkowskian core is stronger. This implies that the photon ring of black holes with Minkowskian core accounts for a higher proportion of the observed total flux and is easier to observe.

\begin{figure}
	\centering
	\subfigure[Schwarzschild]{
		\begin{minipage}[t]{0.33\linewidth}
			\centering
			\includegraphics[width=2.1in]{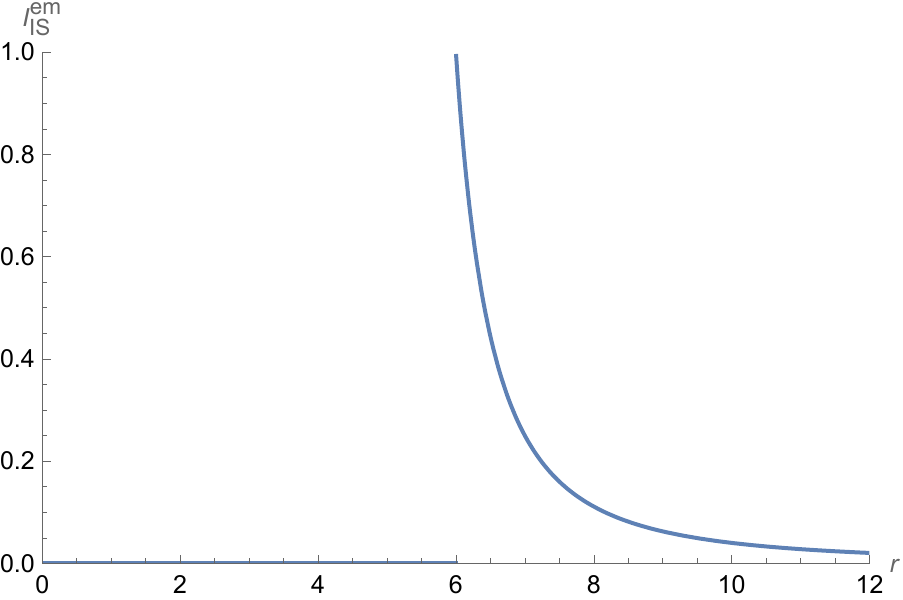}\\
			\vspace{0.1cm}
			\includegraphics[width=2.1in]{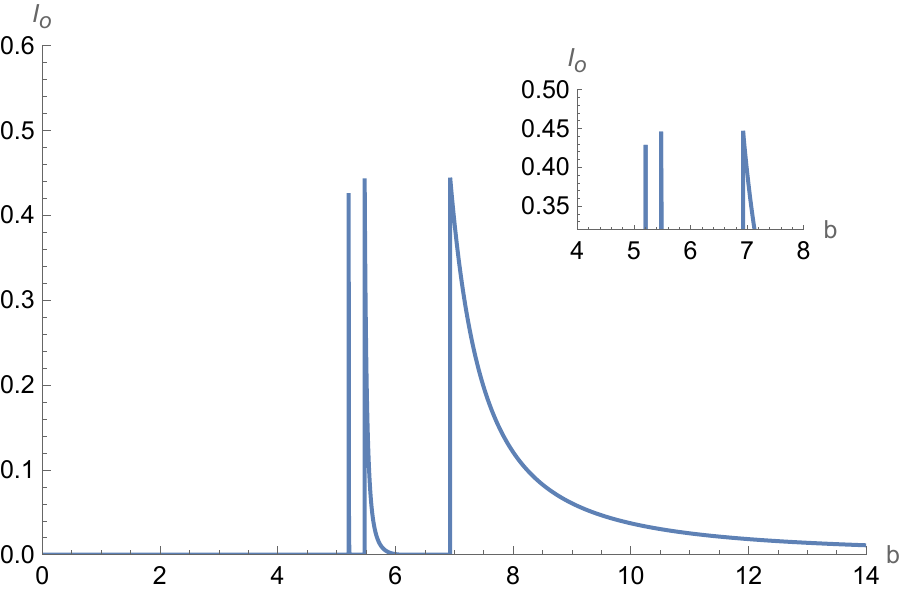}\\
			\vspace{0.1cm}
                \includegraphics[width=2.1in]{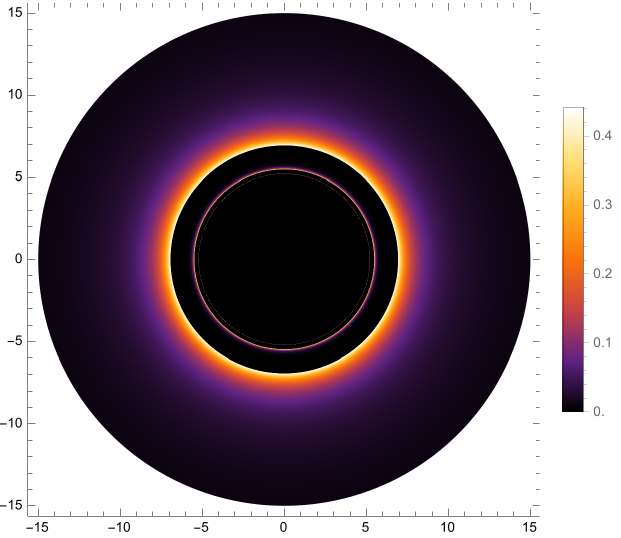}\\
			\vspace{0.1cm}
		\end{minipage}%
	}%
	\subfigure[dS Core]{
		\begin{minipage}[t]{0.33\linewidth}
			\centering
			\includegraphics[width=2.1in]{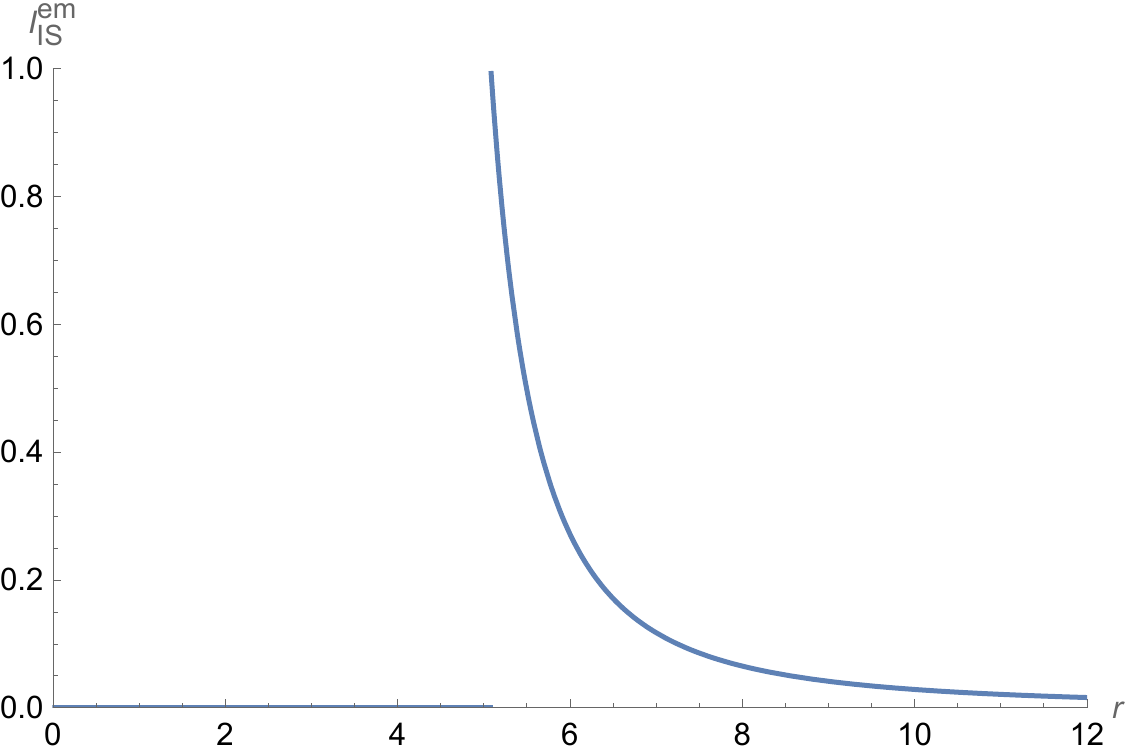}\\
			\vspace{0.1cm}
			\includegraphics[width=2.1in]{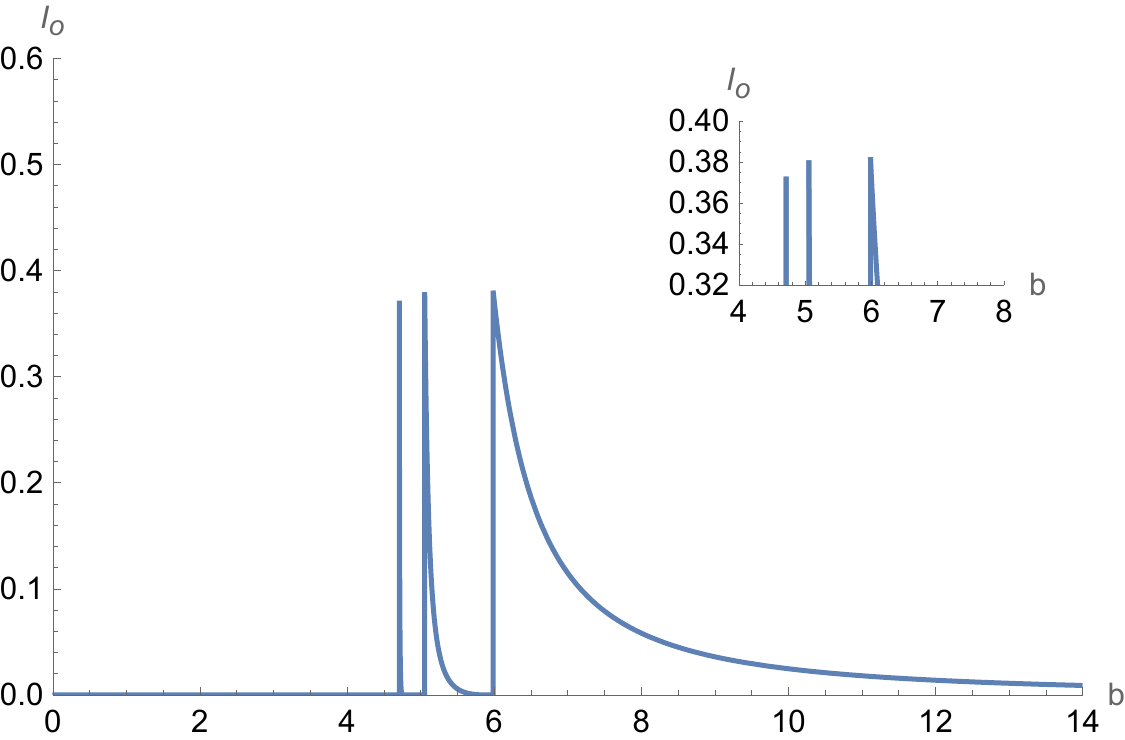}\\
			\vspace{0.1cm}
                \includegraphics[width=2.1in]{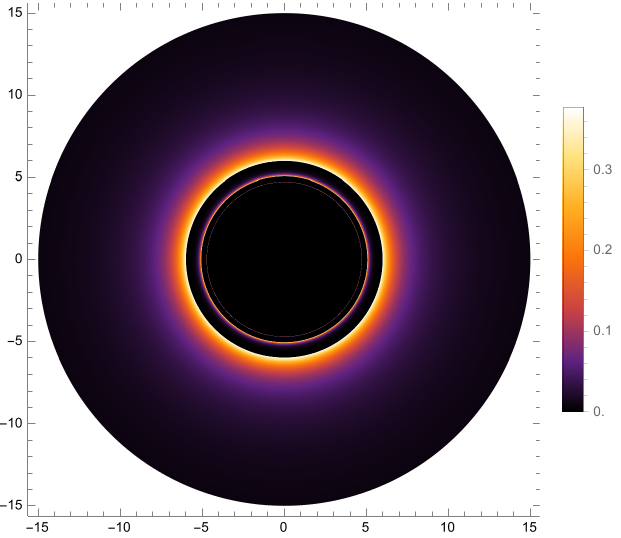}\\
			\vspace{0.1cm}
		\end{minipage}%
	}%
	\subfigure[Minkowskian Core]{
		\begin{minipage}[t]{0.33\linewidth}
			\centering
			\includegraphics[width=2.1in]{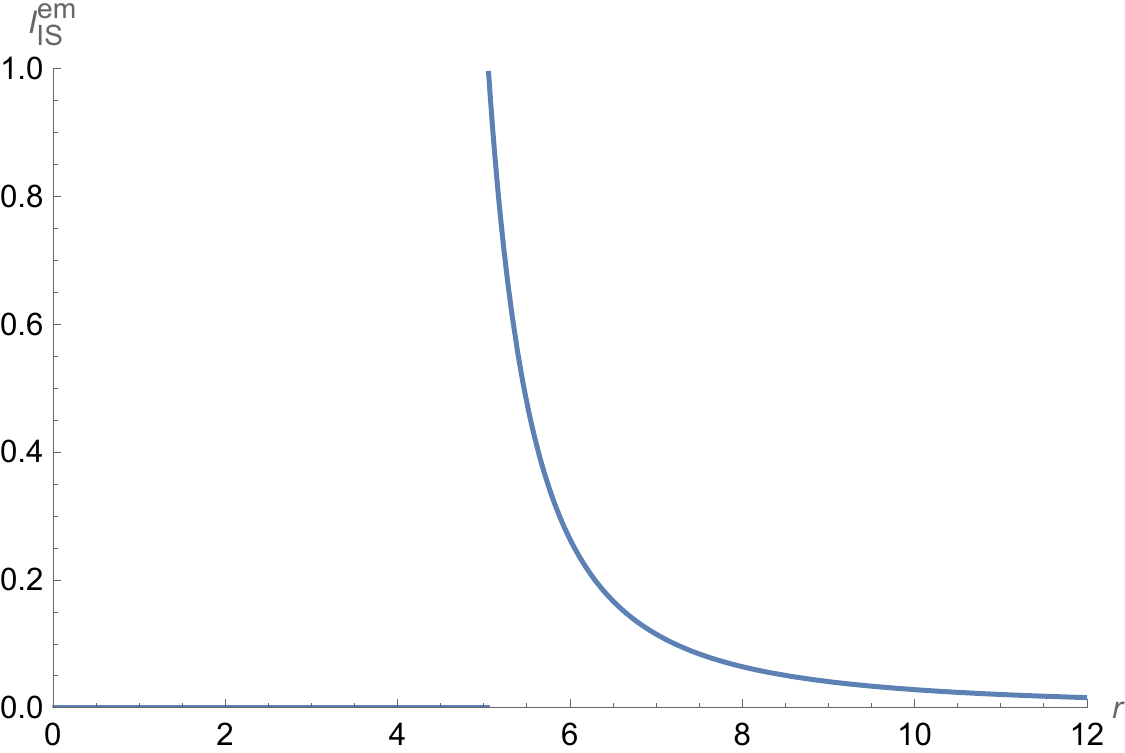}\\
			\vspace{0.1cm}
			\includegraphics[width=2.1in]{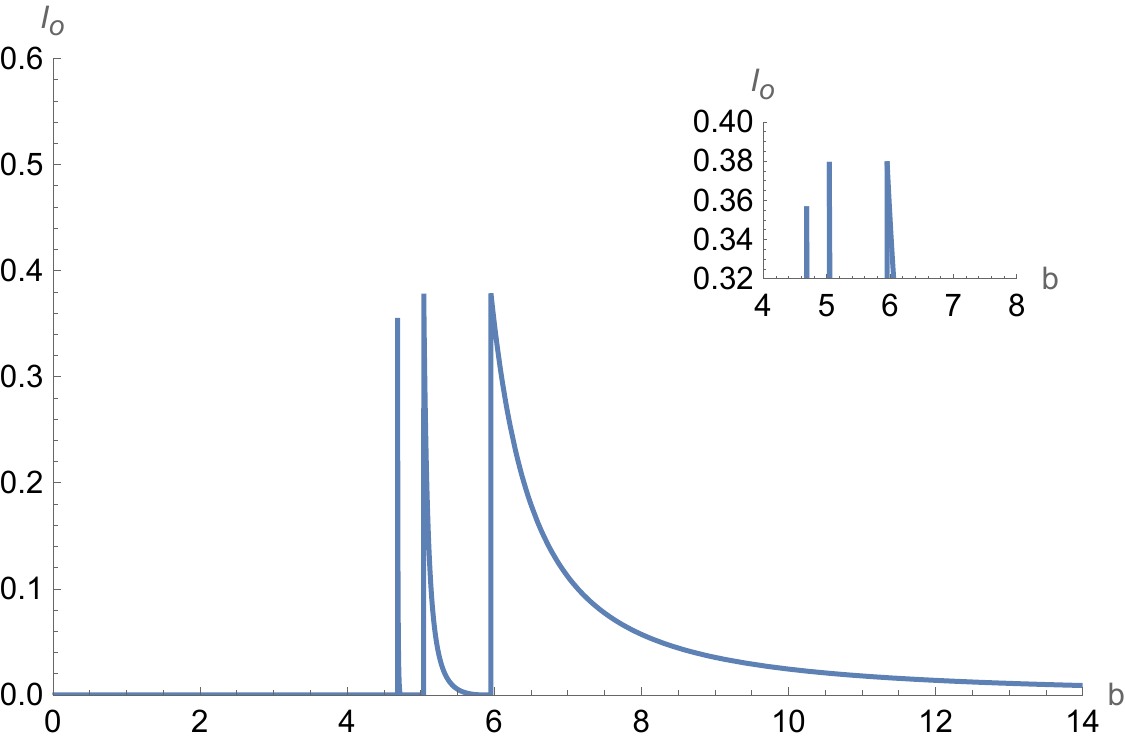}\\
			\vspace{0.1cm}
                \includegraphics[width=2.1in]{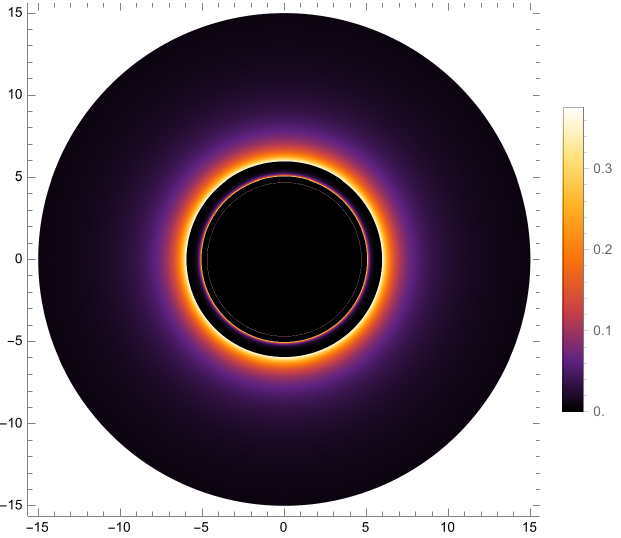}\\
			\vspace{0.1cm}
		\end{minipage}
	}
	\centering
	\caption{Observational appearances  of different black holes under the ISCO emission model. The top row represents the variation of emission intensity with respect to $r$ under the ISCO emission model. The middle row represents the intensity received by the observer, and the bottom row represents the two-dimensional observed image. Each column from left to right shows the results for  Schwarzschild black hole, Bardeen black hole with a dS core, and the regular black hole with Minkowskian core, respectively. We fix $\alpha_0=0.73$. }
	\vspace{-0.2cm}
	\label{fig.ISBD}
\end{figure}

Next we compare the images of Schwarzschild black hole, Bardeen black hole and the black hole with Minkowskian core surrounded by thin accretion disk. Firstly we consider the ISCO emission model described in Eq.(\ref{EQ.Mod1}). The result is illustrated in Fig.(\ref{fig.ISBD}).
We remark that although the observed flux comes from the direct emission, the locations and heights of their peaks are different for these three types of black holes. 
The second row in Fig.(\ref{fig.ISBD}) demonstrates the intensity received by the observer. It is noticed that the position and the value of peaks are different in three types of black holes. In particular, Schwarzschild black hole exhibits significant differences in comparison with the other two regular black holes with $\alpha_0=0.73$. The peak value of Schwarzschild black hole is higher than that of the other two regular black holes, and its position has a larger impact parameter such that its black disk looks bigger than that of regular black holes with the same mass, which might be used to distinguish Schwarzschild black hole from regular black holes. As far as two regular black holes are concerned, we find the peak value of intensity for regular black hole with Minkowskian core is smaller, and the position of the peak appears earlier. This potentially provides a way for us to distinguish between these two types of black holes in this emission model.

\par
\begin{figure}
	\centering
	\subfigure[Schwarzschild]{
		\begin{minipage}[t]{0.33\linewidth}
			\centering
			\includegraphics[width=2.1in]{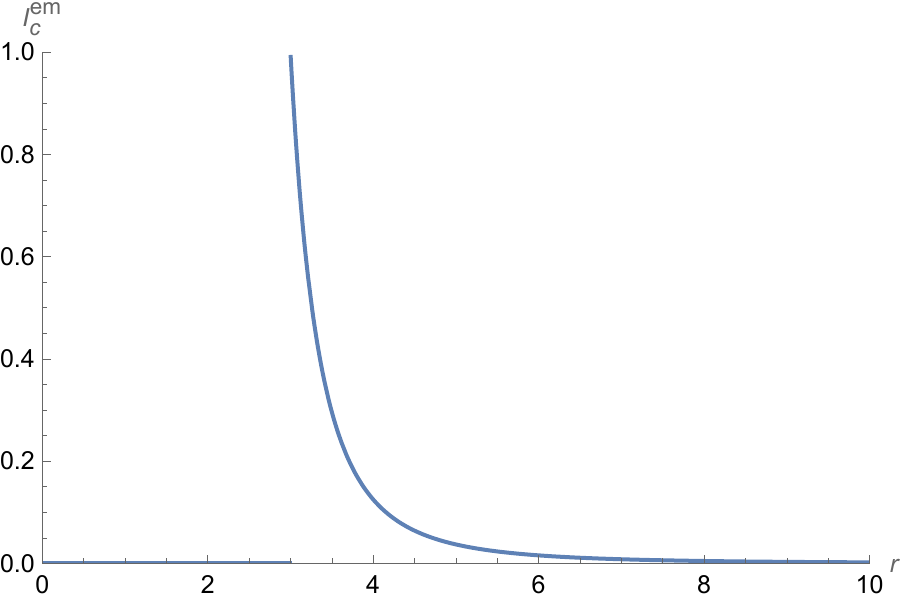}\\
			\vspace{0.1cm}
			\includegraphics[width=2.1in]{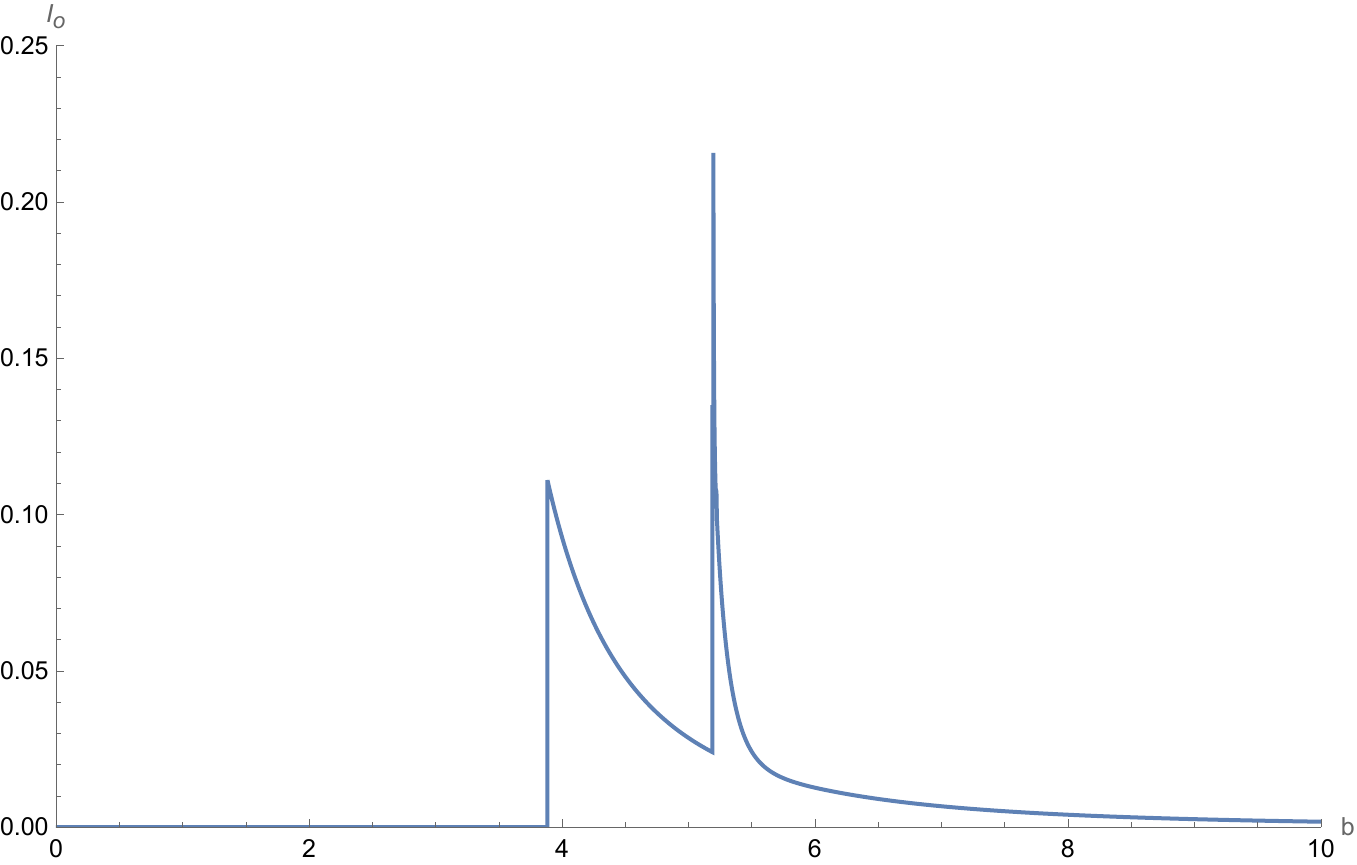}\\
			\vspace{0.1cm}
                \includegraphics[width=2.1in]{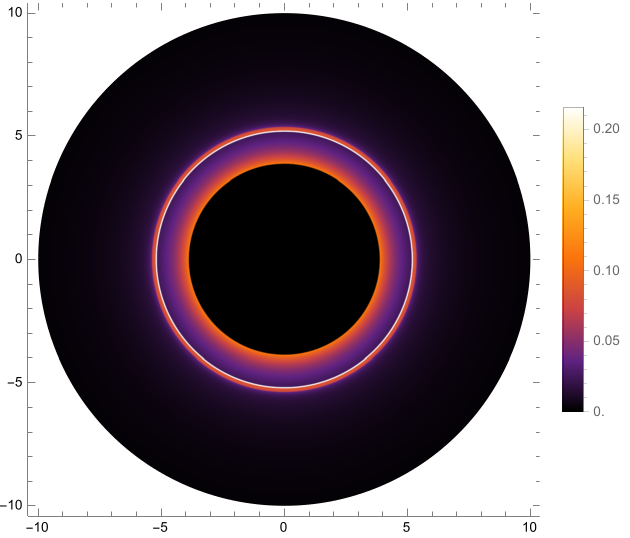}\\
			\vspace{0.1cm}
		\end{minipage}%
	}%
	\subfigure[dS Core]{
		\begin{minipage}[t]{0.33\linewidth}
			\centering
			\includegraphics[width=2.1in]{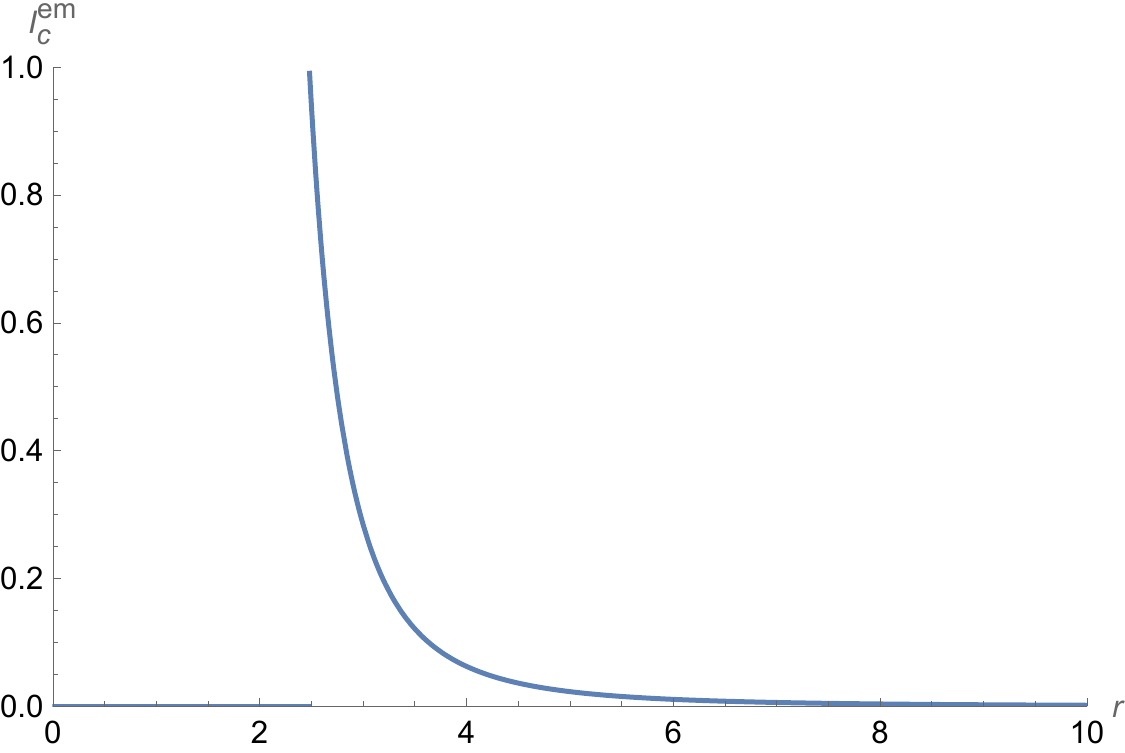}\\
			\vspace{0.1cm}
			\includegraphics[width=2.1in]{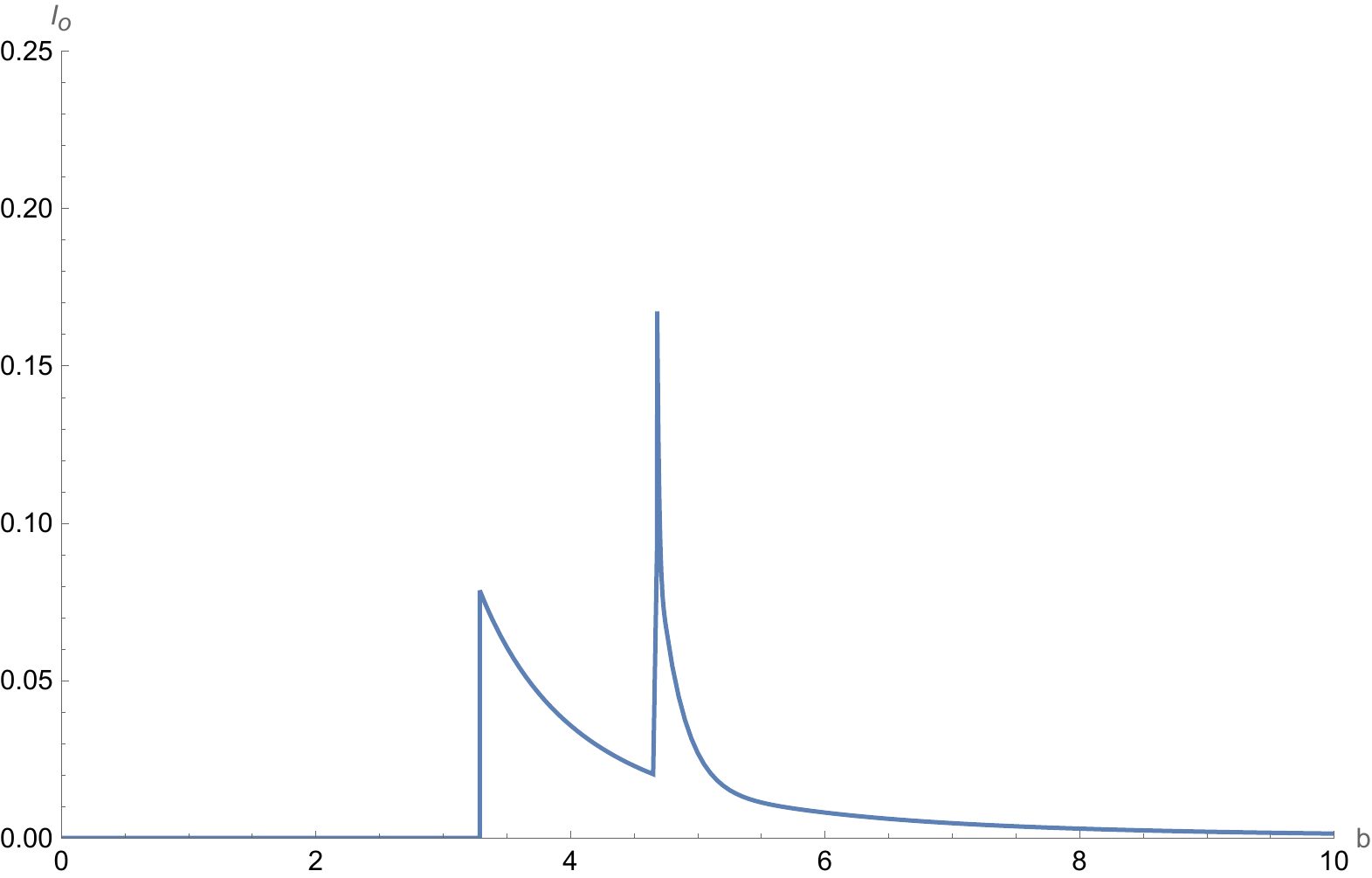}\\
			\vspace{0.1cm}
                \includegraphics[width=2.1in]{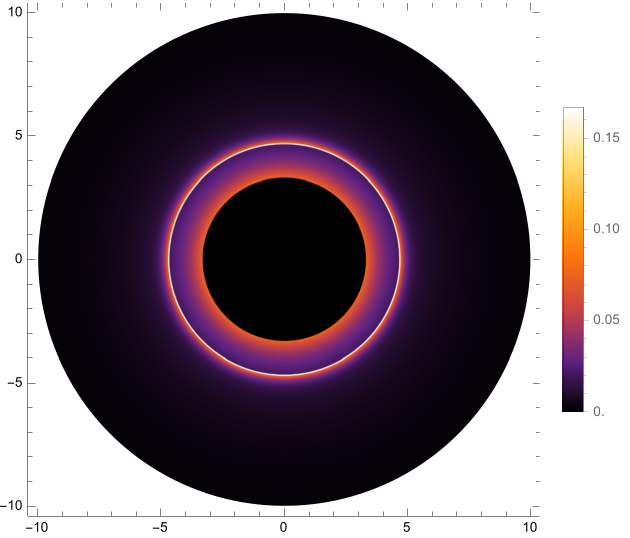}\\
			\vspace{0.1cm}
		\end{minipage}%
	}%
	\subfigure[Minkowskian Core]{
		\begin{minipage}[t]{0.33\linewidth}
			\centering
			\includegraphics[width=2.1in]{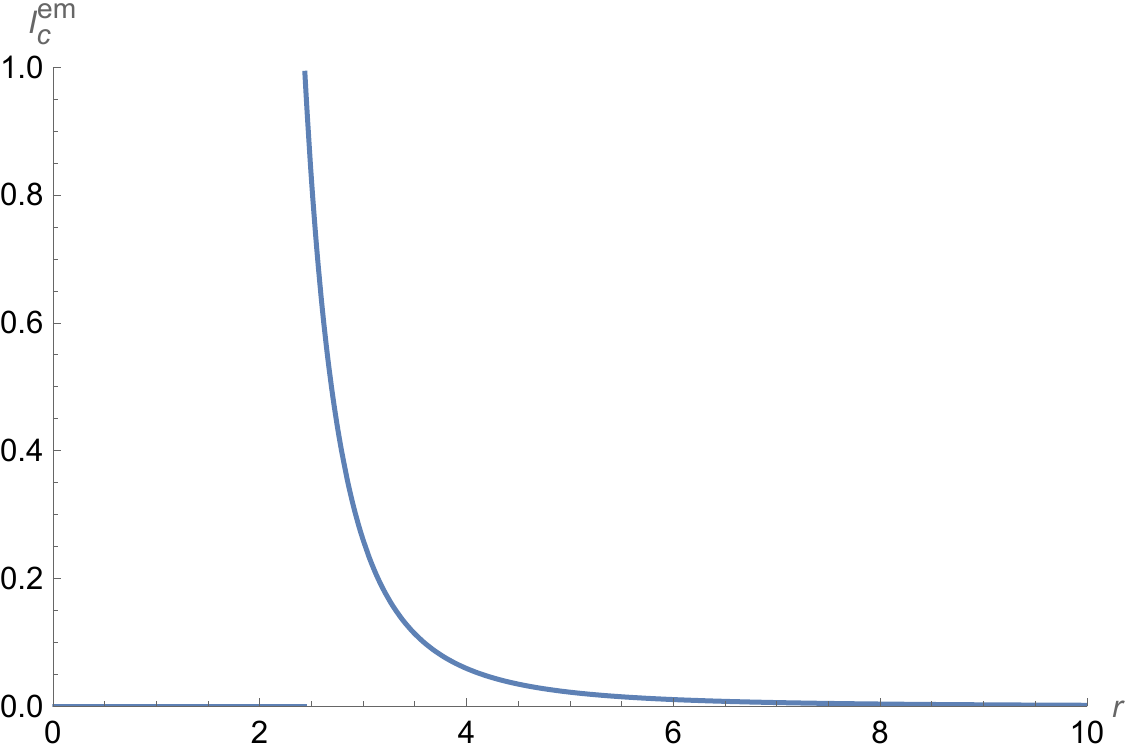}\\
			\vspace{0.1cm}
			\includegraphics[width=2.1in]{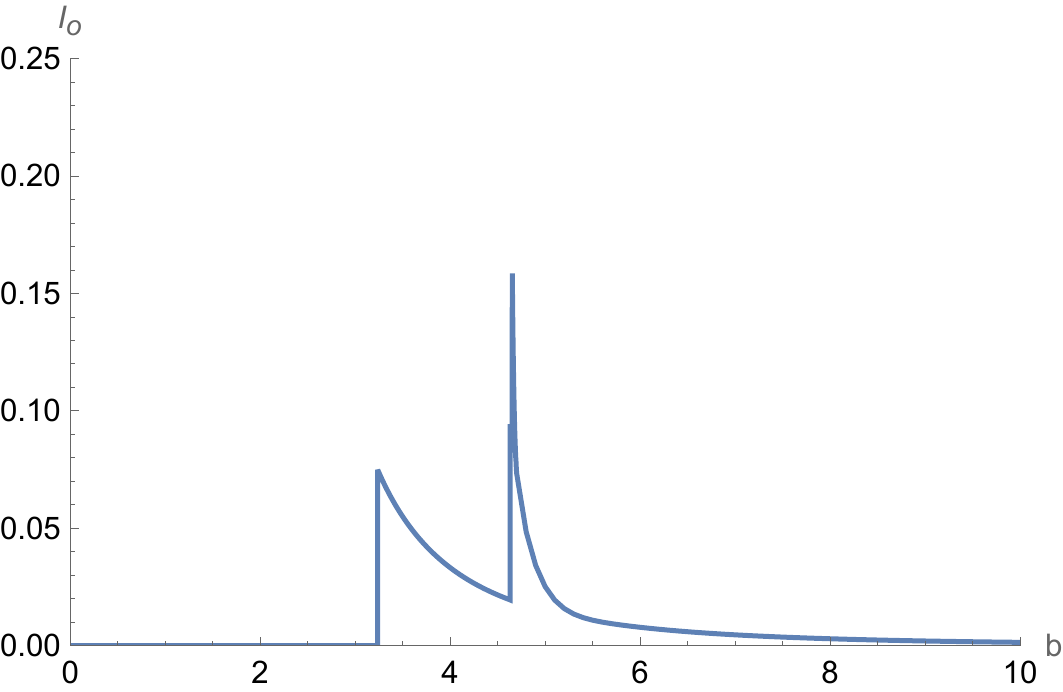}\\
			\vspace{0.1cm}
                \includegraphics[width=2.1in]{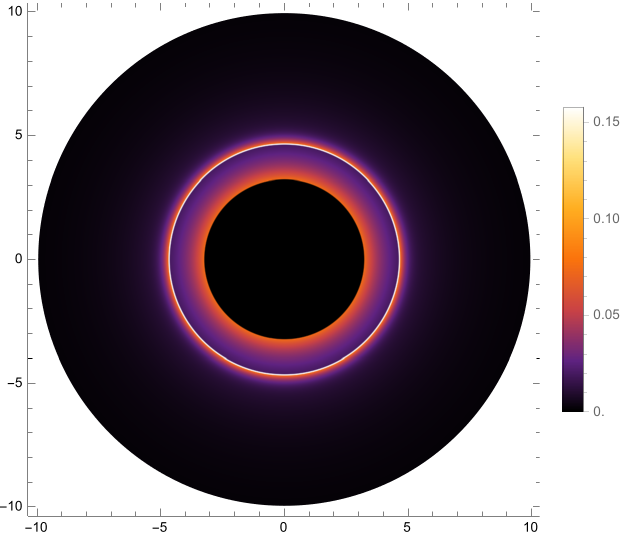}\\
			\vspace{0.1cm}
		\end{minipage}
	}
	\centering
	\caption{The Observational characteristics of different black holes under the photon sphere emission model. The top row represents the variation of emission intensity with respect to $r$ under the photon sphere emission model. The middle row represents the intensity received by the observer, and the bottom row represents the two-dimensional observed image. Each column from left to right shows the results for  Schwarzschild black hole, Bardeen black hole with a dS core, and the regular black hole with Minkowskian core, respectively. We fix $\alpha_0=0.73$. 
 }
	\vspace{-0.2cm}
	\label{fig.PEBD}
\end{figure}

 Next, we compare their images generated by the photon sphere emission model described by Eq.(\ref{Eq.Mod2}). More details are illustrated in Fig.(\ref{fig.PEBD}). For all three types of black holes, the peaks of lensed ring and photon ring coincide and the main source of observed flux comes from the direct emission. The observational intensity of the lensed rings and photon rings for black hole with a Minkowskian core and that with a dS core is significantly less than that of the lensed ring of a Schwarzschild black hole, For Bardeen black hole with a dS core and black hole with a Minkowskian core, the first peak corresponds to the direct emission and exhibits very similar position and intensity. However, the peak values of the lensed ring and photon ring are different. Similarly, the peak values of black holes with a Minkowskian core are lower than those of black holes with a dS core. Therefore, in this model, studying the peak values corresponding to the lensed ring and photon ring is important for distinguishing between these two types of black holes.

\begin{figure}
	\centering
	\subfigure[Schwarzschild]{
		\begin{minipage}[t]{0.33\linewidth}
			\centering
			\includegraphics[width=2.1in]{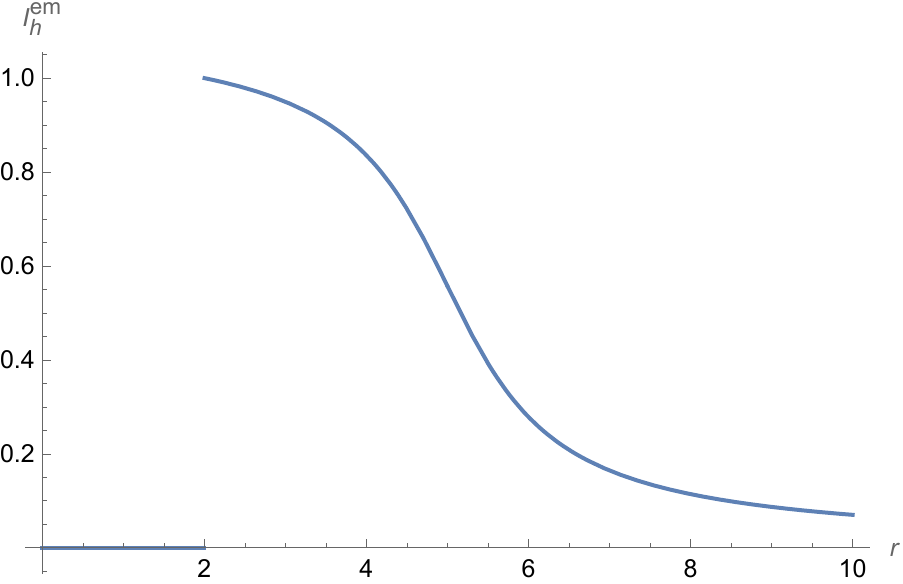}\\
			\vspace{0.1cm}
			\includegraphics[width=2.1in]{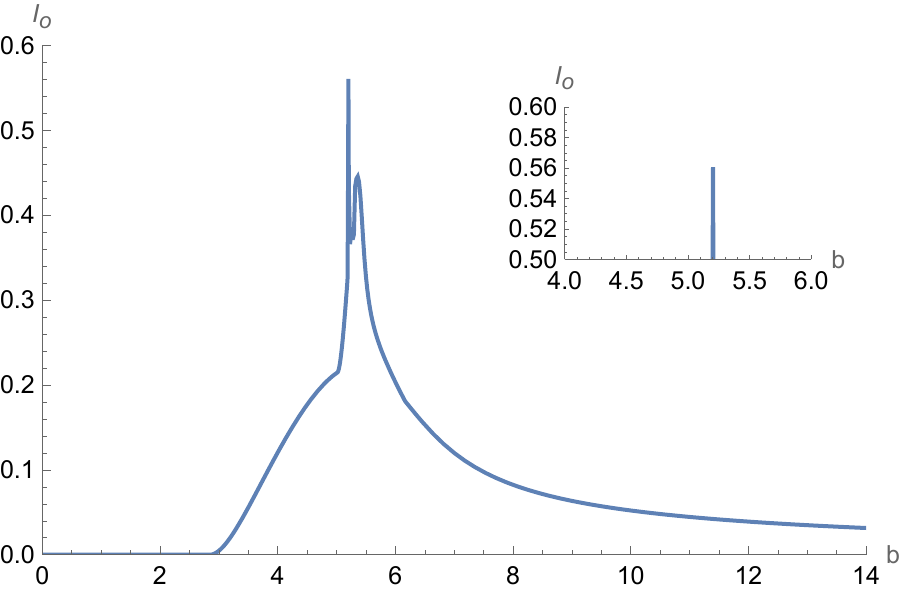}\\
			\vspace{0.1cm}
                \includegraphics[width=2.1in]{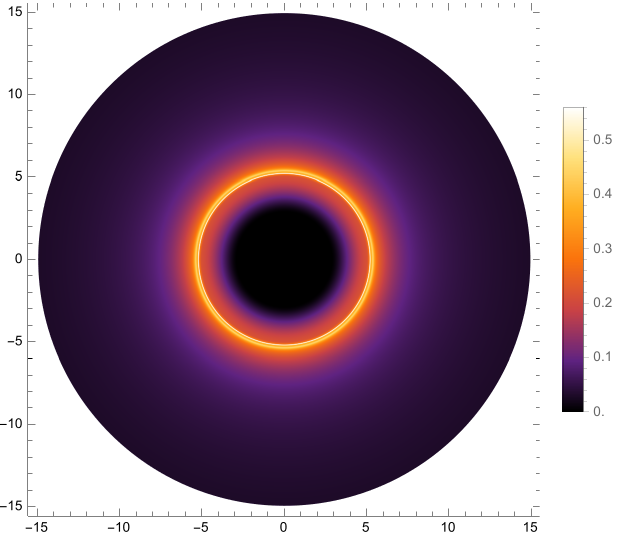}\\
			\vspace{0.1cm}
		\end{minipage}%
	}%
	\subfigure[dS Core]{
		\begin{minipage}[t]{0.33\linewidth}
			\centering
			\includegraphics[width=2.1in]{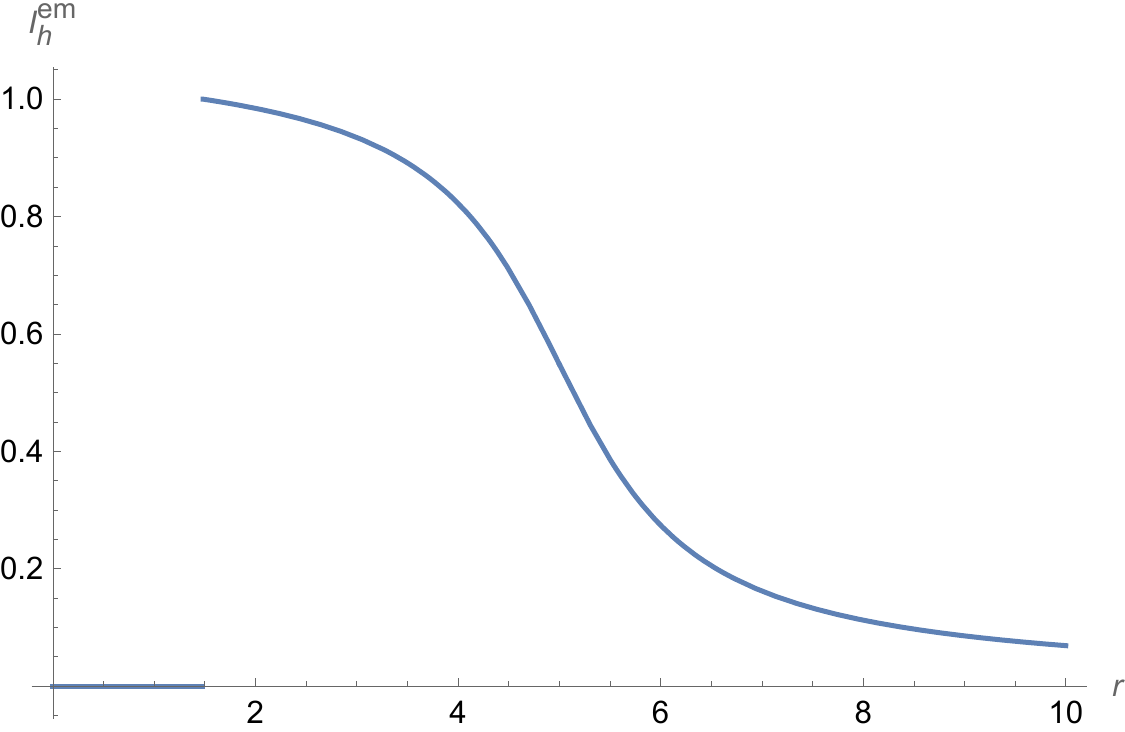}\\
			\vspace{0.1cm}
			\includegraphics[width=2.1in]{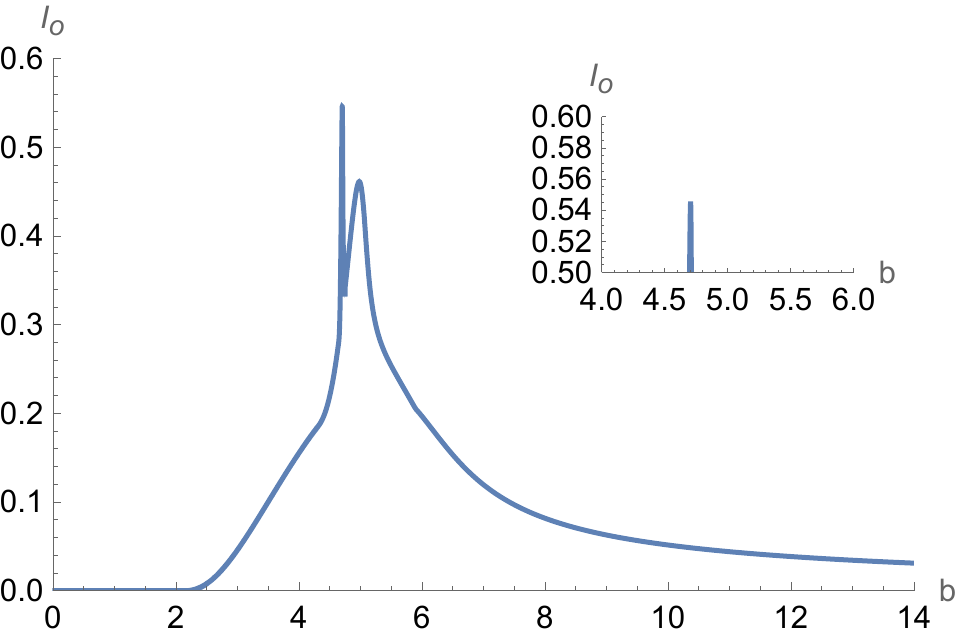}\\
			\vspace{0.1cm}
                \includegraphics[width=2.1in]{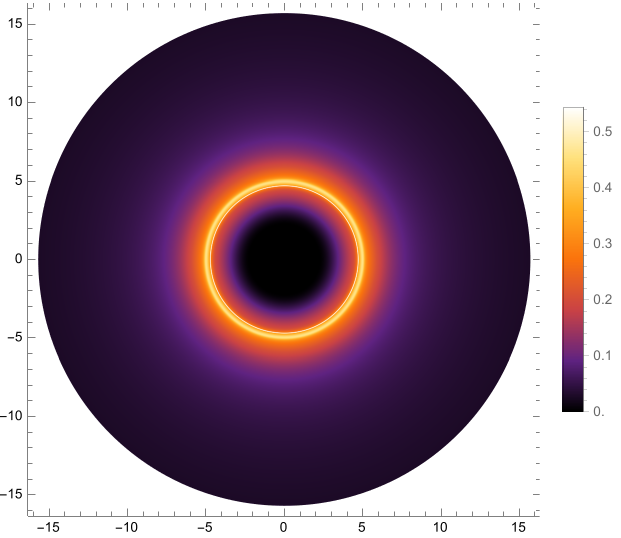}\\
			\vspace{0.1cm}
		\end{minipage}%
	}%
	\subfigure[Minkowskian Core]{
		\begin{minipage}[t]{0.33\linewidth}
			\centering
			\includegraphics[width=2.1in]{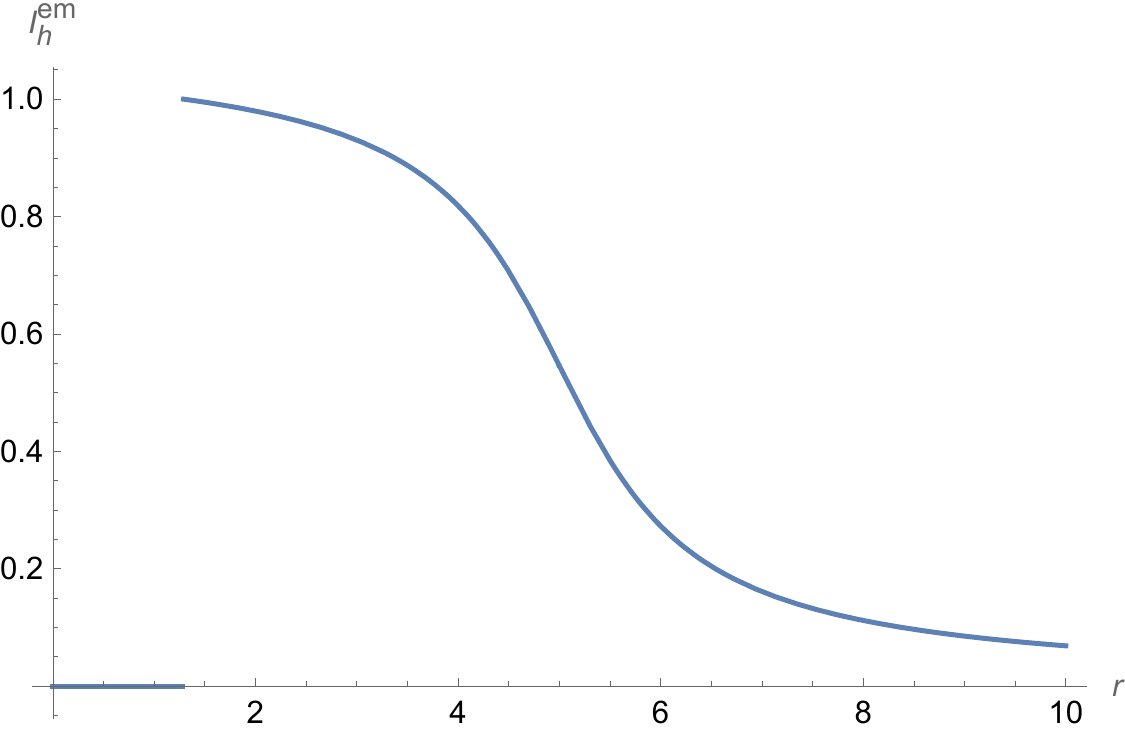}\\
			\vspace{0.1cm}
			\includegraphics[width=2.1in]{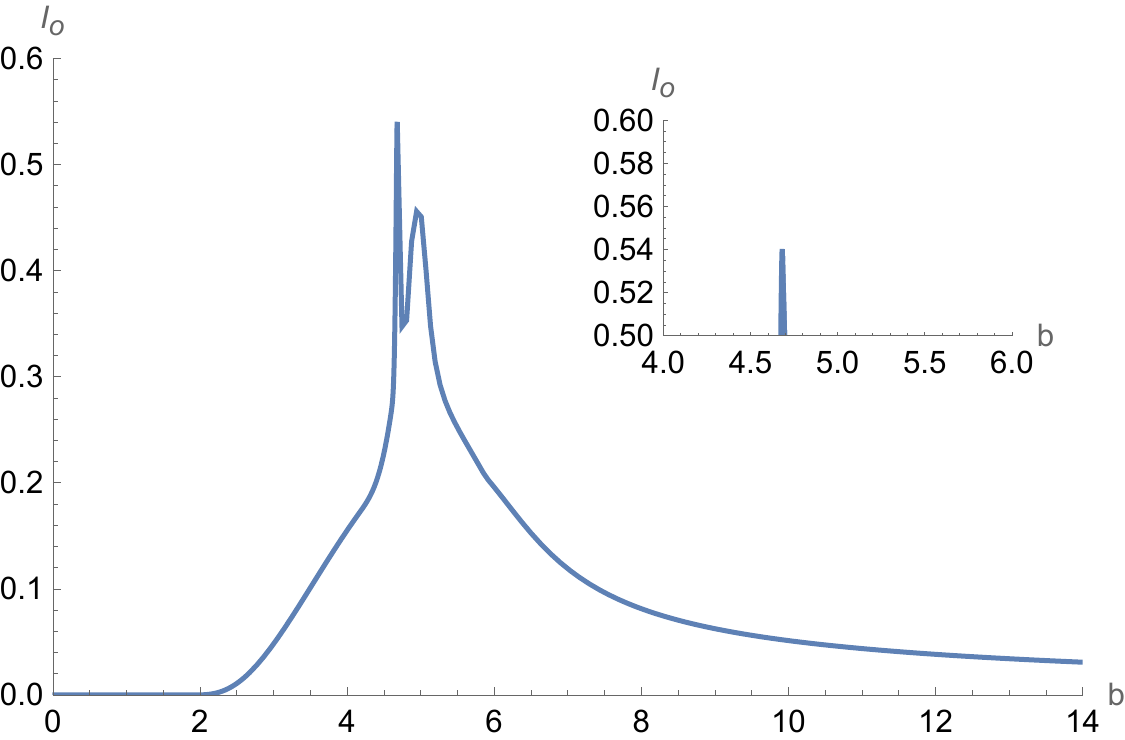}\\
			\vspace{0.1cm}
                \includegraphics[width=2.1in]{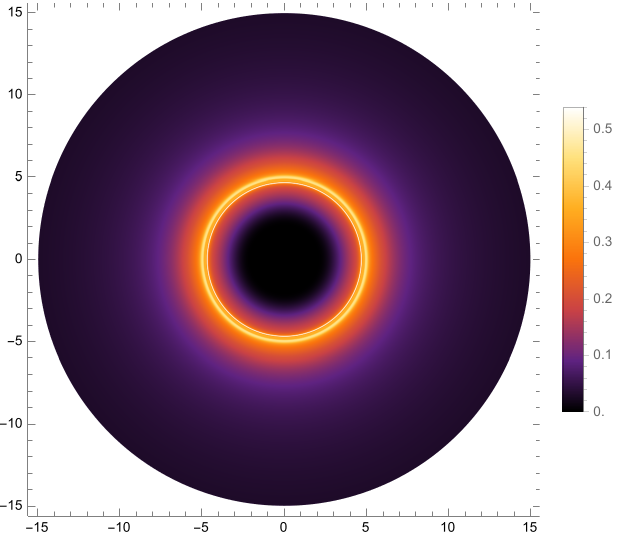}\\
			\vspace{0.1cm}
		\end{minipage}
	}
	\centering
	\caption{The Observational appearances of different black holes under the horizon emission model. The top row represents the variation of emission intensity with respect to $r$ under the horizon emission model. The middle row represents the intensity received by the observer, and the bottom row represents the two-dimensional observed image. Each column from left to right shows the results for  Schwarzschild black hole, Bardeen black hole with a dS core, and the regular black hole with Minkowskian core, respectively. We fix $\alpha_0=0.73$.
 }
	\vspace{-0.2cm}
	\label{fig.EVBD}
\end{figure}

 Finally, we compare  the observational characteristics of these three types of black holes generated by the horizon emission model described by Eq.(\ref{Eq.Mod3}), and present our results in Fig.(\ref{fig.EVBD}). 
 In this model, the peak of the lensed ring is wider than that of photon ring, but the main source of observed flux still comes from the direct emission. For Schwarzschild black holes, the position of intensity peaks differs significantly from that for regular black holes, and thus the location of rings can be used to distinguish Schwarzschild black hole from the other two types of black holes. As for Bardeen black hole, the positions of their two peaks are very similar to those of black holes with Minkowskian cores. However, Bardeen black holes may be distinguished from black holes with Minkowskian cores by the higher observed intensity of their photon ring to an observer.

\section{CONCLUSION AND DISCUSSION}
In this paper, we have investigated the shadow and the optical features of a new sort of regular black holes with sub-Planckian curvature surrounded by a thin accretion disk. Specifically, for $x=2/3$ and $n=2$, we have analyzed  the null geodesics of massless particles around these regular black holes. We have derived the effective potential and obtained the radius of photon sphere and the critical impact parameter. It is found that when the deviation parameter $\alpha_0$ increase, the radius of photon sphere and critical impact parameter decrease. Therefore, the radius of black hole shadow corresponding to the critical impact parameter shrinks as well. Secondly, we have defined three types of light trajectories based on the number of times they pass through the accretion disk, which are denoted as the direct emission, lensed ring and photon ring respectively. We have further obtained the range of these three types of light trajectories and plotted them by the ray tracing code. It is found that the width of photon ring and lensed ring increases with the increase of $\alpha_0$. We have also investigated the transfer function of these three types of light rays in details. It turns out that the slope of the direct emission tends to 1, which means that it is the redshift source and therefore its contribution to the observed flux is dominant, while the slope of the transfer function corresponding to the lensed ring is large, which means that the lensed ring does not contribute much to the observed flux. The largest slope is from the photon ring, which tends to infinity, implying that its contribution to the observed flux is the minimal, and the observed photon ring is a highly demagnetized image. We have found that the deviation parameter $\alpha_0$ suppresses the slope of the transfer function of the lensed ring and photon ring, which means that as $\alpha_0$ increases, the lensed ring and photon ring are more easily observed. Finally, we have investigated the optical characteristics of such regular black holes with sub-Planckian curvature under three toy emission models. Under the model starting from the ISCO emission, the intensity exhibits three peaks corresponding to the photon ring, the lensed ring and the direct emission, respectively. Since the range of $b$ corresponding to the photon ring and the lensed ring is small, they do not contribute much to the total flux, and the direct emission accounts for the major part of the observed flux. It is noticed that the value of all three peaks decreases as $\alpha_0$ increases, and thus the region of the maximal brightness in  two-dimensional observational map becomes darker. In the model with emission from the photon sphere, the intensity peaks of the lensed ring and the photon ring are highly overlapped and indistinguishable, and their peaks are much higher than the direct emission peaks. Nevertheless, the direct emission involves in a larger range of impact parameter $b$, so the direct emission accounts for the major part of the observed flux. As $\alpha_0$ increases, the corresponding intensity peaks drop down significantly such that the maximal observed brightness of 2D image decreases. Finally, in the model where the emission starts from the horizon, the direct emission is still the main contributor to the observed flux, and the increase of the parameter $\alpha_0$ causes a decrease of the maximal observed brightness of the peak. 
\par
Finally, we have compared the images of three different black holes surrounded by the accretion disk, namely Schwarzschild black hole, Bardeen black hole and the new regular black hole with Minkowskian core. First, we have analyzed the behavior of light rays around three types of black holes and found that the halo and photon rings of black hole with Minkowskian core are wider than those with dS core, which we believe is caused by the stronger attraction of black holes with Minkowskian core to photons. We have also discussed the transfer functions of these black holes and found that the deviation parameter $\alpha_0$ tends to suppress the demagnification factor in the second and third transfer function more obviously for regular black holes with Minkowskian core, which means that the lensed ring and photon ring could be more easily to be observed. Finally, we have compared the observed phenomena of three types of black holes under three emission models. It is found that for all three types of black holes, the contribution of the coalescence and photon rings to the observed flux is small, and the main source of the observed flux comes from the direct emission. In all toy models, Schwarzschild black hole has a larger size of central dark region than the remaining two black holes with the same mass. The observed intensity as well as the peaks of intensity is higher. For the black hole with Minkowskian core, its central dark region size is smaller than that of the black hole with dS core, and the peaks of intensity are smaller. The above results indicates that different types of black holes produce different images and shadows indeed, and our investigation on this topic has provided more theoretical foundation for distinguishing different sorts of black holes by astronomical observation.

For simplicity, throughout this paper we have only considered the regular black holes with spherical symmetry. It is expected to extend the above analysis to the rotating regular black holes with Minkowskian core.  
\section*{Acknowledgments}

We are very grateful to Meng-He Wu, Zhong-Wen Feng and Xiao-Mei Kuang for helpful discussions. 
This work is supported in part by the Natural Science Foundation
of China under Grant No.~12035016 and 12275275. It is also supported by Beijing Natural Science Foundation under Grant No. 1222031, and by Sichuan Youth Science and Technology Innovation Research Team with Grant No.~21CXTD0038, by the National Natural Science Foundation of China with Grant No. 11903025 and the National Natural Science Foundation of Sichuan Province with Grant No. 2022NSFSC1833 and No. 2023NSFSC1352.

\end{document}